\pdfoutput=1

\documentclass[usenatbib]{mnras}

\usepackage[T1]{fontenc}
\usepackage{ae,aecompl}

\usepackage{graphicx}
\usepackage{color}
\usepackage{txfonts}
\usepackage{tabularx}
\usepackage{here}
\usepackage{float}
\usepackage{threeparttable}
\usepackage{tabu}
\usepackage{dcolumn}
\usepackage{enumitem}
\usepackage{booktabs}

\newcommand{\oi}{[O\,{\sc i}]}
\newcommand{\oii}{[O\,{\sc ii}]}
\newcommand{\oiii}{[O\,{\sc iii}]}
\newcommand{\oiv}{[O\,{\sc iv}]}

\newcommand{\NI}{[N\,{\sc i}]}
\newcommand{\nii}{[N\,{\sc ii}]}
\newcommand{\niii}{N\,{\sc iii}]}

\newcommand{\sii}{[S\,{\sc ii}]}
\newcommand{\siii}{[S\,{\sc iii}]}
\newcommand{\siv}{[S\,{\sc iv}]}

\newcommand{\hei}{He\,{\sc i}}
\newcommand{\heii}{He\,{\sc ii}}

\newcommand{\neii}{[Ne\,{\sc ii}]}
\newcommand{\neiii}{[Ne\,{\sc iii}]}
\newcommand{\neiv}{[Ne\,{\sc iv}]}
\newcommand{\nev}{[Ne\,{\sc v}]}

\newcommand{\arii}{[Ar\,{\sc ii}]}
\newcommand{\ariii}{[Ar\,{\sc iii}]}
\newcommand{\ariv}{[Ar\,{\sc iv}]}
\newcommand{\arv}{[Ar\,{\sc v}]}

\newcommand{\clii}{[Cl\,{\sc ii}]}
\newcommand{\cliii}{[Cl\,{\sc iii}]}
\newcommand{\cliv}{[Cl\,{\sc iv}]}
\newcommand{\kriii}{[Kr\,{\sc iii}]}
\newcommand{\kriv}{[Kr\,{\sc iv}]}

\newcommand{\xeiii}{[Xe\,{\sc iii}]}
\newcommand{\xeiv}{[Xe\,{\sc iv}]}

\newcommand{\Siii}{[Si\,{\sc ii}]}

\newcommand{\mgv}{[Mg\,{\sc v}]}
\newcommand{\mgiv}{[Mg\,{\sc iv}]}
\newcommand{\feii}{[Fe\,{\sc ii}]}
\newcommand{\feiii}{[Fe\,{\sc iii}]}
\newcommand{\kiv}{[K\,{\sc iv}]}
\newcommand{\ci}{C\,{\sc i}}
\newcommand{\cii}{C\,{\sc ii}}
\newcommand{\ciii}{C\,{\sc iii}}
\newcommand{\civ}{C\,{\sc iv}}
\newcommand{\rbiv}{[Rb\,{\sc iv}]}

\newcommand{\ha}{H$\alpha$}
\newcommand{\hb}{H$\beta$}

\newcommand{\hi}{H\,{\sc i}}

\newcommand{\fiv}{[F\,{\sc iv}]}
\newcommand{\fv}{[F\,{\sc v}]}
\newcommand{\pii}{[P\,{\sc ii}]}
\newcommand{\piii}{[P\,{\sc iii}]}

\usepackage{color} 

\newcommand{\te}{$T_{\rm e}$}
\newcommand{\Ne}{$n_{\rm e}$}

\title[Physical properties of the PN J900]
{Physical properties of the fluorine and neutron-capture element rich PN 
Jonckheere900\thanks{Based on observations made
with Korea Astronomy and Space Science Institute
(KASI) BOAO 1.8 m telescope under the programme ID:
S12A-B17 (PI: M.~Otsuka).}
}

\author[Otsuka \& Hyung]
{
\begin{minipage}{1.0\linewidth}
 Masaaki Otsuka$^{1}$\thanks{E-mail: otsuka@kusastro.kyoto-u.ac.jp}
 and  
 Siek Hyung$^{2}$
\end{minipage}
\\
\\
  \begin{minipage}{1.0\linewidth}
  $^{1}$Okayama Observatory, Kyoto University, Kamogata, Asakuchi, 
   Okayama, 719-0232, Japan\\
  $^{2}$School of Science Education (Astronomy),
  Chungbuk National University, CheongJu, Chungbuk 28644, Korea
\end{minipage}
}

\begin{document}

\date{}

\pagerange{\pageref{firstpage}--\pageref{lastpage}} \pubyear{2019}

\maketitle

\label{firstpage}

\begin{abstract}
We performed detailed spectroscopic analyses of a young C-rich planetary nebula (PN) Jonckheere900 (J900) 
in order to characterise the properties of the central star and nebula. 
 Of the derived 17 elemental abundances, we present the first determination of eight elemental abundances. 
We present the first detection of the [F\,{\sc iv}]\,4059.9\,{\AA}, [F\,{\sc v}]\,13.4\,{\micron}, 
and [Rb\,{\sc iv}]\,5759.6\,{\AA} lines in J900. 
J900 exhibits a large enhancement of F and neutron-capture elements Se, Kr, Rb, and Xe. We investigated the physical 
conditions of the H$_{2}$ zone using the newly detected mid-IR H$_{2}$ lines while also using the the previously measured near-IR H$_{2}$ lines, which indicate warm ($\sim$670\,K) and hot ($\sim$3200\,K) temperature regions. We built the spectral energy distribution (SED) model to be consistent with all the observed quantities. We found that about 67\,\% of all dust and gas components ($4.5\times10^{-4}$\,M$_{\sun}$ 
and 0.83\,M$_{\sun}$, respectively) exists beyond the ionisation front, indicating critical importance of photodissociation regions in understanding stellar mass loss. The best-fitting SED model indicates that the progenitor evolved from an initially $\sim$2.0\,M$_{\sun}$ star which had been in the course of the He-burning shell phase. Indeed, 
the derived elemental abundance pattern is consistent with that predicted by a asymptotic giant branch star nucleosynthesis model for a 2.0\,M$_{\sun}$ star with $Z=0.003$ and partial mixing zone mass of 6.0$\times10^{-3}$\,M$_{\sun}$. Our study demonstrates how accurately determined abundances of C/F/Ne/neutron-capture elements and gas/dust masses help us understand the origin and the internal evolution of the PN progenitors. 
\end{abstract}


\section{Introduction}
\setcounter{footnote}{2}

Planetary nebulae (PNe) represent the final evolutionary 
stage of low-to-intermediate mass stars (initially $1-8$\,M$_{\sun}$). 
During their evolution, such stars lose a large amount 
of their mass, which is atoms nucleosynthesised in the progenitors, 
and molecules and dust grains form in stellar winds. 
The stars become white dwarfs (WDs), and the mass loss is injected into interstellar medium (ISM) 
of their host galaxy. PNe greatly contribute to enriching 
galaxies materially and chemically; they are main suppliers of heavy elements with atomic masses $Z > 30$ 
in galaxies \citep[][references therein]{Karakas:2016ab} as well as of C and C-based dust/molecules.

According to current stellar evolutionary models of low-to-intermediate mass stars, 
such elements with $Z > 30$ (or atomic masses $A > 56$) 
are synthesised by the neutron ($n$)-capture process during the thermal-pulse (TP) asymptotic 
giant branch (AGB) phase \citep[see the reviews by][]{Busso:1999aa,Karakas:2014aa}. 
At low $n$-flux, $n$-capturing is a slow 
process ($s$-process). $n$-capturing is a rapid process ($r$-process) at 
high $n$-flux in massive stars. Free $n$ in low mass stars (initially $\lesssim 4$\,M$_{\sun}$) 
are released mostly by the $^{13}$C($\alpha$,$n$)$^{16}$O reaction in the He-rich intershell, and these $n$ are captured by Fe-peak nuclei. Fe-peak 
nuclei undergo subsequent $n$-captures and $\beta$-decays to transform into heavier elements. The synthesised elements are brought to the stellar surface by the third dredge-up that takes place in the late AGB phase. The AGB elemental abundance yields depend on not only the initial mass and metallicity but also unconstrained parameters such as mixing length, the number of TPs, and the so-called $^{13}$C-pocket mass \citep{Gallino:1998aa,Karakas:2014aa}, 
which is an additional $n$-source and is formed by mixing of the bottom of the H-rich 
convective envelope into the outermost region of the He-rich intershell. 
The $^{13}$C pocket mass is highly sensitive to the enhancement of He-shell burning products such as fluorine (F), Ne, and $s$-elements, as suggested by recent observational and theoretical works \citep[e.g.,][]{Abia:2015aa,Karakas:2018aa,Lugaro:2012aa,Shingles:2013aa,Raai:2012aa}. 
However, the formation of the $^{13}$C pocket and its extent in mass in the He-intershell 
\citep{Karakas:2009aa} is unknown. These elements have been measured in PNe. Hence, 
PN elemental abundances would be effective in constraining internal evolution of the PN progenitors.

After \citet{Pequignot:1994aa} reported the possible detection of 
$s$-elements Se/Br/Kr/Rb/Ba and $r$-element Xe in the PN 
NGC7027, $n$-capture elements have been intensively measured in 
near solar metallicity PNe \citep{Dinerstein:2001aa,Sterling:2002aa,
Sharpee:2007aa,Sterling:2008aa,Miszalski:2013aa,Garcia-Rojas:2015aa,
Madonna:2017aa,Sterling:2017aa,Madonna:2018aa}. 
However, the enhancement of $n$-capture elements is not well understood for low metallicity PNe \citep[e.g.,][]{Otsuka:2009aa,Otsuka:2010aa,Otsuka:2013ab,Otsuka:2015aa}. 
Low-metallicity PNe are generally located in the outer disk of the galaxy, 
assuming all stars evolve in-situ. Such PNe are important for understanding the 
chemical evolution of the Milky Way disk through time-variations of metallicity gradients \citep[e.g.,][]{Maciel:2010aa,Maciel:2013aa,Stanghellini:2010aa,Stanghellini:2018aa}. 
The metallicity gradients can be pinned down by the accurate  determination of elemental abundances of PNe located far from the Galactic centre. For these reasons, we have studied the physical properties of PNe in the outer disk of 
the Milky Way.

Amongst our sample, J900 (PN G194.2+02.5) is an intriguing PN in terms of elemental abundances, molecular gas, and dust. 
J900 has a nebula extended along multiple directions, as can be seen 
in Fig.\,\ref{F-image}. So far, the abundances of the nine elements have been measured \citep[e.g.,][]{Aller:1983aa,Kingsburgh:1994aa,Kwitter:2003aa}. 
\citet{Sterling:2008aa} were the first to add 
the detection of {\kriii}\,2.199\,{\micron} and [Se\,{\sc iv}]\,2.287\,{\micron} lines. Subsequently, \citet{Sterling:2009aa} 
detected several Kr and Xe lines in optical spectra, although they did not calculate Kr and Xe ionic/elemental abundances using these optical emission. \citet{Shupe:1995aa} demonstrated that the molecular 
hydrogen (H$_{2}$) image of the H$_{2}$ $v=1-0$ S(1) line at 2.12\,{\micron} is distributed along the direction perpendicular to the nebular bi-polar axis seen in the optical image. \citet{Huggins:1996aa} tentatively detected the CO $J=2-1$ line at 230.54\,GHz. \citet{Aitken:1982aa} fitted the mid-IR $8-13$\,{\micron} spectrum using the emissivities of the graphite grains and 
the polycyclic aromatic hydrocarbons (PAHs) at 8.65 and 11.25\,{\micron} seen in the Orion Bar. Despite substantial efforts, 
however, the physical properties of J900 are not yet understood well.

In this paper, we analyse the UV-optical high-dispersion echelle spectrum taken using Bohyunsan Optical Astronomy Observatory (BOAO) 1.8-m 
telescope/Bohyunsan Echelle Spectrograph \citep[BOES;][]{Kim:2002aa} and archived spectroscopic and photometric data to investigate the physical properties of 
J900. We organise the next sections as follows. In \S\,\ref{S:obs}, we describe our observations, archival spectra and photometry, and the data reduction. In \S\,\ref{S:ana}, we present our elemental abundance analysis. Here, we report the detection of 
F, Kr, Rb, and Xe forbidden lines in the BOES spectrum and F in the mid-IR \emph{Spitzer} Infrared Spectrograph \citep[IRS;][]{Houck:2004aa} spectrum. We examine the physical conditions of H$_{2}$ lines. 
In \S\,\ref{S:cloudy}, we construct the spectral energy distribution (SED) model using the photoionisation code {\sc Cloudy} v.13.05 
\citep{Ferland:2013aa} to accommodate all the derived quantities. In \S\,\ref{S:dis}, we discuss the origin 
and evolution of J900 by comparison of the derived nebular/stellar properties with AGB nucleosynthesis models, post-AGB evolutionary 
models, and other PNe. Finally, we summarise the present work in \S\,\ref{S:summary}.

\section{Dataset and reduction} \label{S:obs}

  \subsection{BOAO/BOES UV-optical spectrum}

\begin{figure}
   \centering
   \includegraphics[width=\columnwidth,clip]{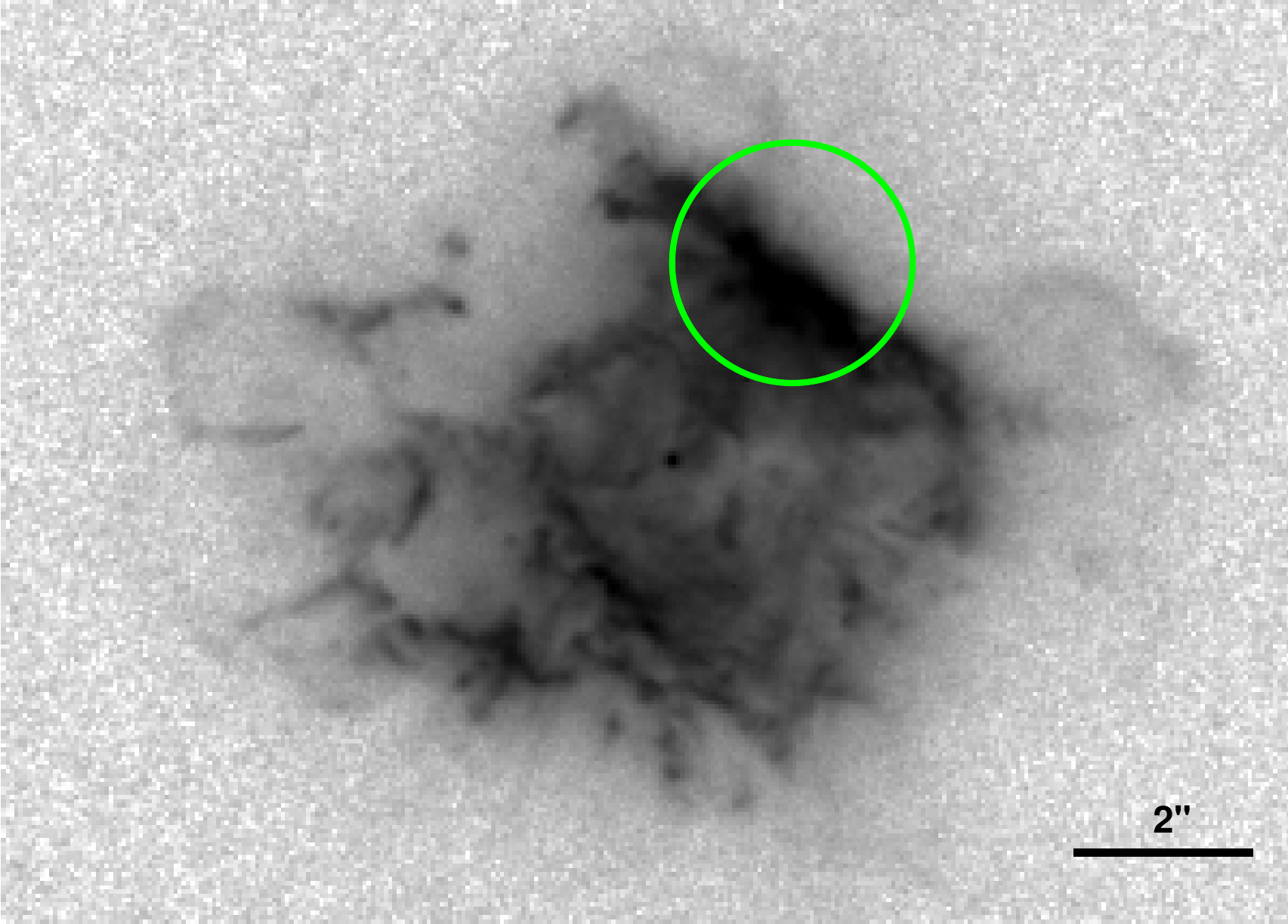}
\vspace{-10pt}
   \caption{The \emph{HST}/WFPC2 PC F658N image of J900.
   North is up and east is left. The green circle indicates 
 the dimension and location of BOES's fibre on this PN (central coordinate
 is RA(2000.0) = 06:25:57.2 and DEC(2000.0) = +17:47:29.5).
    }
   \label{F-image}
   \end{figure}

 We secured the $3700-9300$\,{\AA} spectra
 using the fibre-fed BOES attached to the 1.8-m telescope at BOAO, S. Korea on 2012 February 13.
 BOES has a single 2K$\times$4K pixel E2V-4482 detector. We put a fibre entrance with 2.8$''$ aperture at a position as shown in Fig.\,\ref{F-image}. We employed a single
 7200\,sec to register the faint emission lines and a 300\,sec exposure
 to avoid the saturation for  the strongest lines such as {\oiii} and {\ha}. 
 The seeing was $\sim$2.1$''$, the sky condition was stable,
 but sometimes thin clouds were passing. For flux calibration,
 we observed HR\,5501 at airmass = 1.23, while
 the airmass of J900 was $1.09-1.24$.
 For wavelength calibration and detector sensitivity correction,
 we took Th-Ar comparison and instrumental flat frames, respectively.
  
  We reduced the data using NOAO/{\sc IRAF} 
  echelle reduction package in a standard manner, including
  over-scan subtraction, scattered light
  subtraction between echelle orders. 
  We measured the average spectral resolution ($R$) of 42\,770 using the full
  width at half maximum (FWHM) of 2825 Th-Ar comparison
  lines. We removed 
  sky emission lines using ESO UVES sky emission spectrum 
  after Gaussian convolution to matching BOES spectral resolution. 
The signal-to-noise ratio for continuum varies $\sim$$2-8$ with an average $\sim$4 over $3720-9220$\,{\AA}.

\subsection{\emph{IUE} UV, \emph{AKARI}/IRC, and \emph{Spitzer}/IRS mid-spectra}

   \begin{figure*}
    \centering
    \includegraphics[width=0.80\textwidth,clip]{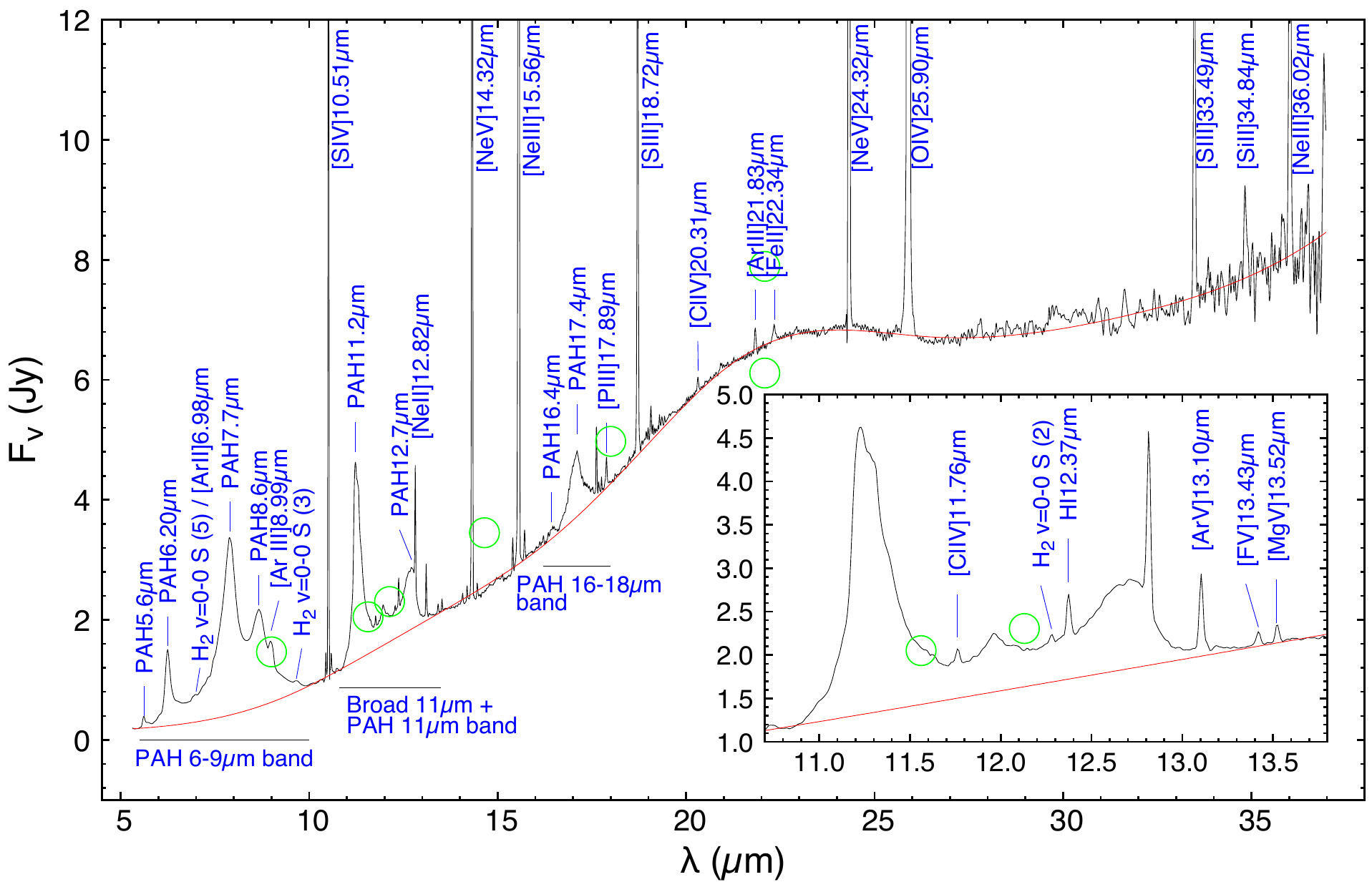}
\vspace{-5pt}
    \caption{The {\sl Spitzer}/IRS spectrum of J900. 
    The flux density is scaled up to matching with
    \emph{AKARI}, \emph{WISE}, and \emph{MSX} photometry (green circles).
    As a guide, we plot the continuum fit by 6-th order spline
    function (red line). The identified atomic lines, H$_{2}$, and
    PAH emission are indicated by the blue lines with labels. The
    $6-9$\,{\micron}, 11\,{\micron}, and $16-18$\,{\micron} PAH bands 
    are also indicated by the horizontal lines. The inset displays the closeup of the
    $10.7-13.8$\,{\micron} spectrum in order to show the detected weak
    atomic gas emission such as [F\,{\sc v}]\,13.43\,{\micron}.
    }
   \label{F-irsspec}
   \end{figure*}

 We chose the following dataset of the
 International Ultraviolet Explorer (\emph{IUE}):  
 SWP07965, 08677, and 53870 (these data cover $1150-1978$\,{\AA}, $R$$\sim$300) 
 and LWR06938 and 07429 (these cover $1852-3348$\,{\AA}, $R$$\sim$400)
  from Mikulski Archive for Space Telescopes (MAST). For each SWP
  and LWR datum, we did average combining, so we connected these two resultant spectra into the single $1150-3348$\,{\AA} spectrum.

 We utilised the \emph{AKARI}/Infrared Camera \citep[IRC;][]{Onaka:2007aa} $2.5-5.0$\,{\micron} spectrum ($R$$\sim$120) as presented in \citet{Ohsawa:2016aa}. We also analyse the \emph{Spitzer}/IRS spectra taken with 
 the Short-Low (SL, $5.2-14.5$\,{\micron}, $R$$\sim$60 -- 130), 
 Short-High (SH, $9.9-19.6$\,{\micron}, $R$$\sim$600),
 and Long-High modules (LH, $18.7-37.2$\,{\micron}, $R$$\sim$600). 
We downloaded the following Basic Calibrated Data (BCD):
 AORKEY 16921344, 20570368, 21458944, 24400640, 27018752, 27019008,
 28538880, 28539136, 33777152, and 33777408. We reduced these BCD
 using the data reduction packages
 {\sc SMART} v.8.2.9 \citep{Higdon:2004aa} and
 {\sc IRSCLEAN} v.2.1.1. We normalised the flux density 
 of the SL-, SH-, and LH-spectra in the overlapping wavelengths, and 
 obtained a single $5.2-37.2$\,{\micron} spectrum. Then, we scaled 
 the flux density of this spectrum to matching with
 \emph{AKARI}, \emph{WISE}, and \emph{MSX} photometry (green circles) 
 by a factor of 1.461 (see \S\,\ref{S-phot} and \ref{S-hb} for details). 
 The resultant spectrum with
 identifications of detected atomic and molecular gas emission lines is
 displayed in Fig.\,\ref{F-irsspec}.

  \subsection{Interstellar extinction correction}
 \label{S-flux}

 \begin{table}
   \renewcommand{\arraystretch}{0.80}
 \caption{Calculated $c$({\hb}) values in J900 using
 {\hi} Balmer (B$n_{\rm u}$, $n_{\rm u}$: the upper principal quantum number) and Paschen (P$n_{\rm u}$) 
series.
 }
 \centering
 \begin{tabular}{@{}l@{\hspace{5pt}}l@{\hspace{5pt}}
  c@{\hspace{8pt}}l@{\hspace{5pt}}l@{\hspace{5pt}}c@{}}
 \midrule
  Line &$\lambda_{\rm lab}$ ({\AA})&$c$({\hb})&Line
  &$\lambda_{\rm lab}$ ({\AA})&$c$({\hb})\\
 \midrule
  B10 & 3797.90   & 5.66(--1) $\pm$ 4.63(--2) & P14 & 8598.39& 5.44(--1) $\pm$ 3.71(--3) \\ 
  B9  & 3835.38   & 4.94(--1) $\pm$ 2.59(--2) & P13 & 8665.02& 5.62(--1) $\pm$ 4.89(--3) \\ 
  B3  & 6562.80   & 5.49(--1) $\pm$ 9.14(--3) & P12 & 8750.47& 5.48(--1) $\pm$ 4.11(--3) \\ 
  P15 & 8545.38	  & 5.52(--1) $\pm$ 6.56(--3) & P11 & 8862.78& 5.26(--1) $\pm$ 3.74(--3) \\ 
\midrule
\end{tabular}
\label{T-chb}
 \end{table}

 We measure emission line fluxes by multiple Gaussian component
 fitting. We correct
 the measured line fluxes $F$($\lambda$) to obtain the interstellar
 extinction corrected line fluxes $I$($\lambda$) using the formula;
 \begin{equation}
  I(\lambda) = F(\lambda)~\cdot~10^{c({\rm H\beta})(1 + f(\lambda))},
   \label{eq-1}
 \end{equation}
 \noindent where $f$($\lambda$) is the interstellar
 extinction function at $\lambda$ computed by the reddening law of
 \citet{Cardelli:1989aa}. Here, we adopt the average $R_{V} = 3.1$ in the Milky Way.
 $c$({\hb}) is the reddening coefficient defined as $\log$$F$({\hb})/$I$({\hb}).

 For the BOES spectrum, we calculate $c$({\hb})
 values (Table\,\ref{T-chb}) 
 by comparison of the observed Balmer and Paschen line ratios to {\hb} 
 with their theoretical values of \citet{Storey:1995aa} for the Case B 
 assumption in an electron temperature {\te} of 10$^{4}$\,K 
 and an electron density {\Ne} of 7000\,cm$^{-3}$ 
 (from the ratios of {\ariv}\,$F$(4711\,{\AA})/$F$(4740\,{\AA})
 and {\cliii}\,$F$(5517\,{\AA})/$F$(5537\,{\AA})), 
 and we adopt the average $c$({\hb}) value of $0.54 ~\pm~ 0.02$.
 For the \emph{IUE} spectrum, we adopt $c$({\hb})$ = 0.57 ~\pm~ 0.06$ calculated by
 comparison of the {\heii}\,$F$(1640\,{\AA})/$F$(2733\,{\AA})
 with the theoretical ratio (30.81) by \citet{Storey:1995aa} under the same assumption 
 applied for the BOES spectrum.
 Due to negligibly small reddening, we do not correct the extinctions for the measured emission line fluxes and flux densities at 
the \emph{AKARI} and \emph{Spitzer} wavelengths.

 \subsection{Line flux normalisation}

  In Appendix Table\,\ref{T-emission}, we list the identified lines. 
The successive columns are laboratory wavelengths,  
ions responsible for the line,  extinction parameters $f(\lambda)$, line intensities $I$($\lambda$) on the scale of  $I$({\hb}) = 100, and 1-$\sigma$ uncertainties of $I(\lambda)$. For the lines in the \emph{IUE} spectra, 
we normalise the line fluxes with respect to the He\,{\sc ii}\,1640\,{\AA}, perform reddening correction, and multiply a constant factor of 235.1, which is determined from the theoretical Case B $I$(1640\,{\AA})/$I$(4685\,{\AA}) ratio \citep[6.558,][]{Storey:1995aa} and the observed $I$({\heii}\,4685\,{\AA})/$I$({\hb}) of 35.85/100 
to express $I$({\hb}) = 100.
  For the lines in the \emph{Spitzer} spectrum, we normalise the line fluxes with
  respect to the line at 12.37\,{\micron}, which is the complex of
  {\hi}\,12.37\,{\micron} ($n = 6-7$, $n$ is the quantum number)
  and 12.39\,{\micron} ($n = 8-11$).
  Then, we multiply all the normalised line fluxes by 1.043 
  in order to express $I$({\hb}) = 100 since 
  the theoretical Case B $I$(12.37/12.39\,{\micron})/$I$({\hb}) of 
  1.043/100 \citep{Storey:1995aa}. 
  Similarly, we normalise the measured line fluxes in the \emph{AKARI} 
  spectrum with respect to the {\hi}\,4.05\,{\micron}, and then multiply 
  7.781 in order to express $I$({\hb}) = 100 as the theoretical 
  Case B $I$({\hi}\,4.05\,{\micron})/$I$({\hb}) of 7.781/100 
  \citep{Storey:1995aa}.

\subsection{Image and photometry data}
\label{S-phot}

 We collected the data taken from
 the Hubble Space Telescope (\emph{HST})/Wide Field and Planetary Camera2 (WFPC2), Isaac Newton Telescope (INT) 2.5-m/Wide Field Camera
 (WFC), United Kingdom Infra-Red Telescope (UKIRT) 3.8-m/Wide Field Camera (WFCAM),
  Midcourse Space Experiment \citep[\emph{MSX};][]{Egan:2003aa}, 
  Wide-field Infrared Survey Explorer \citep[\emph{WISE};][]{Cutri:2014aa}, 
 \emph{AKARI}/IRC \citep{Ishihara:2010aa}, and Far Infrared Surveyor (FIS) \citep{Yamamura:2010aa}. The detailed procedures are as follows.
 
  We downloaded the raw Sloan $g$ and $r$ band images
  taken with the WFC mounted on the 2.5-m INT from the Cambridge Astronomical Survey
  Unit (CASU) Astronomical Data Centre. We reduced these data
  using NOAO/{\sc IRAF} following a standard manner (i.e., bias subtraction,
  flat-fielding, bad-pixel masking, cosmic-ray removal, detector
  distortion correction, and sky subtraction), and performed  PSF fitting
  for the surrounding star subtraction. And then, we performed  
  aperture photometry for J900 and the standard star
  SA\,113-260 and Feige\,22 \citep{Smith:2002aa}. Similarly, 
we downloaded the reduced $J$, $H$, and $K$ images
  taken with the WFCAM mounted on the 3.8-m UKIRT from the UKIRT WFCAM Science Archive (WSA).
  We measured$J$, $H$, and $K$ band magnitudes of J900 based
  on our own photometry of 77 nearby field stars, and  
  converted these respective instrumental magnitudes
  into the $J$, $H$, and $K$ band magnitudes at the system of
  the Two Micron All Sky Survey \citep[2MASS;][]{Skrutskie:2006aa}. 

In Appendix Table\,\ref{T-photo}, we list the reddening corrected flux density
 $I_{\lambda}$ using eq\,(\ref{eq-1}) and $c$({\hb})$ = 0.54 ~\pm~ 0.02$ 
 (\S\,\ref{S-flux}). We skip extinction correction in 
 the photometry bands at the wavelengths longer than \emph{WISE} W1 band.
 For \emph{HST}/F555W, we perform photometry of the 
central star of the PN (CSPN) and the central star plus nebula.
  All the flux densities in the other bands are
 the sum of the radiation from the CSPN and nebula.

\subsection{H\mbox{$\beta$} flux of the entire nebula}

\label{S-hb}
  
  Using a constant factor of $1.461 ~\pm~ 0.169$, 
  we scale the \emph{Spitzer}/IRS spectrum to match with 
  \emph{WISE} W3/W4 bands, \emph{MSX} 12.13/14.65\,{\micron} 
  bands, and \emph{AKARI}/IRC S9W/L18W bands 
  of \citet{Ishihara:2010aa}. {\hi}\,12.37+12.39\,{\micron} line flux is
  $(2.322 ~\pm~ 0.285)$(--13)\,erg\,s$^{-1}$\,cm$^{-2}$, where 
  $A(-B)$ means $A\times10^{-B}$ hereafter.
  Thus, we obtain the extinction free {\hb} line flux of the entire nebula
  $(2.225 ~\pm~ 0.288)$(--11)\,erg\,s$^{-1}$\,cm$^{-2}$ using the
  theoretical Case B $I$({\hb})/$I$(12.37+12.39\,{\micron}) ratio.

  \section{Results} \label{S:ana}
  
  \subsection{Plasma diagnostics}
 \label{S-plasma}

   \begin{figure}
    \centering
   \includegraphics[width=\columnwidth,clip]{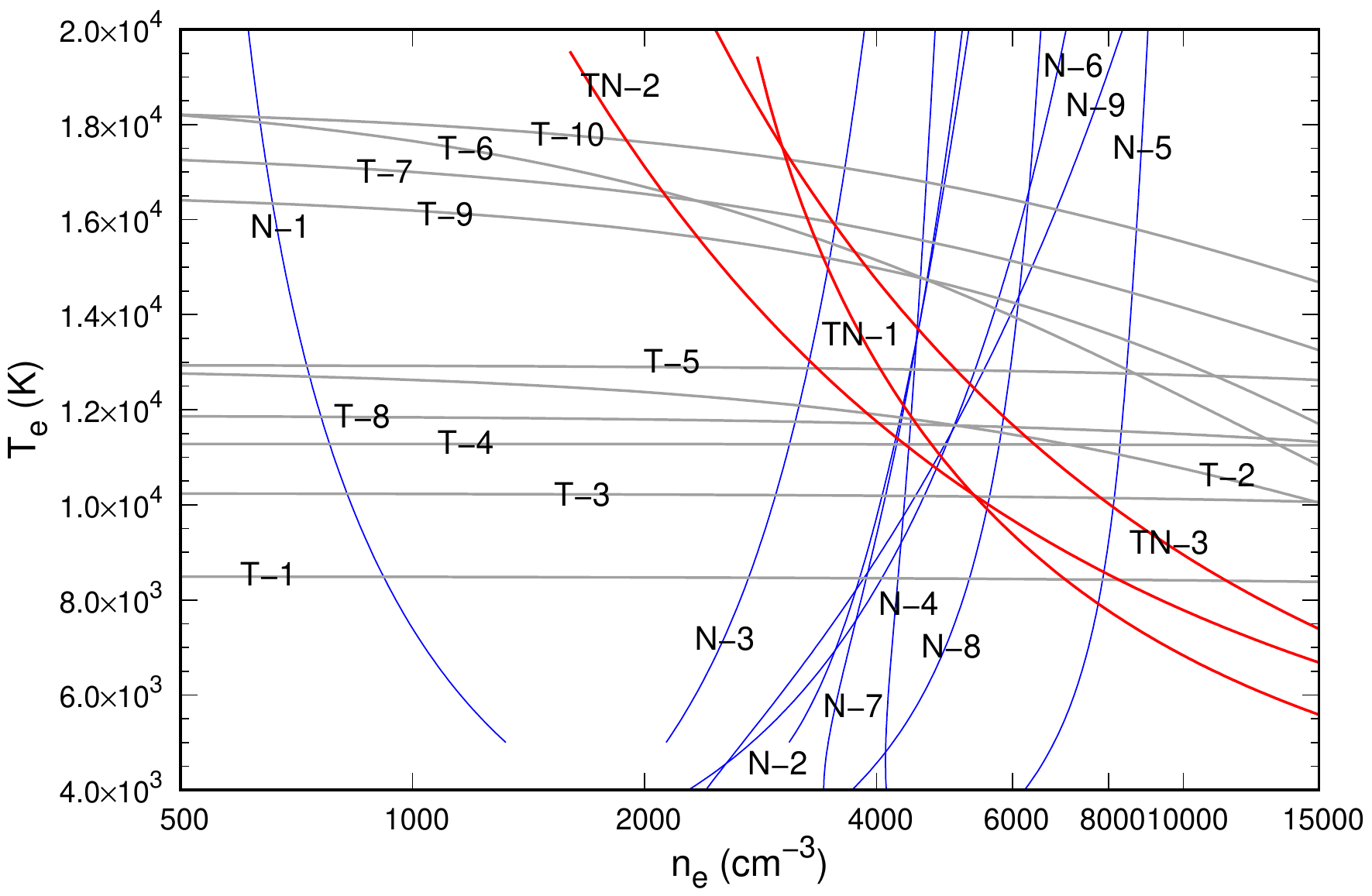}
   \vspace{-15pt}
   \caption{$n_{\rm e} - T_{\rm e}$ curves generated by the
   plasma diagnostic line ratios listed in Table\,\ref{T-neTe}. The blue,
   grey, and red lines are sensitive to {\Ne}, {\te}, and {\te}/{\Ne},
   respectively. The ID of each curve is the same used in
   Table\,\ref{T-neTe}.}
   \end{figure}

\begin{table}
\centering
\renewcommand{\arraystretch}{0.80}
\caption{Summary of plasma diagnostics of J900. 
\label{T-neTe}}
\begin{tabularx}{\columnwidth}{@{}@{\extracolsep{\fill}}
l@{\hspace{4pt}}l@{\hspace{-1pt}}D{p}{\pm}{-1}@{\hspace{-6pt}}D{p}{\pm}{-1}@{}}
    \midrule
 ID & CEL {\Ne}--diagnostic line & \multicolumn{1}{c}{Value}
 & \multicolumn{1}{c}{Result (cm$^{-3}$)}   \\ 
\midrule
N-1 & {\NI}\,5197\,{\AA}/5200\,{\AA}    & 1.346~p~0.053 & 920~p~110 \\ 
N-2 & {\sii}\,6717\,{\AA}/6731\,{\AA}   & 0.584~p~0.015 & 4360~p~490 \\ 
N-3 & {\oii}\,3726\,{\AA}/3729\,{\AA}   & 1.773~p~0.067 & 3290~p~390 \\ 
N-4 & {\siii}\,18.7\,{\micron}/33.5\,{\micron}  & 2.886~p~0.245 & 4660~p~590 \\ 
N-5 & {\cliii}\,5517\,{\AA}/5537\,{\AA} & 0.683~p~0.020 & 8400~p~710 \\ 
N-6 & {\cliv}\,11.8\,{\micron}/20.3\,{\micron}  & 1.361~p~0.178 & 4470~p~1590 \\ 
N-7 & {\ariv}\,4711\,{\AA}/4740\,{\AA}  & 0.879~p~0.016 & 6090~p~340 \\ 
N-8 & {\nev}\,14.3\,{\micron}/24.3\,{\micron}   & 1.749~p~0.147 & 6500~p~1240 \\ 
\midrule
 ID & CEL {\te}--diagnostic line & \multicolumn{1}{c}{Value}
	 & \multicolumn{1}{c}{Result (K)}    \\ 
\midrule
T-1 & {\oi}\,(6300/63\,{\AA})/5577\,{\AA}&96.719~p~7.203 & 8500~p~200 \\ 
T-2 & {\nii}\,(6548/83\,{\AA})/5755\,{\AA}&57.138~p~1.147 & 12\,060~p~120 \\ 
T-3 & {\siii}\,9069\,{\AA}/6313\,{\AA} & 9.023~p~0.282 & 10\,190~p~140 \\ 
T-4 & {\ariii}\,(7751/7135\,{\AA})/5191\,{\AA} & 125.043~p~9.734 & 11\,290~p~360 \\ 
T-5 & {\oiii}\,(4959/5007\,{\AA})/4363\,{\AA} & 98.464~p~1.587 & 12\,850~p~80 \\ 
T-6 & {\cliv}\,(11.8\,{\micron}/20.3\,{\micron})/7531\,{\AA} & 4.031~p~0.292 & 14\,920~p~880 \\ 
T-7 & {\ariv}\,(4711/40\,{\AA})/(7170/7262\,{\AA}) & 25.503~p~0.893 & 15\,120~p~420 \\ 
T-8 & {\neiii}\,15.8\,{\micron}/(3869/3967\,{\AA}) & 0.745~p~0.045 & 11\,640~p~230 \\ 
T-9 & {\arv}\,13.10\,{\micron}/6435\,{\AA} & 8.138~p~0.641 & 14\,140~p~650 \\ 
T-10 & {\neiv}\,(2422/25\,{\AA})/ & 77.386~p~4.338 & 16\,330~p~510 \\ 
    & (4714/15/24/26\,{\AA})\\
\midrule
 ID & CEL {\te}/{\Ne}--diagnostic line & \multicolumn{1}{c}{Value}
	 & \multicolumn{1}{c}{Result (K)}   \\ 
\midrule
TN-1 & {\sii}\,(4069/76\,{\AA})/(6717/31\,{\AA})  & 3.180~p~0.206 & 12\,180~p~1060 \\ 
TN-2 & {\oii}\,(3726/29\,{\AA})/(7320/30\,{\AA})  & 11.195~p~0.244 & 13\,000~p~240 \\ 
TN-3 & {\siii}\,(18.7/33.5\,{\micron})/9069\,{\AA} & 1.798~p~0.097 & 13\,920~p~960 \\ 
\midrule
      &RL {\te}--diagnostic line   & \multicolumn{1}{c}{Value}
	 &\multicolumn{1}{c}{Result (K)}\\
	     \midrule 
 &{\hei}\,7281\,{\AA}/6678\,{\AA} & 0.204~p~0.006 & 9290~p~280 \\
 &$[I_{\lambda}(8194\,\mbox{\AA}) - I_{\lambda}(8169\,\mbox{\AA})]/I$(P11)
    & 0.022~p~0.003 & 9140~p~2540 \\ 
\midrule
\end{tabularx}
\end{table}

   In Fig.\,\ref{T-neTe}, we plot {\Ne}--{\te} curves
   using the collisionally excited line (CEL) ratios listed in Table\,\ref{T-neTe}. 
   The adopted effective collision strengths $\Psi$($T$) and transition probabilities $A_{ji}$
  ($i,j$: energy level, $E_{j} > E_{i}$) are listed in Appendix Table\,\ref{atomf}. 
  For {\te}({\oiii}) (ID: T-5) and {\te}({\oii}) (TN-2) curves, 
   we subtract the recombination contributions of O$^{3+}$ and 
   O$^{2+}$ to the {\oiii}\,4363\,{\AA} and 
    the {\oii}\,7320/30\,{\AA} using eqs.\,(2) and (3) of
   \citet{Liu:2000aa}, respectively; the corrected 
   $I$({\oiii}\,4363\,{\AA}) and $I$({\oii}\,7320/30\,{\AA}) are
   $14.281 ~\pm~ 0.219$ and $10.606 ~\pm~ 0.149$, respectively. 
   Using the multiwavelength spectra 
   allows us to trace {\Ne} and {\te} values in the regions throughout 
   from the hot plasma close to the CSPN (e.g., {\nev}
   with ionisation potential (IP) = 97.1\,eV) to the cold and
   dusty photodissociation regions (PDRs).

   We determine the {\Ne} and {\te} values
   at the intersection of {\Ne}(X) and {\te}(X) or {\te}(X)/{\Ne}(X) diagnostic curves for an ion X. The
   {\te}({\oi}) and {\Ne}({\NI}) values have been determined from their
   curves; two {\te}({\siii}) values are derived based on the {\Ne}({\siii} curve; and the {\Ne}({\cliii}) is determined by the adoption of the average {\te} value between two {\te}({\siii}) ones. For {\te}({\ariii}), we utilised the other similar IP {\Ne}({\cliii}) curve. The {\te}({\oiii}) is determined from the {\Ne}({\cliv}) curve; the {\te}({\neiii}) is from the
   {\Ne}({\ariv}) curve; and the {\te}({\arv}) and {\te}({\neiv}) are found from the {\Ne}({\nev}).

   Our derived CEL {\te} and {\Ne} are comparable 
   to previous works, e.g., \citet{Kwitter:2003aa} and 
   \citet{Kingsburgh:1992aa,Kingsburgh:1994aa}: 
   {\te}({\oiii}) = 11\,600\,K, 
   {\te}({\siii}) = 13\,000\,K,
   {\te}({\nii}) = 11\,500\,K, 
   {\te}({\oii}) = 10\,300\,K, 
   {\te}({\sii}) = 7800\,K, and 
   {\Ne}({\sii}) = 3600\,cm$^{-3}$ (10\,$\%$ uncertainty) 
   by \citet{Kwitter:2003aa}; 
   {\te}({\oiii}) = 12\,000$^{+400}_{-200}$\,K, 
   {\Ne}({\oii}) = 3980$^{+790}_{-650}$\,cm$^{-3}$,
   {\Ne}({\ariv}) = 8240$^{+7000}_{-3600}$\,cm$^{-3}$ by    
   \citet{Kingsburgh:1992aa,Kingsburgh:1994aa}. 
   
   We compute {\te}({\hei}) using the
   {\hei}\,$I$(7281\,{\AA})/$I$(6678\,{\AA}) ratio by adopting the effective 
   recombination coefficient $\alpha_{ji}$ listed in Appendix Table~\ref{T-R-atomic}.
   We adopted $\alpha_{ji}$ of \citet{Benjamin:1999aa} for {\Ne} = 10$^{4}$\,cm$^{-3}$. {\te}(PJ) is calculated 
   from the Paschen continuum discontinuity
   using eq.\,(7) of \citet{Fang:2011aa} and the 
   ionic He$^{+}$ and He$^{2+}$ abundances (Appendix Table\,\ref{T-CEL-RL-abund}) 
   under the derived {\te}({\hei}).
   Due to high uncertainty of {\te}(PJ), we calculate the 
   recombination line (RL) C$^{2+,3+,4+}$ abundances by adopting {\te}({\hei}).

  \subsection{\label{S-abund} Ionic abundance derivations}

  \begin{table*}
   \centering
     \renewcommand{\arraystretch}{0.80}
\caption{Comparison of our derived ionic abundances in J900 with 
\citet[AC83]{Aller:1983aa}, \citet[KB94]{Kingsburgh:1994aa},
  and \citet[K03]{Kwitter:2003aa}. {\te} adopted for calculations of each ionic abundance in our work, KB94, and K03 are listed in columns 
(3), (6), and (8), 
   respectively. 
   }
\begin{tabularx}{\textwidth}{@{}@{\extracolsep{\fill}}lD{p}{\pm}{-1}
  D{p}{\pm}{-1}cccD{p}{\pm}{-1}D{p}{\pm}{-1}@{}}
\midrule
Ion& \multicolumn{1}{c}{Ours} & \multicolumn{1}{c}{Adopted {\te} (K)} &
 \multicolumn{1}{c}{AC83} & \multicolumn{1}{c}{KB94}  &
 \multicolumn{1}{c}{Adopted {\te} (K)}
 & \multicolumn{1}{c}{K03} & \multicolumn{1}{c}{Adopted {\te} (K)} \\
 (1)&
 (2)&
 \multicolumn{1}{c}{in Ours (3)}&
 (4)&
 (5)&
 \multicolumn{1}{c}{in KB94 (6)}&
 (7)&
 \multicolumn{1}{c}{in K03 (8)}\\
\midrule
He$^{+}$  & 7.74(-2) ~p~ 9.95(-3) & 9290 ~p~ 280 & 6.00(--2) & 8.20(--2) &10\,000?   & 6.26(-2) ~p~ 1.88(-2) & 11\,500 ~p~ 1150 \\ 
He$^{2+}$ & 3.11(-2) ~p~ 2.33(-3) & 9290 ~p~ 280 & 3.80(--2) & 4.02(--2) &10\,000?   & 3.25(-2) ~p~ 9.75(-3) & 11\,500 ~p~ 1150 \\ 
C$^{2+}$ (RL) & 1.05(-3) ~p~ 1.23(-4) & 9290 ~p~ 280 & 9.63(--4) &  &  &  &  \\ 
C$^{2+}$ (CEL)& 8.16(-4) ~p~ 2.24(-4) & 11\,800 ~p~ 490 & 7.68(--4) & 1.09(--3) & 12\,000 &  &  \\ 
C$^{3+}$ (CEL)& 1.10(-4) ~p~ 3.43(-5) & 14\,140 ~p~ 650 & 2.61(--4) & 3.64(--4) & 13\,000 &  &  \\ 
N$^{+}$   & 6.80(-6) ~p~ 1.37(-7) & 12\,060 ~p~ 120 & 4.76(--6) & 9.44(--6) & 9900 & 5.40(-6) ~p~ 1.62(-6) & 11\,500 ~p~ 1150 \\ 
O$^{0}$   & 3.08(-5) ~p~ 2.12(-6) & 8500 ~p~ 200 &  &  &  & 7.21(-6) ~p~ 2.16(-6) & 11\,500 ~p~ 1150 \\ 
O$^{+}$   & 2.47(-5) ~p~ 6.41(-7) & 13\,000 ~p~ 240 & 2.05(--5) & 8.48(--5) & 9900 & 1.88(-5) ~p~ 5.64(-6) & 11\,500 ~p~ 1150 \\ 
O$^{2+}$  & 1.67(-4) ~p~ 1.88(-6) & 12\,850 ~p~ 80 & 2.12(--4) & 2.40(--4) & 12\,000 & 2.14(-4) ~p~ 6.42(-5) & 11\,600 ~p~ 1160 \\ 
Ne$^{2+}$ & 7.17(-5) ~p~ 2.79(-6) & 11\,640 ~p~230 & 4.85(--5) & 5.53(--5) & 12\,000 & 5.34(-5) ~p~ 1.60(-5) & 11\,600 ~p~ 1160 \\ 
Ne$^{3+}$ & 8.71(-6) ~p~ 1.64(-6) & 16\,330 ~p~ 510 & 2.90(--5) &  &   &  &  \\ 
Ne$^{4+}$ & 3.50(-6) ~p~ 2.00(-7) & 16\,300 ~p~ 510 & 3.80(--6) &  &   &  &  \\ 
S$^{+}$   & 2.48(-7) ~p~ 1.28(-8) & 12\,180 ~p~ 1060 & 1.39(--8) & [1.87(--6)] & 9900 & 1.68(-7) ~p~ 5.04(-8) & 11\,500 ~p~ 1150 \\ 
S$^{2+}$  & 1.77(-6) ~p~ 7.57(-8) & 12\,060 ~p~ 550 & 1.29(--6) &  &  & 1.54(-6) ~p~ 4.62(-7) & 13\,000 ~p ~ 1300 \\ 
Cl$^{2+}$ & 3.38(-8) ~p~ 2.19(-9) & 11\,800 ~p~ 490 & 2.40(--8) &  &  &  &  \\ 
Cl$^{3+}$ & 2.11(-8) ~p~ 1.06(-9) & 14\,920 ~p~ 880 & 6.10(--8) &  &   & 2.45(-8) ~p~ 7.35(-9) & 11\,600 ~p~ 1160 \\ 
Ar$^{2+}$ & 6.02(-7) ~p~ 2.79(-8) & 11\,290 ~p~ 360 & 3.05(--7) &  &  & 4.93(-7) ~p~ 1.48(-7) & 11\,600 ~p~ 1160 \\ 
Ar$^{3+}$ & 1.80(-7) ~p~ 5.23(-9) & 15\,120 ~p~ 420 & 6.32(--7) & 1.71(--7) & 13\,000 & 2.57(-7) ~p~ 7.65(-8) & 11\,600 ~p~ 1160 \\ 
Ar$^{4+}$ & 4.72(-8) ~p~ 2.90(-9) & 14\,140 ~p~ 650 & 1.08(--7) & 3.11(--7) & 14\,270 &  &  \\ 
\midrule
   \end{tabularx}
\label{T-ionicB}
  \end{table*}

 \subsubsection{The CEL ionic abundances}
  
  We solve atomic multiple energy-level 
  population models for each ion by adopting 
  {\Ne}--{\te} pair (Appendix Table\,\ref{T-Atene}) to calculate the CEL ionic abundances. 
  In Appendix Table\,\ref{T-CEL-RL-abund}, we summarise the resultant CEL ionic
  abundances with 1-$\sigma$ uncertainty. The first time measurement of 21 species out of our calculated 
  37 CEL ionic abundance is performed in this study.
  When more than one line for a targeting ion 
  are detected, we compute ionic abundances using each line, 
  and then we adopt the average value between the derived ionic abundances. 
  The last line in boldface shows the adopted value.

  Sixteen CEL ionic abundances calculated by previous works are
  summarised in Table\,\ref{T-ionicB}. There are five measurements
  including ours since the pioneering work by \citet{Aller:1983aa}, who
  used the image-tube scanner and \emph{IUE} spectra and adopted
  $c$({\hb}) = 0.83 (no information on the adopted {\Ne} and {\te} though).
  The measurements of \citet{Kwitter:2003aa} (their adopted
  $c$({\hb}) = 0.48) based on the $3600-9600$\,{\AA}
  spectrum reasonably agree with ours, except for O$^{0}$, S$^{2+}$, and
  Ar$^{3+}$. The difference in these ionic abundances is mainly
  due to the adopted {\te} as listed in the column (6) of Table\,\ref{T-ionicB}.  O$^{0}$, S$^{2+}$, and Ar$^{3+}$ are 
  ($1.25 ~\pm~ 0.14$)(--5), ($2.39 ~\pm~ 0.22$)(--6), and 
  ($2.91 ~\pm~ 0.12$)(--7), which are close to the relevant values of
  \citet{Kwitter:2003aa}. The difference between
  \citet{Kingsburgh:1994aa} and ours is attributable to the adopted $c$({\hb})  (0.80). 
  In particular, the difference $c$({\hb}) affects the intensity 
  corrections of the lines in the UV wavelength. 
  \citet{Kingsburgh:1994aa} reported
  that S$^{+}$ abundance was determined using
  {\sii}\,6717/31\,{\AA} lines, but we were not able to find the information from their
  Table\,3, so we could not find which caused the difference in their derived S$^{+}$ and elemental S
  abundances. The previous works calculated 
  the CEL ionic abundances using very limited 
  numbers of the {\Ne}--{\te} pairs, whereas we select {\Ne}--{\te} optimised  for each ionic abundance. 
  Thus, we conclude that our derived CEL ionic
  abundances are more reliable than ever.

  \subsubsection{Detection of F, Rb, Kr, and Xe lines and their abundances \label{S-flo}}

   \begin{figure*}
\centering
\begin{tabular}{c@{\hspace{2pt}}c@{\hspace{2pt}}c}
\includegraphics[width=0.28\textwidth,clip]{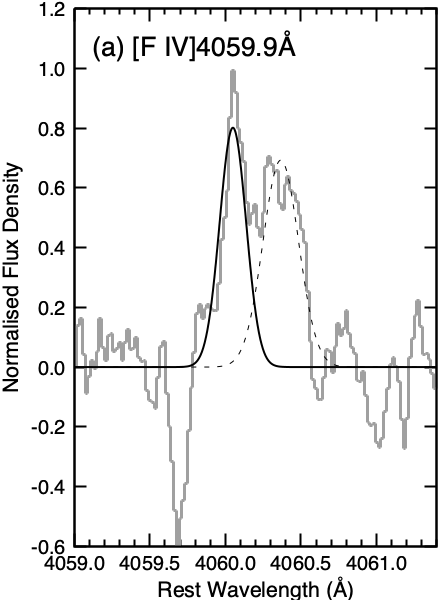}&
\includegraphics[width=0.28\textwidth,clip]{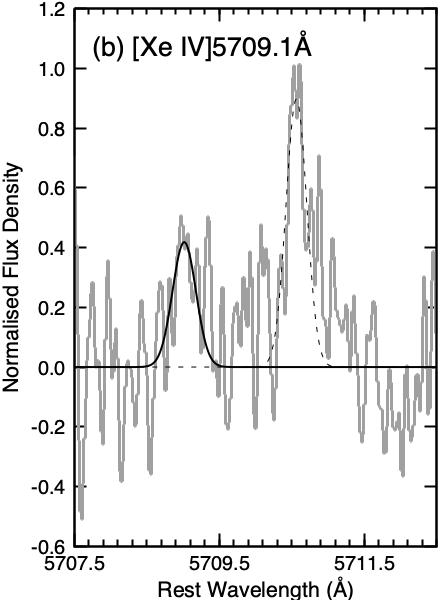}&
\includegraphics[width=0.28\textwidth,clip]{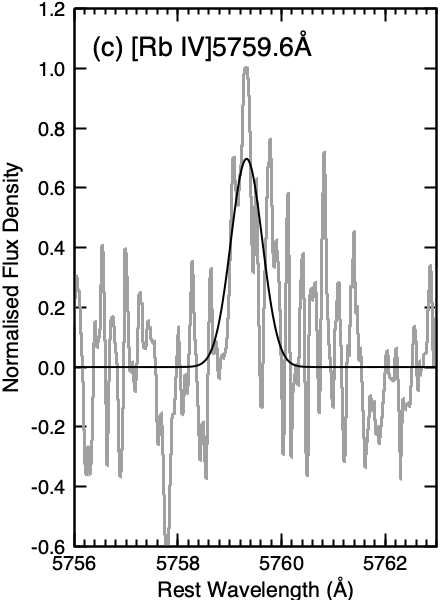}\\
\includegraphics[width=0.28\textwidth,clip]{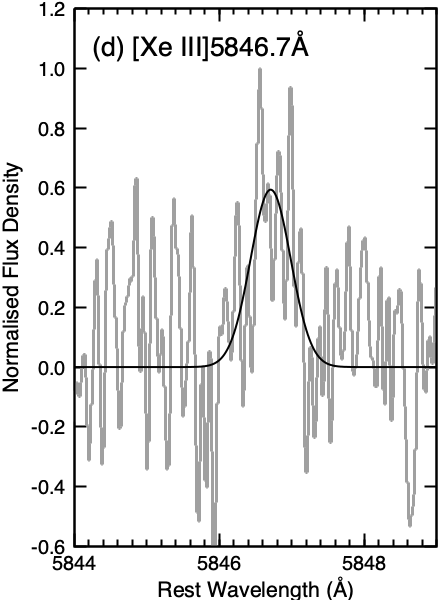}&
\includegraphics[width=0.28\textwidth,clip]{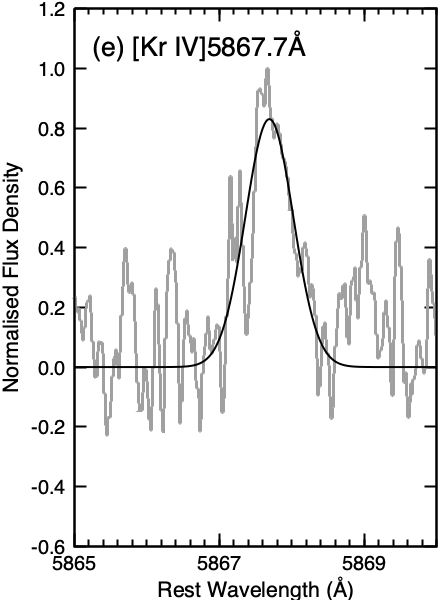}&
\includegraphics[width=0.28\textwidth,clip]{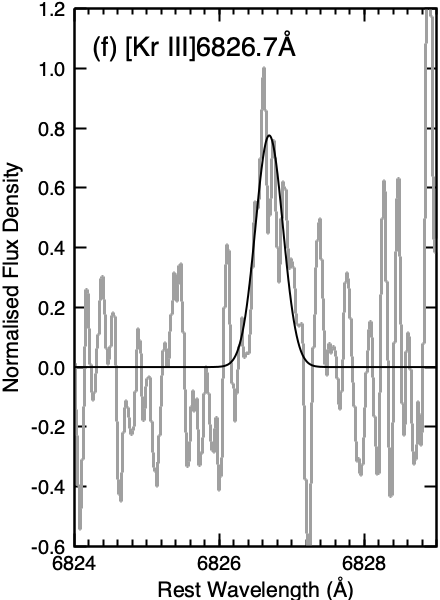}
\end{tabular}
  \caption{The identified emission lines of F, Kr, Rb, and Xe. The grey and black lines 
   are the observed spectrum and the Gaussian fitting result, respectively. The dotted line profile 
    in panels (a) and (b) is the fitting result of N\,{\sc ii}\,4060.20\,{\AA} and 5710.77\,{\AA} lines, 
   respectively. The contribution from He\,{\sc ii} lines to {\rbiv}\,5759.6\,{\AA} and {\xeiii}\,5846.7\,{\AA} 
   is subtracted. Contamination of skylines to {\kriii}\,6826.7\,{\AA} is subtracted out. 
   }
   \label{F-spro}
\end{figure*}

  F abundances have been calculated in only 16 PNe, so far 
  \citep{Zhang:2005ab,Otsuka:2008aa,Otsuka:2011aa,Otsuka:2015aa}. 
The FWHM of the detected line centred at 
  $\lambda_{\rm c} = 13.42$\,{\micron} is 0.025\,{\micron},
  so we consider
  $[$Mn\,{\sc vi}$]$\,13.40\,{\micron}, 
  {\hei}\,13.42-13.44\,{\micron}, and 
  {\fv}\,13.43\,{\micron} lines as candidates. The possibility of 
  $[$Mn\,{\sc vi}$]$\,13.40\,{\micron} and He\,{\sc i} 
  has been ruled out as our {\sc Cloudy} model does not predict these 
  Mn and He\,{\sc i} lines with detectable intensity in the 
  \emph{Spitzer}/IRS spectrum. Our {\sc Cloudy} model can 
  reproduce the observed {\fiv} and {\fv} line intensities with a consistent F
  abundance, simultaneously (we demonstrate in \S\,\ref{S:cloudy}). 
  Thus, the identification of 
  {\fv}\,13.43\,{\micron} and {\fiv}\,4059.9\,{\AA} 
  lines (Fig.\,\ref{F-spro}(a)) is likely to be 
  appropriate and then the {\fv}\,13.43\,{\micron} line 
  would be firstly identified in J900.

  Rb has been discovered in only a handful of PNe 
  \citep{Sterling:2016aa,Mashburn:2016aa,Madonna:2017aa,Madonna:2018aa} 
  since the first measurement of \citet{Garcia-Rojas:2015aa}. 
  We also detect  {\xeiv}\,5709.1\,{\AA} (Fig.\,\ref{F-spro}(b)) and 
 {\rbiv}\,5759.56\,{\AA} lines (Fig.\,\ref{F-spro}(c)) in J900 for the first time. Besides, we confirm the presence of 
{\xeiii}\,5846.7\,{\AA} (Fig.\,\ref{F-spro}(d)),
{\kriv}\,5867.7\,{\AA} (Fig.\,\ref{F-spro}(e)), and 
{\kriii}\,6826.7\,{\AA} (Fig.\,\ref{F-spro}(f)) firstly done by 
  \citet{Sterling:2009aa}.  

  In order to accurately calculate Rb$^{3+}$, we subtract the {\heii} line contribution 
  to the {\rbiv}\,5759.55\,{\AA} using the theoretical ratio of the {\heii}\,$I(5759.64\,\mbox{\AA})/I(5852.67\,\mbox{\AA})$ \citep[0.39;][]{Storey:1995aa}. 
Fig.\,\ref{F-spro}(c) shows 
  the residual spectrum after subtracting {\heii}\,5759.64\,{\AA} out. 
  For J900, \citet{Sterling:2009aa} reported the detection of above Xe and Kr lines, but 
  the Xe$^{2+,3+}$ and Kr$^{3+}$ had never been calculated. Similarly, 
  we subtract the contribution of {\heii}\,5846.66\,{\AA} 
  using the theoretical ratio of {\heii}\,$I(5846.66\,\mbox{\AA})/I(5857.26\,\mbox{\AA})$ \citep[0.914;][]{Storey:1995aa}. The residual spectrum after this process 
  is shown in Fig.\,\ref{F-spro}(d). 
  
\citet{Sterling:2008aa} calculated Kr$^{2+}$ to be ($1.28 ~\pm~ 0.61$)(--9)
  under {\te}({\oiii}) = $11\,600 ~\pm~ 1160$\,K and 
  {\Ne} = $3600 ~\pm~ 720$\,cm$^{-3}$ using the {\te}-insensitive fine-structure line 
  {\kriii}\,2.199\,{\micron}. Perhaps, more careful treatment for their Kr$^{2+}$ value is necessary as 
  \citet{Sterling:2008aa} pointed out that the line at 2.199\,{\micron} 
  would be the complex of the {\kriii}\,2.199\,{\micron} and
  H$_{2}$\,$v=3-2$\,S(3) at 2.201\,{\micron} lines. 
  Despite this, our derived Kr$^{2+}$ using the {\te}-sensitive {\kriii}\,6826.70\,{\AA} line (Fig.\,\ref{F-spro}(f)) is consistent with the value of \citet{Sterling:2008aa}.
  This unexpected outcome implies that the H$_{2}$\,$v=3-2$\,S(3) 
  contribution to the {\kriii}\, 2.199\,{\micron} 
  is negligible, suggesting that intensity measurement of both of Kr 
  lines and our adopted {\te} for Kr$^{2+}$ using {\kriii}\,6826.70\,{\AA} is reasonable.

  \subsubsection{The RL ionic abundances}

  The RL He$^{+,2+}$ and C$^{2+,3+,4+}$  are summarised in Appendix Table\,\ref{T-CEL-RL-abund}. We adopt
  {\te}({\hei}) and {\Ne} = 10$^{4}$\,cm$^{-3}$. 
  \citet{Aller:1983aa} did not give information on which He\,{\sc i,ii} lines they used. 
  In the earlier studies, \citet{Kingsburgh:1994aa} used 3 {\hei}/1 {\heii} lines
  and \citet{Kwitter:2003aa} seems to use the {\hei}\,5875\,{\AA}
  and {\heii}\,4686\,{\AA} lines only. 
  Meanwhile, we use
  the 9 {\hei} and 12 {\heii} lines to determine the final He$^{+}$ and He$^{2+}$, in order to minimise the 
  uncertainty of their ionic abundances.

    In Appendix Table\,\ref{T-CEL-RL-abund}, our final RL C$^{2+,3+,4+}$ 
  and associated standard deviation errors are given. The RL C$^{2+}$
  using C\,{\sc ii}\,4267\,{\AA} only had been measured in
  \citet{Aller:1983aa} and \citet{Kingsburgh:1994aa}, whereas our work newly
  provides RL C$^{3+,4+}$. Our derived RL He$^{+,2+}$ and
  C$^{2+}$ agree with the previous measurements
  (Table\,\ref{T-ionicB}).

  The degree of the discrepancy between RL and CEL ionic/elemental abundances, 
  i.e., abundance discrepancy factor (ADF) defined as the ratio of 
  the RL to CEL abundances, is derived. The ADF(C$^{2+}$) of $1.29 ~\pm~ 0.38$ is consistent with both 
  the value calculated from \citet[][1.25, see Table\,\ref{T-ionicB}]{Aller:1983aa} and 
  the average ADF(C$^{2+}$) amongst 6 C-rich PNe displaying the CEL C/O ratio $> 1$ 
  in \citet[][$1.41 ~\pm~ 0.58$]{Delgado-Inglada:2014aa}.

\subsection{Elemental abundances}

 \begin{table*}
    \renewcommand{\arraystretch}{0.80}
  \caption{Elemental abundances of J900. $\epsilon({\rm X}) = 12 + \log\,({\rm
  X/H})$, where $\log$\,H is 12. AC83, KB94, K03, and SD08 mean
  \citet{Aller:1983aa}, \citet{Kingsburgh:1994aa},
  \citet{Kwitter:2003aa}, and \citet{Sterling:2008aa},
  respectively. S derived by \citet{Kingsburgh:1994aa} seems to be
  overestimated (\S\,\ref{S-abund}).
  The last columns are the derived abundances with
  respect to the solar abundances, defined as [X/H] =
  $\epsilon({\rm X})({\rm Ours}) - \epsilon({\rm X})_{\sun}$, except for Se. [Se/H]
  = $\epsilon({\rm Se})({\rm AD08}) - \epsilon({\rm X})_{\sun}$. We take
  $\epsilon$(X)$_{\sun}$ of \citet{Lodders:2010aa} in the column (8) and those of \citet{Asplund:2009aa} in the column (9).}
 \begin{tabularx}{\textwidth}{@{}@{\extracolsep{\fill}}lD{p}{\pm}{-1}D{p}{\pm}{-1}
  D{.}{.}{-1}
  D{.}{.}{-1}
  D{p}{\pm}{-1}
  D{p}{\pm}{-1}
  D{p}{\pm}{-1}
  D{p}{\pm}{-1}@{}}
\midrule
  \multicolumn{1}{c}{X}  &
  \multicolumn{1}{c}{X/H(Ours)}   &
  \multicolumn{1}{c}{$\epsilon({\rm X})$(Ours)}&
  \multicolumn{1}{c}{$\epsilon({\rm X})$(AC83)}&
  \multicolumn{1}{c}{$\epsilon({\rm X})$(KB94)}&
  \multicolumn{1}{c}{$\epsilon({\rm X})$(K03)}&
  \multicolumn{1}{c}{$\epsilon({\rm X})$(SD08)}&
  \multicolumn{1}{c}{[X/H]$_{\rm Lo}$(Ours)}   &
  \multicolumn{1}{c}{[X/H]$_{\rm As}$(Ours)}   \\
  \multicolumn{1}{c}{{\rm (1)}}&
      \multicolumn{1}{c}{{\rm (2)}}&
	  \multicolumn{1}{c}{{\rm (3)}}&
	      \multicolumn{1}{c}{{\rm (4)}}&
		  \multicolumn{1}{c}{{\rm (5)}}&
		      \multicolumn{1}{c}{{\rm (6)}}&
			  \multicolumn{1}{c}{{\rm (7)}}&
			       \multicolumn{1}{c}{{\rm (8)}}&
			         \multicolumn{1}{c}{{\rm (9)}}\\  
\midrule
\multicolumn{1}{c}{He} & 1.08(-1) ~p~ 1.02(-2) & 11.04 ~p~ 0.04 &10.99 &11.09&11.00 ~p~ 0.13 &&+0.11 ~p~ 0.05 &+0.11 ~p~ 0.04\\ 
\multicolumn{1}{c}{C(CEL)} & 1.09(-3) ~p~ 2.60(-4) & 9.04 ~p~ 0.10 &9.10  &9.31 &&&+0.65 ~p~ 0.11 &+0.57 ~p~ 0.12\\ 
\multicolumn{1}{c}{C(RL)} & 2.01(-3) ~p~ 7.08(-4) & 9.30 ~p~ 0.15  &     &      & &&+0.91 ~p~ 0.16 & +0.87 ~p~ 0.16\\ 
\multicolumn{1}{c}{N} & 6.93(-5) ~p~ 1.82(-5) & 7.84 ~p~ 0.11 & 7.77&7.64&8.01 ~p~ 0.09&&-0.02 ~p~ 0.17 & -0.03 ~p~ 0.12\\ 
\multicolumn{1}{c}{O} & 2.95(-4) ~p~ 1.11(-5) & 8.47 ~p~ 0.02 & 8.50&8.59&8.55 ~p~ 0.06&&-0.26 ~p~ 0.07 &-0.26 ~p~ 0.05\\ 
\multicolumn{1}{c}{F} & 1.45(-7) ~p~ 2.45(-8) & 5.16 ~p~ 0.07 & & & &&+0.72 ~p~ 0.09 & +0.70 ~p~ 0.31\\ 
\multicolumn{1}{c}{Ne} & 8.89(-5) ~p~ 3.26(-6) & 7.95 ~p~ 0.02 &7.91 &8.00 &7.95 ~p~ 0.09 &&-0.10 ~p~ 0.10&-0.02 ~p~ 0.10\\ 
\multicolumn{1}{c}{Mg} & 2.22(-5) ~p~ 3.17(-6) & 7.35 ~p~ 0.06 & & & &&-0.19 ~p~ 0.09 &-0.29 ~p~ 0.07\\ 
\multicolumn{1}{c}{Si} & 9.33(-6) ~p~ 3.07(-6) & 6.97 ~p~ 0.14 & & & &&-0.56 ~p~ 0.16 &-0.58 ~p~ 0.15\\ 
\multicolumn{1}{c}{P} & 3.06(-7) ~p~ 3.45(-8) & 5.49 ~p~ 0.05 &  & & &&+0.04 ~p~ 0.07 &+0.02 ~p~ 0.06\\ 
\multicolumn{1}{c}{S} & 3.43(-6) ~p~ 1.26(-7) & 6.53 ~p~ 0.02 & 6.38&[7.22]&6.33 ~p~ 0.16&&-0.63 ~p~ 0.03 &-0.63 ~p~ 0.03\\ 
\multicolumn{1}{c}{Cl} & 7.47(-8) ~p~ 1.45(-9) & 4.80 ~p~ 0.03 &5.09&    &4.59 ~p~ 0.10&& -0.45 ~p~ 0.07 &-0.43 ~p~ 0.30\\ 
\multicolumn{1}{c}{Ar} & 8.94(-7) ~p~ 3.29(-8) & 5.95 ~p~ 0.02 &6.07&5.49&6.08 ~p~ 0.09&& -0.55 ~p~ 0.10 &-0.49 ~p~ 0.13\\ 
\multicolumn{1}{c}{K} & 2.73(-8) ~p~ 4.51(-9) & 4.44 ~p~ 0.07 & &&&&-0.67 ~p~ 0.08 &-0.64 ~p~ 0.12\\ 
\multicolumn{1}{c}{Fe}  & 6.69(-7) ~p~ 3.20(-8) & 5.83 ~p~ 0.21 & &&&&-1.63 ~p~ 0.22 &-1.71 ~p~ 0.21\\ 
\multicolumn{1}{c}{Se}&&&&&&3.65 ~p~ 0.25&+0.32 ~p~ 0.25&+0.34 ~p~ 0.25\\
\multicolumn{1}{c}{Kr} & 7.11(-9) ~p~ 7.40(-10) & 3.85 ~p~ 0.05 & &&&3.80 ~p~ 0.30 &+0.57 ~p~ 0.09 &+0.60 ~p~ 0.08\\ 
\multicolumn{1}{c}{Rb} & 7.28(-10) ~p~ 2.16(-10) & 2.86 ~p~ 0.13 & &&&&+0.48 ~p~ 0.13 &+0.32 ~p~ 0.16\\ 
\multicolumn{1}{c}{Xe} & 2.61(-9) ~p~ 4.59(-10) & 3.42 ~p~ 0.08 & &&&&+1.15 ~p~ 0.11 & +1.18 ~p~ 0.10\\
\midrule
\end{tabularx}
\label{T-elements}
 \end{table*}

We derive the elemental abundances by applying ionisation correction
factor ICF(X) for element X. ICFs recover the ionic abundances
in the unseen ionisation stages in the wavelength coverage
by the \emph{IUE}, BOES, \emph{AKARI}/IRC, and \emph{Spitzer}/IRS spectra. The abundance of the
element X, which is the number density ratio with respect to hydrogen, 
is expressed by the following equation.
\begin{equation}
{\rm X/H} = {\rm ICF(X)}~\times\Sigma_{\rm m}~({\rm
 X}^{\rm m+}/{\rm H}^{+}).
\end{equation}
Except for He, C(CEL), O, Ne, S, and Ar (whose ICFs are 1), 
we determine ICF(X) based on the IP of the element X. 
To determine each ICF definition and ICF values 
for each element, we refer to the mean ionisation fraction predicted in the 
{\sc Cloudy} model of the highly-excited C-rich PN NGC6781 by \citet{Otsuka:2017aa}, 
simply because NGC6781 is very similar to J900 in many respects, including the 
effective temperature of the central star ($\sim$120\,000\,K), nebula shape 
(bipolar nebula), and dust and molecular gas features (C-dust dust and PAH/H$_{2}$), 
suggesting that the ionisation fraction would also be similar between the two objects. 
Note that we do not apply the empirical ICF values determined in NGC6781 to J900.

We define that C is the sum of C$^{+,2+,3+,4+}$. 
For the CEL C, we correct C$^{4+}$ using ICF(C) = $1.26 ~\pm~ 0.24$ calculated 
from eq.\,(A25) of \citet{Kingsburgh:1994aa}. For the RL C, 
we correct C$^{+}$ using ICF(C) = ($1 - {\rm (C^{+}/C)_{CEL}}$)$^{-1}$. 
ICF(N) of $1.54 ~\pm~ 0.06$ corresponds to the O/(O$^{+}$ + O$^{2+}$), 
where the O is the sum of O$^{+,2+,3+}$. 
The theoretical model by \citet{Otsuka:2017aa} predicted that 
the F, Ne, and Mg ionic fractions are very similar, so  
we determine ICF(F,Mg) = $7.28 ~\pm~ 1.02$ from the
Ne/(Ne$^{3+}$ + Ne$^{4+}$) ratio. Assuming that Cl is the sum of
Cl$^{+,2+,3+,4+}$, we determine ICF(Cl) = $1.06 ~\pm~ 0.06$ from Ar/(Ar$^{+}$ + Ar$^{2+}$ + Ar$^{3+}$). 
As the fractional ionic stages of Si, P, and K are similar to
that of S in the model of \citet{Otsuka:2017aa}, 
we adopt ICF(Si) of $10.16 ~\pm~ 2.88$ from the S/S$^{+}$ ratio; ICF(P) is $1.70
~\pm~ 0.09$ from the S/(S$^{+}$ + S$^{2+}$) ratio; and ICF(K) = $4.96 ~\pm~ 0.23$
from the Ar/Ar$^{3+}$ ratio, respectively. 
ICF(Fe) of $11.96 ~\pm~ 0.55$ corresponds to the O/O$^{+}$. 
Since the cross-section and recombination coefficients for Kr, Xe, and Rb are 
not fully determined yet, we cannot use the ICF(Kr,Xe,Rb) values predicted by photoionisation models. 
Thus, the use of the ICF(Kr,Rb,Xe) based on the ionisation potential is currently best.
ICF(Kr) is $1.70 ~\pm~ 0.05$ calculated using eq.\,(4) of \citet{Sterling:2015aa}. 
ICF(Xe) is as same as ICF(Kr). ICF(Rb) is $1.76 ~\pm~ 0.07$ from the O/O$^{2+}$ ratio by following \citet{Sterling:2016aa}. 

In Appendix Table\,\ref{T-elements}, we summarise the derived elemental abundances
(X/H, second column), $\epsilon$(X) corresponding
to  12 + $\log$\,(X/H) (third column), and the abundances relative to the Sun [X/H] (last two columns).

\citet{Delgado-Inglada:2015aa} concluded that Cl and Ar abundances are the
best metallicity indicators based on chemical abundance analysis of
20 Galactic PNe. They argued that their abundances indicate the
composition in the primordial gas clouds where the PN
progenitors were born. Following their reasoning and the derived average [Cl,Ar/H] abundance (--0.50), we conclude 
that J900 was born in an environment with the metallicity $Z \sim 0.005$.

Potassium (K) is a volatile element ($\sim$50\,\% condensation temperature is $\sim$1000\,K) 
considered to be synthesised mainly via the O-burning in massive stars. The evolution of 
K in the Galaxy seems to be similar to that of $\alpha$-elements, according to 
\citet{Takeda:2009aa}. The observed [K/H] abundance is comparable to the observed abundances 
of heavy $\alpha$-elements and Cl. Our derived extremely small [Fe/H] does not represent the metallicity of
J900, judging from the derived [Cl,Ar/H]. 
We will discuss elemental abundances by comparison with the AGB nucleosynthesis models later.

\subsection{Physical conditions of the H$_{2}$ lines}

\label{S-H2}

  \begin{figure}
   \includegraphics[width=\columnwidth,clip]{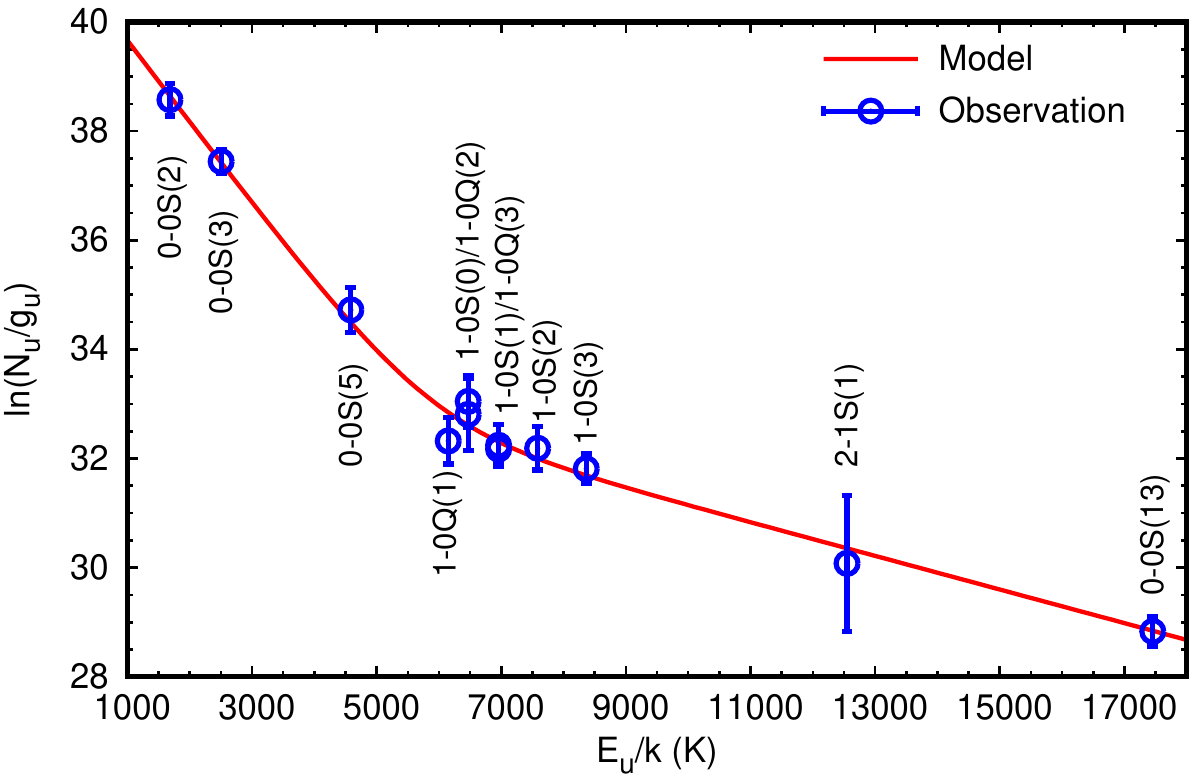}
\vspace{-15pt}
   \caption{Excitation diagram of H$_{2}$ in
   J900. The H$_{2}$ lines in near-IR are taken
   from \citet{Hora:1999aa}. The H$_{2}$ intensities used in this plot are listed in
   Table\,\ref{T-H2}. The red line is the best fit curve to the
   observed data with two temperatures = $667 ~\pm~ 54$\,K and
   $3247 ~\pm~ 268$\,K. See text for details. 
   }
   \label{F-h2}
  \end{figure}

   \begin{table}
    \renewcommand{\arraystretch}{0.80}
   \caption{The H$_{2}$ line intensity in J900. H99 means \citet{Hora:1999aa}.}
    \label{T-H2}
\begin{tabularx}{\columnwidth}{@{}@{\extracolsep{\fill}}
    lD{.}{.}{-1}D{p}{\pm}{-1}l@{}}
     \midrule
     Line &\multicolumn{1}{c}{$\lambda$ ({\micron})} &\multicolumn{1}{c}{$I($H$_{2}$) (erg\,s$^{-1}$\,cm$^{-2}$\,sr$^{-1}$)}&{Source}\\
     \midrule
H$_{2}$ $v=0-0$\,S(2)&	12.293	&1.81(-5) ~p~ 5.41(-6)&This work\\
H$_{2}$ $v=0-0$\,S(3)&	9.664	&9.72(-5) ~p~ 2.10(-5)&This work\\
H$_{2}$ $v=0-0$\,S(5)&	6.910	&7.29(-5) ~p~ 2.98(-5)&This work\\
H$_{2}$ $v=0-0$\,S(13)&	3.846	&1.70(-5) ~p~ 6.72(-6)&This work\\
H$_{2}$ $v=1-0$\,Q(3)&	2.423	&3.83(-5) ~p~ 1.49(-5)&H99\\
H$_{2}$ $v=1-0$\,Q(2)&	2.413	&2.23(-5) ~p~ 1.06(-5)&H99\\
H$_{2}$ $v=1-0$\,Q(1)&	2.407	&2.77(-5) ~p~ 1.17(-5)&H99\\
H$_{2}$ $v=2-1$\,S(1)&	2.248	&8.51(-6) ~p~ 1.06(-5)&H99\\
H$_{2}$ $v=1-0$\,S(0)&	2.223	&1.60(-5) ~p~ 1.06(-5)&H99\\
H$_{2}$ $v=1-0$\,S(1)&	2.122	&5.11(-5) ~p~ 1.17(-5)&H99\\
H$_{2}$ $v=1-0$\,S(2)&	2.033	&2.66(-5) ~p~ 1.06(-5)&H99\\
H$_{2}$ $v=1-0$\,S(3)&	1.957	&7.34(-5) ~p~ 2.02(-5)&H99\\
    \midrule
   \end{tabularx}
   \end{table}

It is necessary to determine the physical conditions of the regions giving rise to the H$_{2}$ lines in order to infer 
    hydrogen density ($n_{\rm H}$) in the cold PDRs beyond the ionisation 
    front. These PDRs are critically important in assessing how the progenitor ejected its 
dust and gas mass. Hence, we investigate H$_{2}$ temperature $T$(H$_{2}$) and column density 
    $N$(H$_{2}$) under the local thermodynamic equilibrium (LTE) using 
    excitation diagram (Fig.\,\ref{F-h2}) based on the mid-IR H$_{2}$ lines 
    $v=0-0$\,S(2)\,12.29\,{\micron}, S(3)\,9.66\,{\micron}, 
    and S(5)\,6.91\,{\micron} in the \emph{Spitzer}/IRS spectrum 
    (Fig.\,\ref{F-irsspec}), and S(13)\,3.85\,{\micron} in the \emph{AKARI}/IRC spectrum, 
    and the near-IR H$_{2}$ lines by \citet{Hora:1999aa}.
 
The H$_{2}$ column density in the upper state $N_{u}$ is written as 
 \begin{equation}
  N_{u} = \frac{4 \pi I({\rm H_{2}})}{A} \cdot \frac{\lambda}{hc},
   \label{h2_eq1}
 \end{equation}
 \noindent where $I$(H$_{2}$) is the H$_{2}$ line intensity
 in erg\,s$^{-1}$\,cm$^{-2}$\,sr$^{-1}$ (Table\,\ref{T-H2}),
 $A$ is the transition probability taken from \citet{Turner:1977aa},
 $h$ is the Planck constant, and
 $c$ is the speed of light. In LTE, the Boltzmann equation relates
 $N_{u}$ to the $T$(H$_{2}$) and $N$(H$_{2}$) via
 \begin{equation}
  {\rm ln}\,\bigl(\frac{N_{u}}{g_{u}}\bigr) = -\frac{E_{u}}{kT({\rm H}_{2})} 
   + {\rm ln}\,\bigl[N({\rm H}_{2}) \cdot \frac{hcB}{2kT({\rm H}_{2})}\bigr],
   \label{h2_eq2}
 \end{equation}
 \noindent where $g_{u}$ is the vibrational degeneracy,
 $E_{u}$ is the energy of the excited level taken from
 \citet{Dabrowski:1984aa}, $k$ is the Boltzmann constant, and
 $B$ is the rotational constant (60.81\,cm$^{-1}$).

Fig.\,\ref{F-h2} shows the plot of $\ln\,(N_{u}/g_{u})$ versus $E_{u}/k$, based on  eq.\,(\ref{h2_eq2}). Note that the plot consists of two regimes relative to the x-axis $E_{u}/k \sim 6000$\,K, perhaps representing the two different temperature components: 
 the derived $T({\rm H}_{2})$ and 
 $N$(H$_{2}$) for the warm temperature component ($E_{u}/k \lesssim 6000$\,K)  are
 $667 ~\pm~ 54$\,K and $(1.14 ~\pm~ 0.29)$(+19)\,cm$^{-2}$, and 
 those for the hot component ($E_{u}/k \gtrsim 6000$\,K) are $3247 ~\pm~ 268$\,K 
and $(5.38 ~\pm~ 1.37)$(+16)\,cm$^{-2}$, respectively. When we fit the H$_{2}$ lines at 6.91, 9.68, and 
12.28\,{\micron} only, we obtain 
$T$(H$_{2}$) = $755 ~\pm~36$\,K  and $N$(H$_{2}$) = $(8.85 ~\pm~ 0.39)$(+18)\,cm$^{-2}$, respectively. 
We interpret that the H$_{2}$ lines with $E_{u}/k \gtrsim 6000$\,K emit in the regions just outside
  the ionised nebula. There, the fast central star wind interacts with the slowly expanding AGB mass-loss.
 Meanwhile, H$_{2}$ lines with $E_{u}/k \lesssim 6000$\,K would emit in the outermost of such wind interaction
  regions. Accordingly, the wind-wind interaction is anticipated to create the regions with 
  a temperature gradient varying from $\sim$3300 to $\sim$700\,K. 
  
  Excluding the mid-IR H$_{2}$ data and applying single temperature fitting towards the near H$_{2}$ data, we obtain $T$(H$_{2}$) of 
 $2736 ~\pm~ 703$\,K for the hot component, which is in agreement with \citet{Hora:1999aa} who 
 estimated $T$(H$_{2}$) of $2200 ~\pm~ 100$\,K with a single temperature component fitting. \citet{Isaacman:1984aa} and \citet{Hora:1999aa} concluded that H$_{2}$ is emitted by thermal shocks rather than UV fluorescence. Our excitation diagram supports their conclusion.

\subsection{PAH and Dust features}
\label{S-dust}

\citet{Aitken:1982aa} fit $8-13$\,{\micron} spectra 
using the emissivities of graphite grains and PAHs. 
Graphite grains have a broad feature around 30\,{\micron}. 
However, the \emph{Spitzer}/IRS spectrum does not display such a broad feature. Thus, 
the carrier of the dust continuum (indicated by the red line in Fig.\,\ref{F-irsspec}) would be amorphous carbon.

J900 exhibits several prominent broad emission features 
in $6-9$\,{\micron}, 11\,{\micron}, and $15-20$\,{\micron} overlying the dust continuum. They would be attributed to PAHs composing of a large number of C-atoms \citep[$>$50;][]{Peeters:2017aa}. The emission bands at 6.2, 7.7\,{\micron}, and 8.6\,{\micron} are charged PAHs, while that at 
11.3\,{\micron} is neutral PAH \citep{Peeters:2017aa}. 

Following the classification of \citet{Matsuura:2014aa}, respective PAH $6-9$\,{\micron} and 11\,{\micron} band profiles would be divided into classes $\mathcal{B}$ and $\alpha$, where Galactic and Large Magellanic PNe have been grouped \citep[e.g.,][]{Otsuka:2015ab,Bernard-Salas:2009aa}. 
J900 displays the 16.4 and 17.4\,{\micron} emission features in the PAH $15-20$\,{\micron} band \citep[][and reference therein]{Boersma:2010aa}. Through a detailed analysis of the PAH $15-20$\,{\micron} band, \citet{Boersma:2010aa} reached conclusions that (1) 16.4/17.4/17.8/18.9\,{\micron} emission features do not correlate with each other and their carriers are independent molecular species and (2) the PAH 16.4\,{\micron} features only has a tentative connection with PAH 6.2/7.7\,{\micron}. Indeed, we confirm that the $I$(6.2\,{\micron})/$I$(11.2\,{\micron}) and $I$(16.4\,{\micron})/$I$(11.2\,{\micron}) ratios ($0.812 ~\pm~ 0.148$ and $0.013 ~\pm~ 0.002$, respectively, Appendix Table\,\ref{T-emission}) are along a correlation found in Fig.\,5 of \citet{Boersma:2010aa}. Later, \citet{Peeters:2012aa} reported that the PAHs 12.7 and 16.4\,{\micron} emission correlate with the PAHs 6.2 and 7.7\,{\micron}, which arise from species in the same conditions that favour PAH cations. 
Following \citet{Boersma:2010aa} and \citet{Peeters:2012aa}, we conclude 
that the 6.2/7.7/8.6/12.7/16.4\,{\micron} emission arise from charged PAHs and the 11.3\,{\micron} emission arises from 
neutral PAHs. 

The carrier of the 17.4\,{\micron} band is under debate; certainly, mid-IR fullerene C$_{60}$ bands show four emission features at 7.0/8.5/17.4/18.9\,{\micron} \citep[e.g.,][]{Cami:2010aa,Otsuka:2014aa}, but 
three emission features at 7.0/8.5/18.9\,{\micron} are not seen in J900. \citet{Peeters:2017aa} 
suggested charged PAHs, while Fig.\,22 of \citet{Sloan:2014aa} indicates an aliphatic compound alkyne chain, 2,4-hexadiyne, 
displaying a strong emission feature.

The ratio of neutral PAH $I$(3.3\,{\micron})/$I$(11.3\,{\micron}) is directly related to the internal energy of the molecule. Therefore, given the internal energy, the ratio provides the number of C-atoms ($N_{\rm C}$) composing PAH molecules \citep{Allamandola:1987aa}. Using the function of $I$(3.3\,{\micron})/$I$(11.3\,{\micron}) versus $N_{\rm C}$ established in \citet{Ricca:2012aa} and the observed \emph{AKARI}/IRC PAH 3.3\,{\micron} to 
\emph{Spitzer}/IRS 11.3\,{\micron} flux ratio ($0.15 ~\pm~ 0.01$, Appendix Table\,\ref{T-emission}), 
we estimate $N_{\rm C}$ to be $\sim$150. This ratio is consistent with the value measured in a high-excitation PN NGC7027 
\citep[0.13;][]{Otsuka:2014aa}.

\section{Photoionisation modelling }
\label{S:cloudy}

To understand the evolution of J900, we investigate the 
current evolutionary status of the CSPN and the physical conditions of the gas and the dust grains by 
constructing a coherent {\sc Cloudy} photoionisation model to be consistent with 
all the observed quantities of the CSPN and the dusty nebula. 
Below, we explain how to constrain the input parameters and obtain 
the best fit.

\subsection{Modelling approach}

\paragraph*{Incident SED} 

For the incident SED of the CSPN, we adopt a non-LTE stellar atmosphere model grid of  \citet{Rauch:2003aa}. As an initial 
guess value, we adopt an effective temperature ($T_{\rm eff}$) 
of 134\,800\,K calculated using the observed $I$({\heii}\,4686\,{\AA})/$I$({\hb}) ratio and the eq (3.1) established in \citet{Dopita:1991aa}. We adopt the surface gravity $\log\,g$ of 
6.0\,cm\,s$^{-2}$ by referring to the post-AGB evolution models of \citet{Miller-Bertolami:2016aa}. 
We scale the flux density of the CSPN's SED to match with the observed \emph{HST}/F555W magnitude of the CSPN (Table\,\ref{T-photo}).

\paragraph*{Distance $D$}

For comparison between the model predicted values and the observations, we have to fix the distance $D$ towards J900. $D$ ranges from 3.89 to 5.82\,kpc in literature \citep{Kingsburgh:1992aa,Stanghellini:2008aa,Stanghellini:2010aa,Giammanco:2011aa,Frew:2016aa,Stanghellini:2018aa} of which the average $D$ is 4.65\,kpc (1-$\sigma$ = 0.72\,kpc). 
The test model with $D=4.6$\,kpc predicts that the ionisation front 
locates at $\sim$8.5$''$ away from the CSPN whereas the \emph{HST}/F656N image suggests $\sim$3.5$''$. 
After several test model runs with different $D$ values, we find that the models with 
$D$ of $\sim$6.0\,kpc can reproduce the ionisation front radius of $\sim$3.5$''$. Hence, 
we keep $\log\,g =6.0$\,cm\,s$^{-1}$ and F555W magnitude of the CSPN and $D$ of 6.0\,kpc during the iterative 
model fitting, whereas we vary $T_{\rm eff}$ to search for the best-fit model parameters that would reproduce the observational data.

\paragraph*{Nebula geometry and hydrogen density profile \label{S-C-H}}

We adopt the cylindrical geometry with a height of 3$''$. The input radial hydrogen number density profile, $n_{\rm H}(R)$ ($R$ is the distance from the CSPN) in $R \le 3.5''$ is to be determined based on our plasma diagnostics. As the average {\Ne} is 4960\,cm$^{-3}$ amongst {\Ne}({\oii}), {\Ne}({\siii}), {\Ne}({\cliii}), {\Ne}({\cliv}), {\Ne}({\ariv}), and {\Ne}({\nev}), we adopt a constant 4000\,cm$^{-3}$ by taking the reasonable assumption that $n$(H$^{+}$) is {\Ne}/1.15.

Assuming pressure equilibrium at the ionisation front, 
$n_{\rm H}$($R > 3.5''$) is approximately 2{\Ne}{\te}/$T$(H$_{2}$). 
The factor of 2 is because the gas within the ionisation front is totally ionised and so has both electrons and protons. 
The {\sii} lines often indicate the physical conditions of the transition zone between ionised and neutral gases.  Adopting {\Ne}({\sii}) and {\te}({\sii}), 
$n_{\rm H}$($R > 3.5''$) is $\sim$($3.3 - 15.6$)$\times$10$^{4}$\,cm$^{-3}$. We further searched 
for values of $n_{\rm H}$($R > 3.5''$) that can reproduce the observed mid-IR H$_{2}$ line fluxes based on the results of a 
small grid model with a constant $n_{\rm H}$($ \le 3.5''$) = 4000\,cm$^{-3}$ and 
different values of $n_{\rm H}$($R > 3.5''$). As discussed by \citet{Otsuka:2017aa}, it would be difficult 
to reproduce the observed H$_{2}$ line fluxes by the CSPN heating only. Thus, we add a 
constant temperature region in the PDRs using the ``temperature floor'' option, and 
we finally found a model, with $n_{\rm H}(R > 3.5'') = 87\,500$\,cm$^{-3}$, that 
reasonably matches the observed mid-IR H$_{2}$ line fluxes.

By following the method of \citet{Mallik:1988aa}, we calculate the filling factor ($f$), which is the ratio of the root mean square {\Ne} to the forbidden line {\Ne}. We keep the derived $f = 0.75$ throughout model iteration.

\paragraph*{$\epsilon$(X) / Dust grains and PAH molecules}

Elements up to Zn can be handled in {\sc Cloudy}. 
In the model calculations, we updated the originally installed $A_{ji}$ and $\Omega$($T_{\rm e}$) 
of forbidden C/N/O/F/Ne/Mg/P/S/Cl/Ar/K lines with the values 
in Appendix Table\,\ref{atomf}. For the first guess, we adopt the empirically determined $\epsilon$(X) listed 
in Table\,\ref{T-elements}, except for Se, Kr, Rb, and Xe whose atomic data are not installed in 
{\sc Cloudy}. We adopt the CEL $\epsilon$(C) as the representative carbon abundance. For the 
unobserved elements, we adopt the predicted $\epsilon$(X) by the AGB nucleosynthesis model 
of \citet{Karakas:2018aa} for a star of initial metallicity $Z = 0.003$ and 2.0\,M$_{\sun}$.

As J900 is a PN rich in carbon dust (\S\,\ref{S-dust}), we run test 
models to determine which carbon grain's thermal emission can fit the observed mid-IR SED. 
In these tests, we consider the optical data of graphite 
and amorphous carbon provided by \citet{Rouleau:1991aa}. As we explained in \S\,\ref{S-dust}, 
the graphite grain models do not fit 
the observed SED at all due to the emission peak around 30\,{\micron}. 
\citet{Rouleau:1991aa} provided two types of optical constants measured from samples 
``BE'' (soot produced from benzene burned in air) and ``AC'' (soot produced by striking 
an arc between two amorphous carbon electrodes in a controlled Ar atmosphere). 
A comparison between the AC and the BE model SEDs shows that the former gives the better fit to the observed SED. 
Thus, we adopt the spherically shaped AC grains with the radius range $a = 0.005-0.50$\,{\micron} and the $a^{-3.5}$ size distribution.

To simplify the model calculations, we also adopt the spherically shaped PAH molecules. 
Taking into account the discussion on the condition of PAH (\S\,\ref{S-dust}), 
we adopt both the neutral and charged PAH molecules with the radius $a = 3.6-11$(--4)\,$\mu$m and the $a^{-3.5}$ size distribution. 
The optical constants for PAH-Carbonaceous grains are adopted from the theoretical 
work by \citet{2007ApJ...657..810D}. We include the stochastic heating mechanism 
in the model calculations.

\paragraph*{H$_{2}$ molecules \label{S-C-H2}}
Even if the temperature floor option is invoked, 
both the observed near-IR and mid-IR H$_{2}$ lines cannot be simultaneously 
reproduced in {\sc Cloudy}. Thus, in the present model, we aim at reproducing 
the H$_{2}$ $v=0-0$\,S(2)\,12.29\,{\micron}, S(3)\,9.66\,{\micron}, and S(5)\,6.91\,{\micron} line fluxes 
because the majority of the molecular gas is expected to exist in cold dusty PDRs, responsible for these 
H$_{2}$ lines. We select the database of the UMIST chemical reaction network \citep{McElroy:2013aa}.

\begin{table}
\renewcommand{\arraystretch}{0.80}
\caption{
\label{T-modelin}
The adopted and derived parameters in the best model of J900.}
\begin{tabularx}{\columnwidth}{@{}@{\extracolsep{\fill}}l@{\hspace{4pt}}l@{}}
\midrule
CSPN&Values\\ 
\midrule
$T_{\ast}$ / $\log\,g$ / $L_{\ast}$ &129\,640\,K / 6.0\,cm\,s$^{-2}$ / 5562\,L$_{\sun}$\\
$[$Z/H$]$ / $M_{\rm V}$ / $D$     &--0.5 / 2.951 / 6.0\,kpc    \\
\midrule
Nebula&Values\\
\midrule 
$\epsilon$(X)    &He: 10.97, C: 8.92, N: 7.66,  O: 8.47, F: 5.03, Ne: 7.90, \\
                 &Mg: 7.08, Si: 5.97, P: 5.21, S: 6.47, Cl: 4.83, Ar: 5.89, \\
                 &K: 4.30,  Fe: 5.68, Others: \citet{Karakas:2018aa}\\
Geometry         &Cylinder with height = 18\,000\,AU (3$''$)\\
                 &Inner radius = 6377\,AU (1.07$''$)\\
                 &Ionisation boundary = 21\,150\,AU (3.51$''$) \\
                 &Outer radius = 22\,032\,AU (3.65$''$)\\
$n_{\rm H}$($R$) &   4000\,cm$^{-3}$   in $R \le 20\,999$\,AU ($ \le 3.5''$)\\
                     & 87\,500\,cm$^{-3}$ in $R > 20\,999$\,AU ($>3.5''$)\\
Filling factor ($f$)                  &0.75\\
$\log\,I$({\hb})&--10.645\,erg\,s$^{-1}$\,cm$^{-2}$\\
Temperature floor &829\,K\\
Gas mass               &0.825\,M$_{\sun}$\\
\midrule
Dust/PAH &Values\\
\midrule 
Particle size    &PAH neutral \& ionised: $3.6-11$(--4)\,{\micron} \\
                 &AC: $0.005-0.50$\,{\micron}\\
Temperature      &PAH neutral: $126 - 238$\,K, ionised: $124 - 204$\,K\\
                 &AC: $51 - 168$\,K\\ 
Mass             &PAH neutral: 1.29(--5)\,M$_{\sun}$, ionised: 8.90(--6)\,M$_{\sun}$\\
                 &AC: 4.24(--4)\,M$_{\sun}$\\
GDR             &1849\\
\midrule
\end{tabularx}
\end{table}

\paragraph*{Model iteration}

In our modelling, we vary the following 20 parameters: $T_{\rm eff}$, 14 elemental 
abundances (He/C/N/O/F/Ne/Mg/Si/P/S/Cl/Ar/K/Fe), the inner radius of the nebula, temperature floor 
in the PDRs, neutral/ionised PAHs and AC mass fractions. 
The N, F, Mg, Si, P, and Fe abundances are determined using either one or 
two ionised species, accordingly their relatively uncertain ICFs could lead to an incorrect elemental abundance derivation, 
While we do not give the optimised range of these six elemental abundances, 
we permit to vary $\epsilon$(X) for the other elements within $\pm$3-$\sigma$ from the observed value. 
The mass of the cold gas component is more critical than that of the hot component in term of 
mass determination. Therefore, we vary the temperature 
floor in the range between 500 and 900\,K, corresponding to the warm temperature $T$(H$_{2}$) component (\S\,\ref{S-H2}). 
Interactive calculations stop when either \emph{AKARI}/FIS N65\,{\micron} or WIDE-S 90\,{\micron} band fluxes 
reach the observed values. Note that the calculation determines the outer radius of the nebula when it finishes. 
We do not consider the observed \emph{AKARI}/WIDE-L 
145\,{\micron} as a calculation stopping criterion owing to its measurement uncertainty. 
The quality of fit is computed from the reduced-$\chi^{2}$ value calculated from the 
following 196 observational constraints; 144 atomic and 3 mid-IR H$_{2}$ line fluxes relative to both $I$({\hb}) as well as $\log\,I$({\hb}), 
33 broadband fluxes, 14 band flux densities, the ionisation boundary radius, and 
$N$(H$_{2}$) of the warm temperature component. We use the radio flux densities from 
\citet[43\,GHz]{Umana:2008aa}, \citet[30\,GHz]{Pazderska:2009aa}, 
\citet[20 and 6\,GHz]{Isaacman:1984ab}, \citet[5 and 2.7\,GHz]{Milne:1979aa}, and 
\citet[1.4\,GHz]{Vollmer:2010aa} (see Appendix Table\,\ref{T-photo} and refer to these measurements by the bands' 
wavelengths). We define the ionisation bounding radius as the radial distance from the CSPN at which $T_{\rm e}$ drops below 4000\,K.

\subsection{Model results}
\label{S:model-r}

\begin{figure*}
\includegraphics[width=0.7\textwidth,clip]{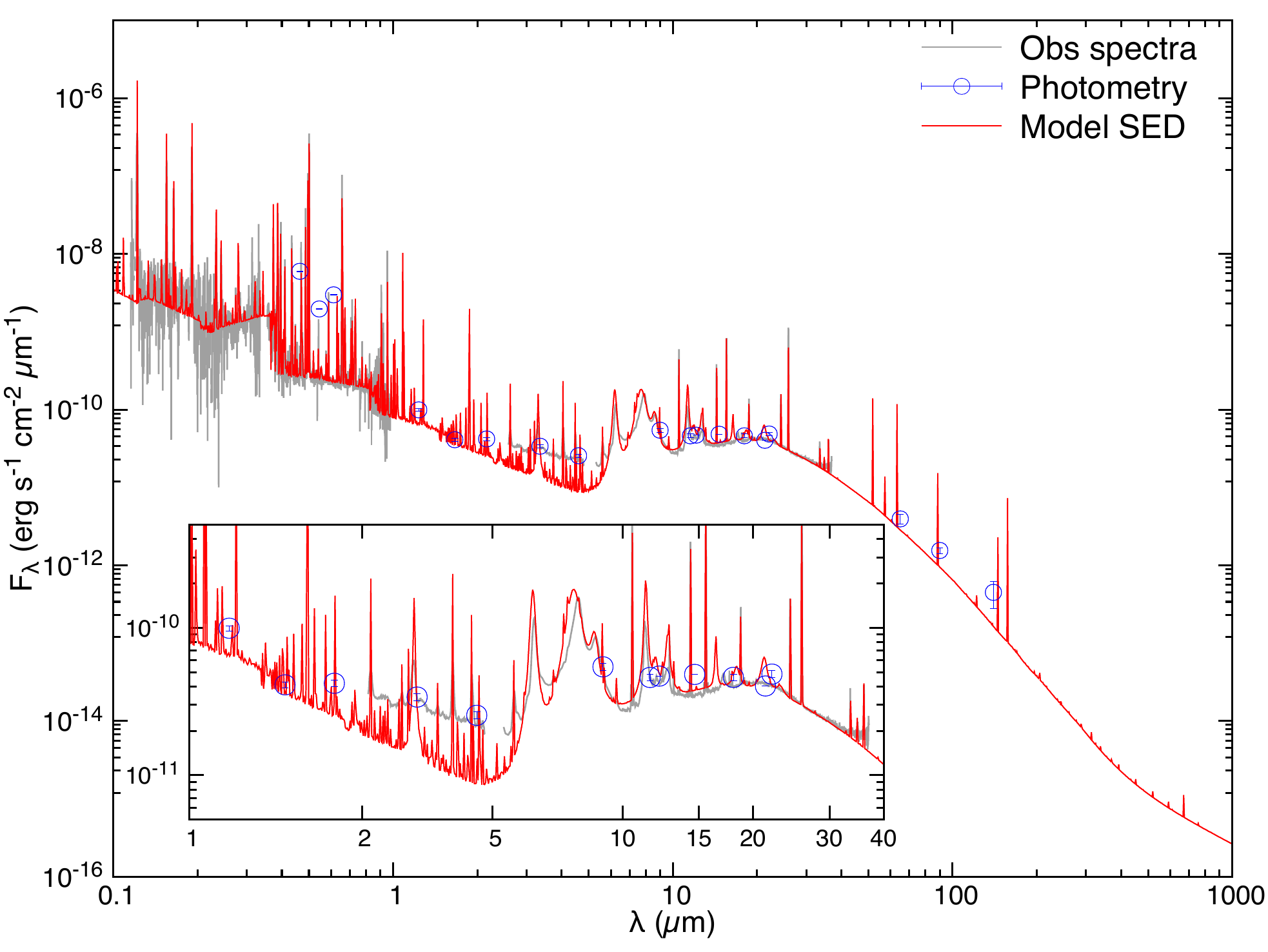}
\vspace{-5pt}
\caption{Comparison of the observations (grey lines and blue circles) with the model SED by our {\sc Cloudy} fitting in 
the range of $0.1-1000$\,{\micron}. We scaled the de-reddened flux density of the observed spectra 
to match with broadband flux densities of the entire nebula listed in Table \ref{T-photo} and 
$I$(He\,{\sc ii}\,1640\,{\AA}) of 5.23(--11)\,erg\,s$^{-1}$\,cm$^{-2}$ for the {\sl IUE} spectrum. 
The inset displays the close-up of $1-40$\,{\micron} SED.}
\label{F-SED}
\end{figure*}

\begin{figure}
\includegraphics[width=\columnwidth,clip]{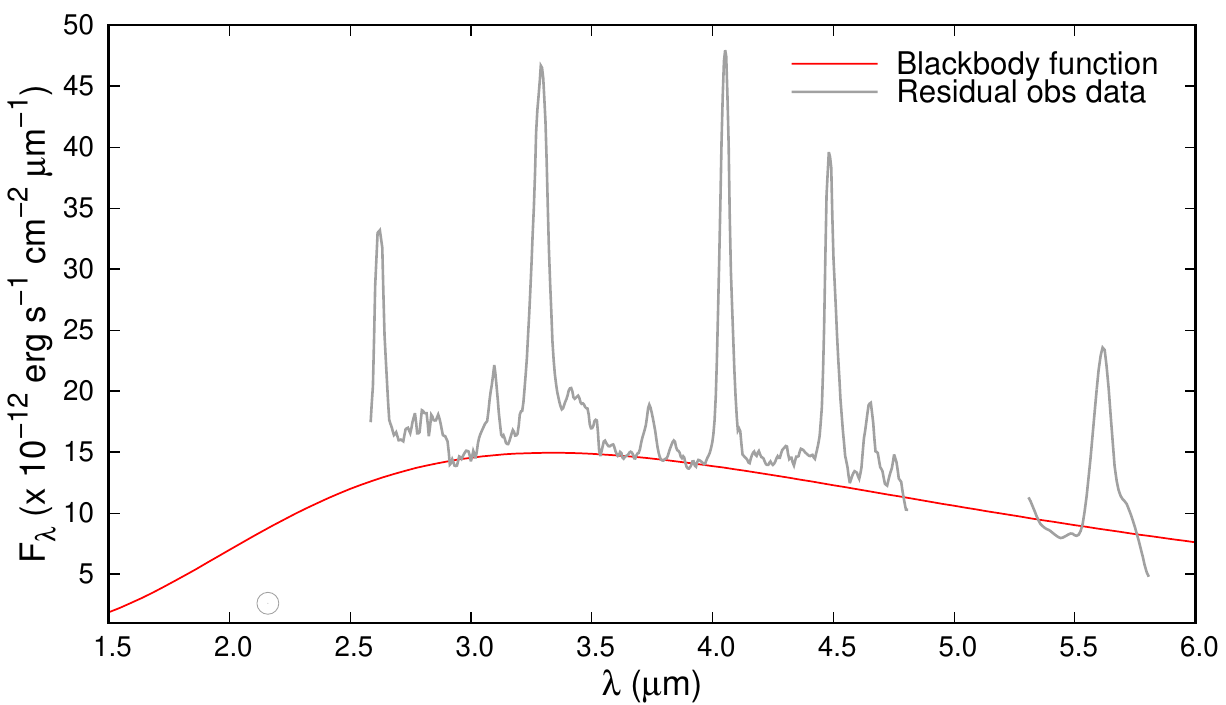}
\vspace{-15pt}
\caption{
Near-IR excess and the blackbody fitting. The grey lines/circle are the residual flux density between the 
observed {\sl AKARI}/IRC and {\sl Spitzer}/IRS spectra and WFCAM 
$K$ photometry and the corresponding values obtained from the {\sc Cloudy} model. The red line is the best fit of 
this near-IR excess with a blackbody function with a single temperature of $870 ~\pm~ 20$\,K.}
\label{F-NIR}
\end{figure}

\begin{figure}
\includegraphics[width=\columnwidth,clip]{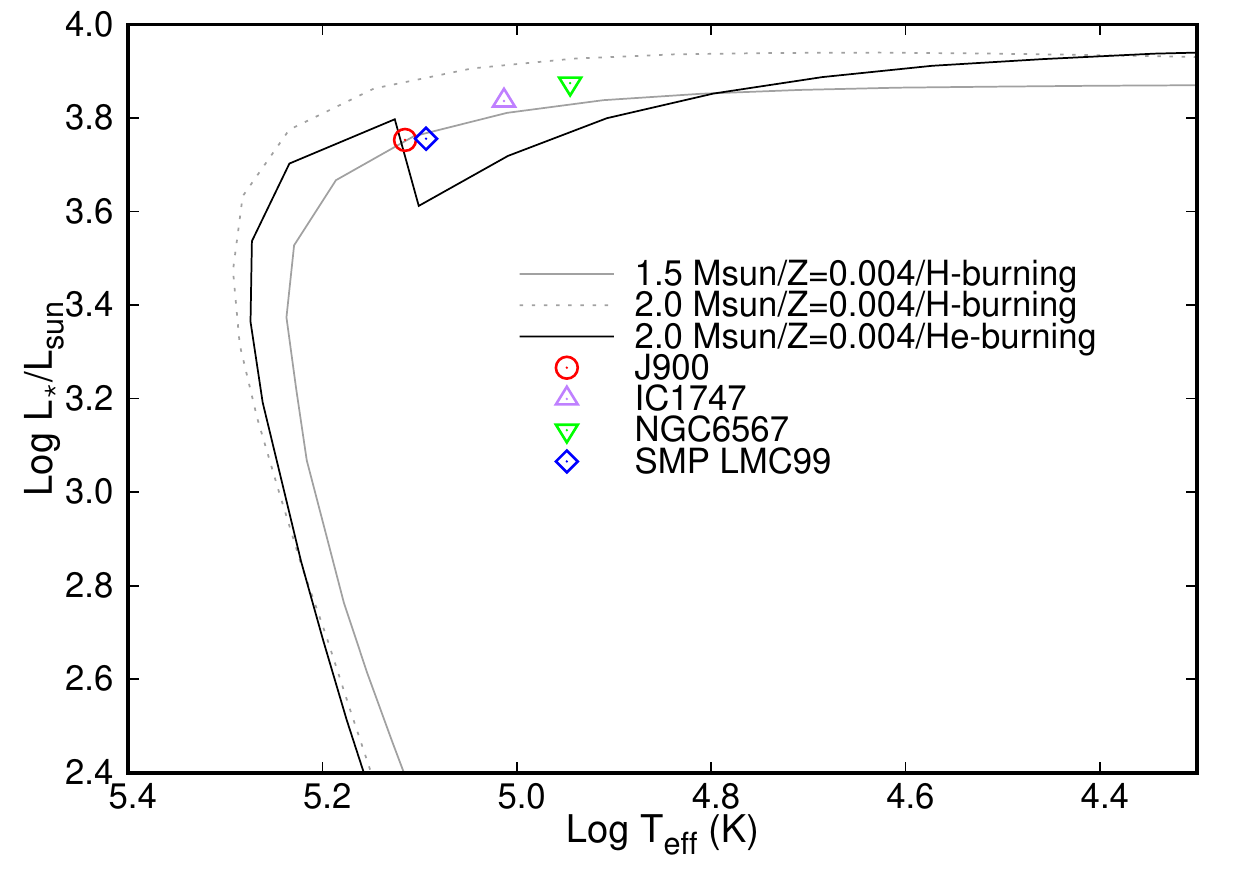}
\vspace{-15pt}
\caption{The location of the CSPN of J900 (red circle) on the post-AGB H-shell and He-shell burning tracks. 
For comparisons, we plot the location of PNe, IC1747, NGC6567, and SMP LMC99 (see \S\,\ref{S:dis}). 
See text for details.
}
\label{F-track}
\end{figure}

The input parameters and the derived physical quantities are summarised in Tables\,\ref{T-modelin}, \ref{T-mass}, and \ref{T-NC}. 
The observed and model predicted line fluxes and 
band fluxes/flux densities are compared in Appendix Table\,\ref{T-cloudy} 
(reduced-$\chi^{2}$ of 27 in the best model). We try to reproduce the observed H$_{2}$ line fluxes at 6.91/9.68/12.28\,{\micron}. 
From such efforts, we have found a reasonable temperature floor from the best-fitting model (829\,K), which is close to 
the empirically derived $T$(H$_{2}$) = $755 ~\pm~ 36$\,K 
in the case of fitting towards the above three H$_{2}$ lines (see \S\,\ref{S-H2}). The best-fitting model predicts the 
H$_{2}$ 6.91, 9.68, and 12.28\,{\micron} line fluxes of 0.555, 0.768, and 0.128 with respect to $I$({\hb}) = 100, respectively. 
Our model succeeds in reproducing H$_{2}$ in the mid-IR, which is the majority of the molecular gas mass. 
Due to such a high floor temperature for CO lines, the predicted $N$(CO) is far from 
the estimated $N$(CO) by \citet{Huggins:1996aa} (Table\,\ref{T-NC}). Fitting the observationally 
estimated $N$(CO) requires a lower floor temperature, accordingly causing lower H$_{2}$ 6.91/9.68\,{\micron} line fluxes and higher $N$(H$_{2}$). 
However, we should take into account that the upper limit of $N$(CO) by \citet{Huggins:1996aa} 
might be too large because their estimate was done using a no-constrained temperature range ($7-77$\,K) 
and the uncertain CO $J=2-1$ line flux only. The predicted radial profiles of $N$(C) and $N$(CO) (not presented here) 
indicate that CO seems to be largely not disassociated. The currently best-fitting model can be revised 
by detecting more CO lines and other molecular gas lines in future observations.

Our model fairly well reproduces the empirically determined $\epsilon$(X) (Table\,\ref{T-Cabund}), except for Mg \& Si. The predicted $\epsilon$(Mg,Si) values are 
7.08 and 5.97,  whereas those empirically determined values were $7.35 ~\pm~ 0.06$ and $6.97 ~\pm~ 0.14$, 
respectively (Table\,\ref{T-elements}). The discrepancy of $\epsilon$(Mg) may come from the ICF(Mg) (3.34 
in the model vs. $7.28 ~\pm~ 1.02$ in the empirical method). By applying the model predicted ICF(Mg) to the empirically determined 
Mg$^{3+,4+}$, we obtain $\epsilon$(Mg) = $7.35 - \log\,(7.35/3.34)$ =  7.01. The large discrepancy of $\epsilon$(Si) must 
be caused by the line emissivity of [Si\,{\sc ii}]\,34.8\,{\micron} which is peaked in the neutral gas 
regions beyond the ionisation front (i.e., $R > 3.5''$). In $\epsilon$(Si) estimation, we assumed 
that the S/S$^{+}$ and Si/Si$^{+}$ ratios are comparable. However, this assumption is irrelevant 
as the model predicts that S/S$^{+}$ and Si/Si$^{+}$ ratios are 8.49 and 4.59, respectively. 
Along with {\te}({\oi}) and {\Ne}({\NI}) for $\epsilon$(Si) and ICF(Si) = 4.59 {from the model}, we obtain a slightly improved $\epsilon$(Si) = $6.24 ~\pm~ 0.07$. 
We will avoid any further discussion for the Si abundance since it involves various uncertainty.

The modelled SED (red line) displays a reasonable fit to the observed one 
(grey lines and blue circles, Fig.\,\ref{F-SED}) except for the 
$\sim$$2-5$\,{\micron} SED (inset of Fig.\,\ref{F-SED}), where the observed 
SED substantially exceeds the modelled one. In Fig.\,\ref{F-NIR}, we show 
the residual flux density between the observed and the corresponding predicted SEDs. 
The NIR-excess as seen in 
Fig.\,\ref{F-NIR} is clearly broad and is not due to the emission of H$_{2}$, PAH, and atomic gas.

The NIR-excess SED in J900 was firstly suggested by \citet{1991A&A...250..179Z}.
This near-IR (NIR) excess can be reproduced by 
an approximate luminosity of 85\,L$_{\sun}$ from the gas shell radius of 19\,AU (0.003$''$) with a single blackbody 
temperature of $870 ~\pm~ 20$\,K.  Therefore, the NIR-excess would be due to either (i) thermal radiation from 
very small dust grains or (ii) normal-sized dust grains in substructures surrounding the CSPN 
(e.g., disc). We rule out the former possibility; as the heat capacity in small-sized grains is smaller than that in large ones, small-sized grains distributed nearby the central star would heat up over the grain sublimation temperature (1750\,K) by the central star radiation. We confirmed this in a test model including AC grains with $a < 0.005$\,$\mu$m. 
The estimated emitting radius seems to support the latter possibility.

In Fig.\,\ref{F-track},  we plot the derived luminosity ($L_{\ast}$) and $T_{\rm eff}$ of the CSPN 
on the post-AGB evolutionary tracks. For comparison, we plot the location of the 
PNe, IC1747, NGC6567, and SMP LMC99, also (see \S\,\ref{S:dis} for details). 
These plots are used to infer the initial mass of the progenitor star as well as its post-AGB evolution. Here, we select the H-shell burning 
post-AGB evolutionary track of stars with initial mass 1.5\,M$_{\odot}$ and 2.0\,M$_{\odot}$ with $Z=0.004$ by 
\citet{Vassiliadis:1994ab}. Additionally, we plot a He-shell burning evolutionary track 
of a star of initial mass 2.0\,M$_{\sun}$ star with $Z=0.004$ generated by a linear extrapolation from 
the 2.0\,M$_{\sun}$/$Z=0.0016$ and 2.0\,M$_{\sun}$/$Z=0.008$ He-burning models of \citet{Vassiliadis:1994ab}. 
The location of the CSPN on these tracks indicates that J900 evolved from a star of initial mass of $\sim$1.5-2\,M$_{\sun}$. 
We will further discuss the origin and evolution of J900 in \S\,\ref{S:dis}.

\begin{table}
\renewcommand{\arraystretch}{0.80}
\caption{
\label{T-Cabund}
Comparison of $\epsilon$(X) determined by our empirical method ($\epsilon$(X)$_{\rm Emp.}$, see Table\,\ref{T-elements}) and {\sc Cloudy} best fitting ($\epsilon$(X)$_{\rm Cloudy}$).
The values in the last column are the difference between them.
}
\centering
\begin{tabularx}{\columnwidth}{@{}@{\extracolsep{\fill}}lD{p}{\pm}{-1}D{.}{.}{-1}D{.}{.}{-1}@{}}
\midrule
X  &\multicolumn{1}{c}{$\epsilon$(X)$_{\rm Emp.}$} &\multicolumn{1}{c}{$\epsilon$(X)$_{\rm Cloudy}$} 
&\multicolumn{1}{c}{$\Delta$($\epsilon$(X)$_{\rm Cloudy}$ -- $\epsilon$(X)$_{\rm Emp.}$)}\\
\midrule
He &11.04 ~p~ 0.04&10.97 &-0.07\\
C  &9.04 ~p~ 0.10&8.92   &-0.12\\
N  &7.84 ~p~ 0.11&7.66   &-0.18\\
O  &8.47 ~p~ 0.02&8.47   &0.00\\
F  &5.16 ~p~ 0.07&5.03   &-0.13\\
Ne &7.95 ~p~ 0.02&7.90   &-0.05\\
Mg &7.35 ~p~ 0.06&7.08   &-0.27\\
Si &6.97 ~p~ 0.14&5.97   &-1.00\\
P  &5.49 ~p~ 0.05&5.21   &-0.28\\
S  &6.53 ~p~ 0.02&6.47   &-0.06\\
Cl &4.80 ~p~ 0.03&4.83   &+0.03\\
Ar &5.95 ~p~ 0.02&5.89   &-0.06\\
K  &4.44 ~p~ 0.07&4.28   &-0.16\\
Fe &5.83 ~p~ 0.21&5.63   &-0.20\\
\midrule
\end{tabularx}
\end{table}

\begin{table}
\renewcommand{\arraystretch}{0.80}
\caption{
\label{T-mass}
Nebular mass component  in J900.}
\centering
\begin{tabularx}{\columnwidth}{@{}@{\extracolsep{\fill}}lcc@{}}
\midrule
Gas species          &within ionisation front&entire nebula\\
\midrule
Ionised atomic  (M$_{\sun}$)&2.61(--1) &2.65(--1)\\
Neutral atomic  (M$_{\sun}$)&1.66(--2) &5.48(--1)\\
Molecules       (M$_{\sun}$)&4.52(--4) &1.11(--2)\\
Total           (M$_{\sun}$)&2.78(--1) &8.25(--1)\\
\midrule
Dust and PAH species &within ionisation front&entire nebula\\
\midrule
AC grains   (M$_{\sun}$)&1.43(--4) &4.24(--4)\\
Ionised PAH (M$_{\sun}$)&8.49(--7) &8.90(--6)\\
Neutral PAH (M$_{\sun}$)&1.24(--6) &1.29(--5)\\
Total       (M$_{\sun}$)&1.45(--4) &4.46(--4)\\
\midrule
GDR &1916&1849\\
\midrule
\end{tabularx}
\end{table}

\begin{table}
\renewcommand{\arraystretch}{0.80}
\caption{
\label{T-NC}
Comparison between the model predicted and observed molecular gas column densities. 
$N$(H$_{2}$) is the value of the warm temperature component. 
$N$(CO) is the empirically determined value by \citet{Huggins:1996aa}.}
\centering
\begin{tabularx}{\columnwidth}{@{}@{\extracolsep{\fill}}lcc@{}}
\midrule
Species X &$\log\,N$(X)(model) (cm$^{-2}$)&$\log\,N$(X)(obs) (cm$^{-2}$)\\
\midrule
H$_{2}$  &18.94&$19.06 ~\pm~ 0.11$\\
CO       &13.29&$<14.95$\\
\midrule
\end{tabularx}
\end{table}

In Table\,\ref{T-mass}, we summarise the nebular mass components and the gas-to-dust mass ratio (GDR) within the ionisation front 
(i.e., PDRs are excluded; see the second column) and in the entire nebula (i.e., PDRs included; see the third column). 
The GDR corresponds to the ratio of (total gas mass) to (AC dust + ionised and neutral PAHs). 
In Table\,\ref{T-NC}, we compare the predicted $N$(H$_{2}$) and $N$(CO) with the observed values. 
To account for the observed H$_{2}$ line fluxes, we added regions with a constant $n_{\rm H} = 87\,500$\,cm$^{-3}$ 
beyond the ionisation front. Therefore, it is not unusual that the majority of the predicted gas mass is a 
neutral atomic gas component. Similarly, most of the AC grains and PAHs are co-distributed in the neutral atomic/molecular gas-rich PDRs. Note that our model revises previous gas mass and GDR estimates. 
It is very important that about 67\,$\%$ of the total gas and AC grains exists beyond the ionisation front.

In earlier studies, \citet{Huggins:1996aa} estimated the ionised atomic gas mass of 1.1(--2)\,M$_{\sun}$ and the molecular 
gas mass of 4.0(--4)\,M$_{\sun}$ using the line flux of the tentative detection of CO $J=2-1$ at 230\,GHz and unconstrained 
CO gas temperature and CO/H$_{2}$ mass ratio under $D = 1.6$\,kpc. 
Adopting $D = 6.0$\,kpc, their estimated masses would increase up to 0.15\,M$_{\sun}$ 
for the ionised atomic gas and 5.63(--3)\,M$_{\sun}$ for the molecular gas. 
However, these values are far from our ``total'' gas mass estimate (i.e., the sum of the ionised/neutral atomic gas and molecular gas masses) 
because they did not consider the neutral gas component. 
\citet{1999A&A...352..297S} derived GDR of 714 which is the ratio of the ionised atomic gas mass to the dust mass. 
They estimated the dust mass by blackbody SED fitting for the IRAS bands with contributions from 
components (e.g., atomic gas emission) other than dust continuum, where they excluded the neutral atomic and molecular gas components. If we exclude these two gas components, the GDR would decrease from 1849 to 585 [= 2.61(--1)\,M$_{\sun}$/4.46(--4)\,M$_{\sun}$], which is comparable with \citet{1999A&A...352..297S}.

In short, our model thoroughly revises the previous gas/dust mass and GDR estimates, 
which allows us to  trace the ejected mass from the progenitor of J900. Furthermore, 
our modelling work implies that we often tend to underestimate the whole mass-loss 
history of the PN progenitors and GDR as well, when PDRs are excluded. 
Our model also seems to remind us of the importance of PDRs in understanding stellar mass loss.

\section{Discussion}
\label{S:dis}

\subsection{Comparison with AGB models}
\label{S-AGB}

\begin{figure}
 \includegraphics[width=\columnwidth]{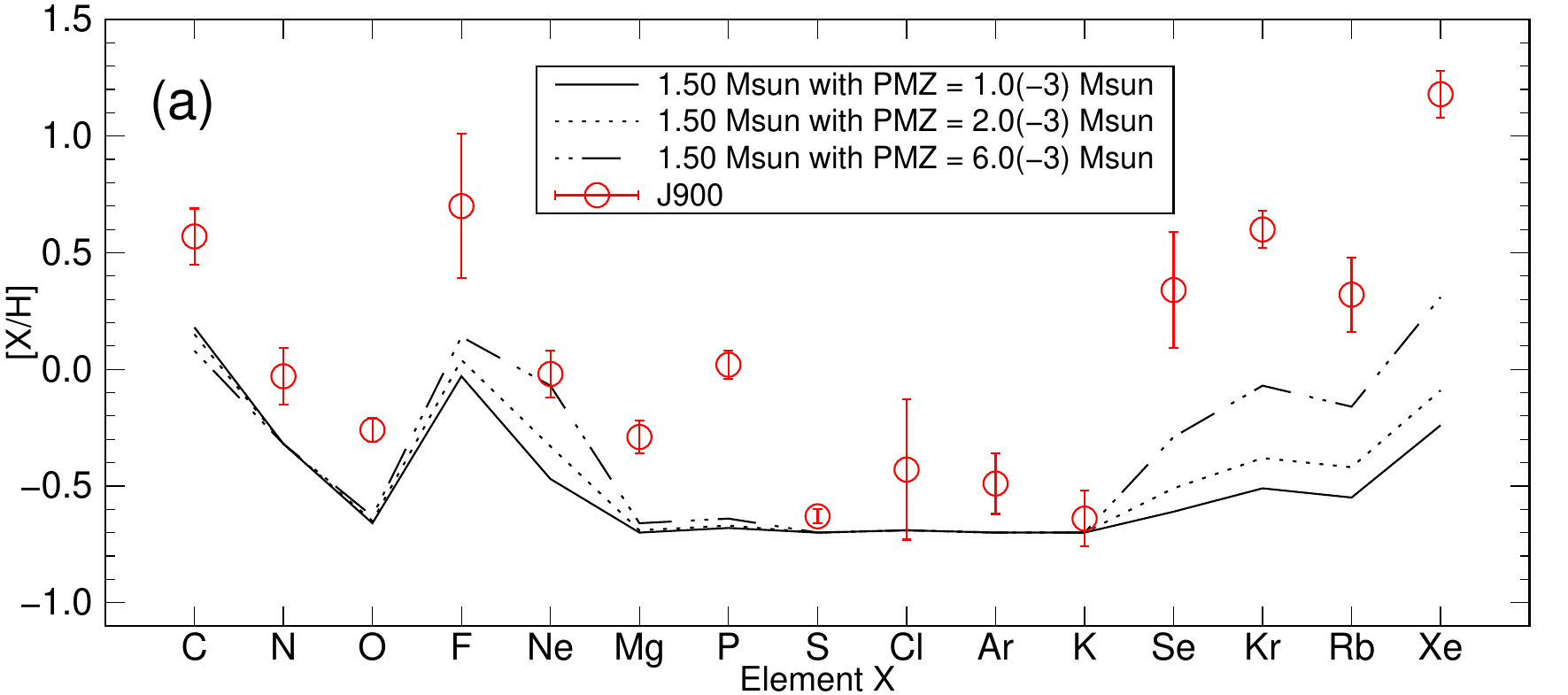}\\
 \includegraphics[width=\columnwidth]{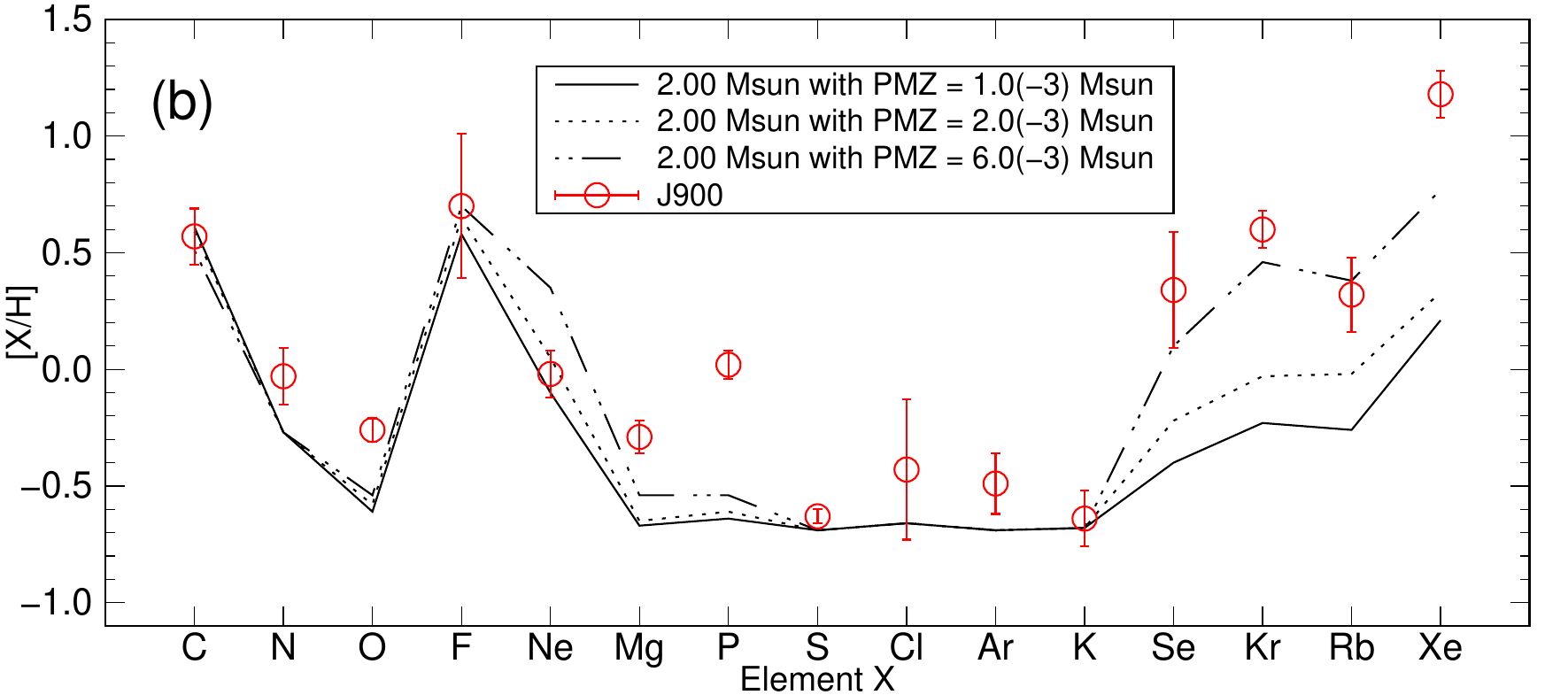}\\
 \includegraphics[width=\columnwidth]{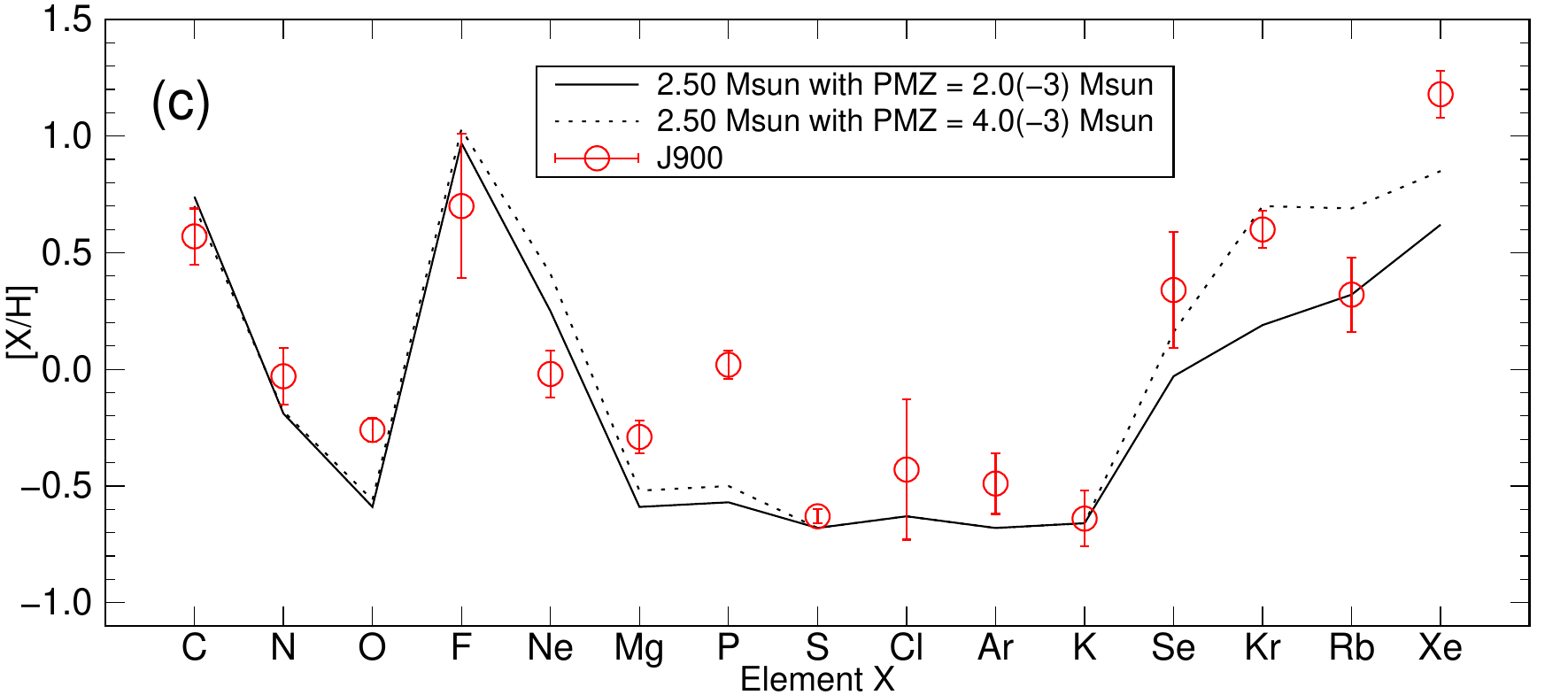}
\vspace{-15pt}
\caption{Comparisons between the observed elemental abundances (red circles) 
and the AGB nucleosynthesis model predictions by \citet{Karakas:2018aa} 
for stars of initial mass of 1.5, 2.0, and 2.5\,M$_{\sun}$ stars with 
different partial mixing zone (PMZ) masses. Here, we adopt the empirically determined values listed 
in column (9) of Table\,\ref{T-elements}.}
\label{F-agb}
\end{figure}

\begin{figure}
 \includegraphics[width=\columnwidth]{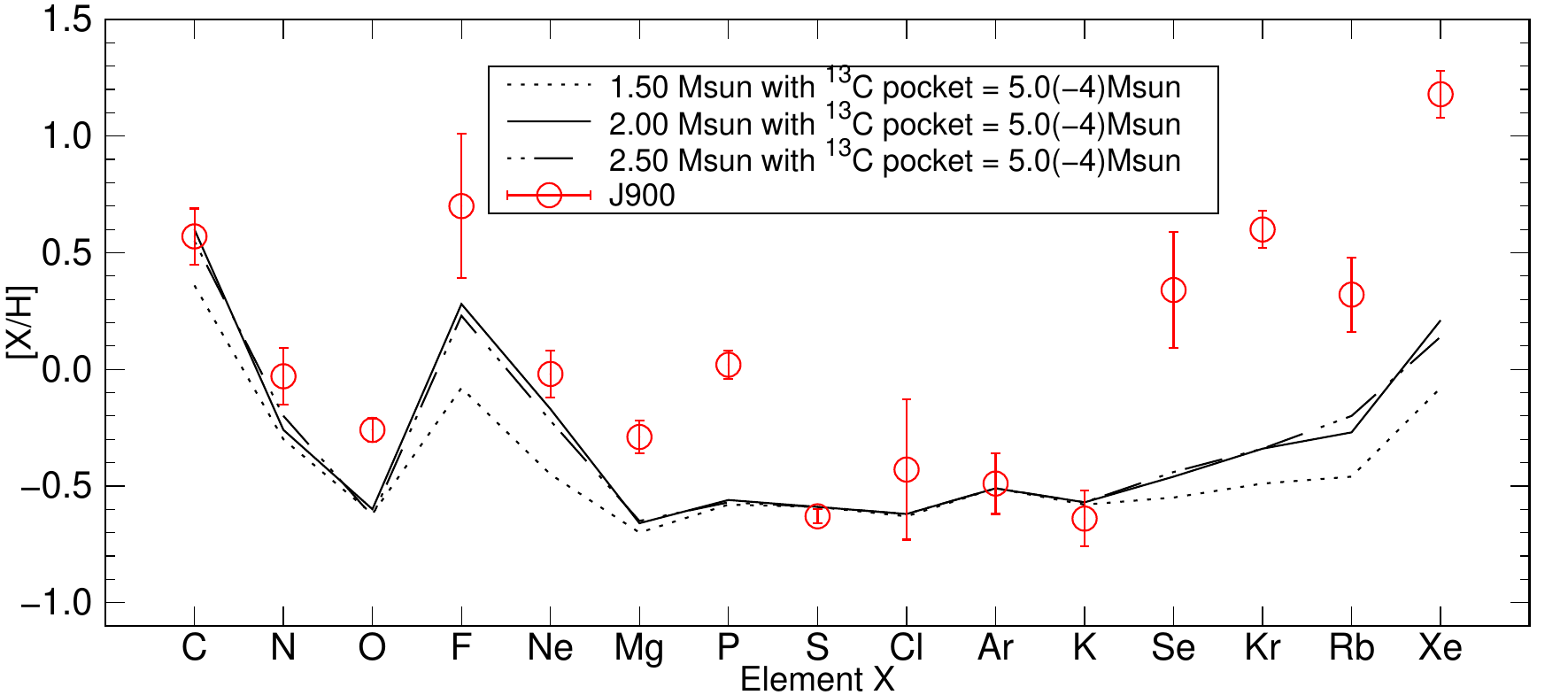}
\vspace{-15pt}
\caption{Comparisons between the observed elemental abundances (red circles) 
and the AGB nucleosynthesis model predictions for tars of initial mass of 1.5, 2.0, 
and 2.5\,M$_{\sun}$ with $Z=0.003$, which are taken from the FRUITY database \citep{Cristallo:2011aa,Cristallo:2015aa}. 
These models adopt the standard $^{13}$C pocket mass for low mass stars \citep[$\sim$5.0$\times10^{-4}$\,M$_{\sun}$;][]{Gallino:1998aa}. 
}
\label{F-agb3}
\end{figure}

\begin{table*}
\centering
\renewcommand{\arraystretch}{0.80}
\caption{Elemental abundances of PNe displaying similar abundance patterns to J900. 
In all the PNe, C abundances are the C CEL values. 
Se and Kr in IC1747 and NGC6567 are from \citet{Sterling:2015aa}. 
The abundances in IC1747 are the average between \citet{Henry:2010aa} and \citet{Wesson:2005aa}. 
C, N, and Ne in SMP LMC99 are from \citet{Leisy:2006aa}, and 
the others are taken from \citet{Mashburn:2016aa}. $\epsilon$(Se) in J900 is from \citet{Sterling:2015aa}.
}
\begin{tabularx}{\textwidth}{@{}@{\extracolsep{\fill}}lcccccccccccccl@{}}
\midrule
PN       
&  $\epsilon$(He)  
&$\epsilon$(C)    
&$\epsilon$(N)
&$\epsilon$(O)
&$\epsilon$(F)       
&$\epsilon$(Ne)  
&$\epsilon$(S)   
&$\epsilon$(Cl)  
&$\epsilon$(Ar)
&$\epsilon$(Se)   
&$\epsilon$(Kr) 
&$\epsilon$(Rb)
&$\epsilon$(Xe)
&Ref.\\
\midrule
IC1747  &11.07 &8.98 &8.14 &8.58 &$\cdots$&8.01&6.66&4.98&6.14&3.68 &$<4.34$\phantom{m}   &$\cdots$    &$\cdots$&(1),(2),(3)\\
NGC6567 &11.01 &8.74 &7.61 &8.30 &$\cdots$&7.45&6.68&4.75&5.67&3.23&$<3.62$\phantom{m}   &$\cdots$    &$\cdots$&(3),(4)\\
SMP LMC99&11.04&8.69 &8.15 &8.42 &$\cdots$&7.62&6.42&$\cdots$&5.95&3.56&3.78      &$<2.62$\phantom{m}     &$\cdots$&(5),(6)\\ 
J900     &11.04 &9.04 &7.84 &8.47 &5.16    &7.95&6.53&4.80&5.95&3.65 &3.85      &2.86        &3.42&(3),(7)\\
\midrule
\end{tabularx}
\begin{description}
\item[References -- ] (1) \citet{Henry:2010aa}; (2) \citet{Wesson:2005aa}; (3) \citet{Sterling:2015aa}; 
(4) \citet{Hyung:1993aa};  (5) \citet{Leisy:2006aa}; (6) \citet{Mashburn:2016aa}; (7) This work.
\end{description}
\label{T-compA}
\end{table*}

In order to verify and further constrain the evolution of the progenitor star, then, we compare the observed and the AGB model predicted $\epsilon$(X).  Here, we select the AGB nucleosynthesis models of \citet{Karakas:2018aa} for the initially 1.5, 2.0, and 2.5\,M$_{\sun}$ stars with $Z=0.003$. The models of \citet{Karakas:2018aa} artificially add a partial mixing zone (PMZ) at the bottom of the convective envelope during each TDU. At the PMZ, protons from the H-envelope are mixed into the He-intershell and then captured by $^{12}$C. Thus, $^{13}$C pocket as an extra $n$-source is formed. In Fig.\,\ref{F-agb}(a) to (c), we plot the observed (red circles with an error bar) and the model predicted abundances (lines and dots). As \citet{Karakas:2018aa} use the scaled solar abundance of \citet{Asplund:2009aa} as the initial abundance, we adopt the empirically determined values listed in the column (9) of Table\,\ref{T-elements} (i.e, [X/H]$_{\rm AS}$). 
It should be noted that empirically determined $\epsilon$(Mg) would be overestimated as we explained in \S\,\ref{S:model-r}; 
adopting the {\sc Cloudy} model predicted ICF(Mg), we obtain [Mg/H]$_{\rm AS}$ = $-0.63 ~\pm~ 0.07$ whose value is in agreement with the AGB model predicted values. In these plots, we find that (1) C, F, Ne, and $n$-capture elements increase as the initial mass is greater, (2) F production is more sensitive to the initial mass than C and Ne productions, and (3) Ne production is 
sensitive to both the initial mass of the progenitor star and the PMZ mass. It is the most important finding that all the observed $n$-capture elemental abundances cannot be explained without PMZ masses and their yields are in large dependence of PMZ mass. Thus, the C$^{13}$ pocket is certainly formed in the progenitor star. Either the 2.0\,M$_{\sun}$ model with PMZ $= 6.0(-3)$\,M$_{\sun}$ or the 2.5\,M$_{\sun}$ model with PMZ $= 4.0(-3)$\,M$_{\sun}$ are in excellent accordance with the observed $\epsilon$(X). The enhancement of Ne is related to $^{22}$Ne, which is formed via the double $\alpha$ capturing by $^{14}$N in $^{13}$C pocket. In $\ge 2.5$\,M$_{\odot}$ models, the He-shell temperature exceeds 300M\,K required to activate the $^{22}$Ne($\alpha$,$n$)$^{25}$Mg reaction during the last few TPs \citep[see their Fig.\,5 of ][]{Karakas:2018aa}. Consequently, as suggested in \citet{Raai:2012aa}, high $n$ density produces more Rb over Kr as seen in Fig.\,\ref{F-agb}(c). From the comparison between the observed and predicted [Rb/Kr], we infer the initial mass to 
be $< 2.5$\,M$_{\sun}$. The derived gas mass (0.838\,M$_{\sun}$, Table\,\ref{T-mass}) and the range of the CSPN mass 
\citep[$0.64-0.69$\,M$_{\sun}$;][]{Vassiliadis:1994ab}, the progenitor is certainly a $>1.5$\,M$_{\sun}$ star.

Note that the observed O abundance in J900 is larger than the predicted O in the models with any PMZ masses. 
The observed [O/(Cl,Ar)] and $\epsilon$(O,Cl) in J900 are along the result of \citet{Delgado-Inglada:2015aa}, 
who demonstrated that O in C-rich PNe is enhanced by $\sim$0.3\,dex in their plots of O/Cl versus $\epsilon$(O) 
and $\epsilon$(Cl). For an explanation of the O enhancement in C-rich PNe, 
\citet{Delgado-Inglada:2015aa} proposed a convective overshooting, which is not considered 
in the AGB models of \citet{Karakas:2018aa}.

There are several other AGB nucleosynthesis models. Of these, we compare the observed abundances with the predictions by the FRUITY models \citep{Cristallo:2011aa,Cristallo:2015aa} as presented in Fig.\,\ref{F-agb3}. 
These models adopt the standard $^{13}$C pocket mass \citep[$\sim$5$\times10^{-4}$\,M$_{\sun}$;][]{Gallino:1998aa}. 
Clearly, any FRUITY models fail to explain the observed 
$n$-capture elements, owing to low $^{13}$C pocket mass. Thus, we realise the importance of the PMZ/$^{13}$C pocket 
in the production of these elements.

The evolutionary age of the progenitors does not make much difference, whether or not the progenitor experienced  H-burning (i.e., evolving to H-rich WD) or He-burning (i.e., to H-poor WD) after the AGB phase. We estimate the current age of the progenitor to be 
$\sim$$1.2$\,Gyr for a star of initial mass of 2.0\,M$_{\sun}$ 
(or $\sim$$0.6$\,Gyr for 2.5\,M$_{\sun}$) with $Z=0.004$
by referring to the H-burning post-AGB evolutionary tracks of \citet{Vassiliadis:1994ab}. 
Our estimate is largely different from \citet{Stanghellini:2018aa}, who classified 
J900 as the older population with its age of $>7.5$\,Gyr in terms of C and N enhancements only.

We summarise this section as follows. We compare the observed 15 elemental abundances with AGB model predictions. As a result, 
we derive the current evolutionary status of the CSPN. J900 is a young C-rich PN together with large enhancements of F and $n$-capture elements, and it evolved from a star of initial mass of $\sim$2.0\,M$_{\sun}$ formed in a low-metallicity environment ($Z\sim0.003$).

\subsection{Comparison with other PNe}

\begin{figure}
\includegraphics[width=\columnwidth]{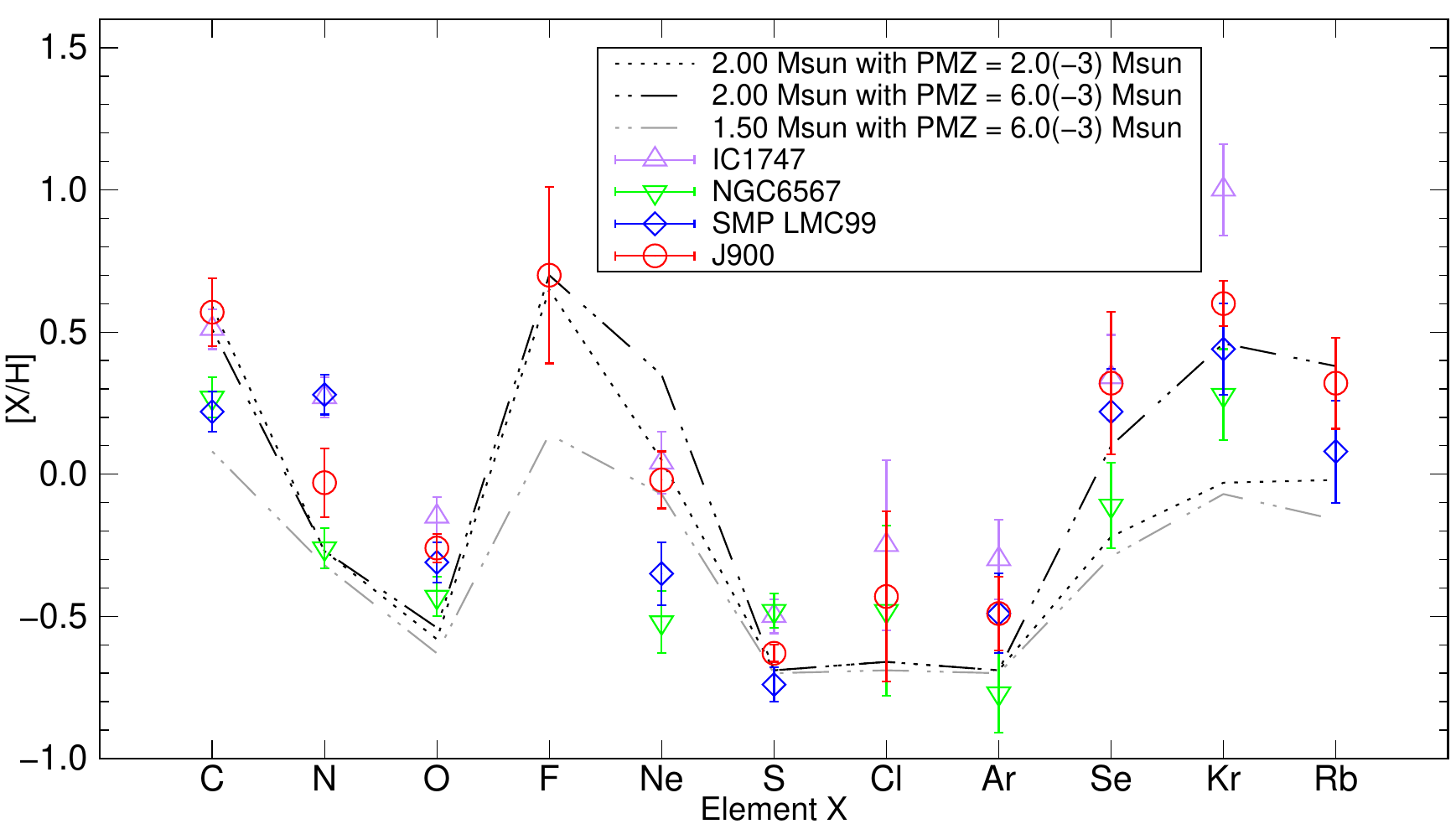}
\vspace{-15pt}
\caption{Comparisons between the observed elemental abundances of PNe (Table\,\ref{T-compA}) 
and the AGB nucleosynthesis model predictions by \citet{Karakas:2018aa} 
for stars of initial mass of 1.5 and 2.0\,M$_{\sun}$ with different PMZ masses.}
\label{F-agb2}
\end{figure}

We compare J900 with PNe IC1747 and NGC6567 in the Milky Way and SMP LMC99 in the Large Magellanic Cloud in order 
to examine the validity of our derived abundances and the possible range of the PMZ mass in J900. 
We select these PNe because their abundances are very similar to J900, and $n$-capture elements are already measured (Table\,\ref{T-compA}).

In Fig.\,\ref{F-track}, we plot their $T_{\rm eff}$ and $L_{\ast}$ on the post-AGB 
evolutionary tracks in order to check the current evolutionary status and 
the initial mass of these three PNe. 
The respective $T_{\rm eff}$ values are 103\,090\,K in IC1747 and 88\,130\,K in NGC6567, which are 
derived using the observed $I$({\heii}\,4686\,{\AA})/$I$({\hb}) ratio and the equation 
(3.1) of \citet{Dopita:1991aa}. In the $L_{\ast}$ calculations for IC1747 (6860\,L$_{\sun}$) and NGC6567 
(7490\,L$_{\sun}$), we utilise the stellar atmosphere model 
of \citet{Rauch:2003aa} with the derived $T_{\rm eff}$, $\log\,g$ = 5.0\,cm\,s$^{-2}$, 
[Z/H] = --0.5, and the following parameters: (1) For IC1747,  $D$ = 3.24\,kpc 
\citep[from GAIA parallax;][]{Gaia:2018aa}, $m_{V}$ of the CSPN 
\citep[16.4;][]{Tylenda:1991aa}, and \mbox{c({\hb}) = 1.00} \citep{Wesson:2005aa}; (2) For NGC6567, 
$D$ = 2.4\,kpc \citep{Phillips:2004aa}, 
{\sl HST} NICMOS1 $F_{\lambda}$ of 6.42(--12)\,erg\,s$^{-1}$\,cm$^{-2}$\,{\micron}$^{-1}$ at the F108N 
band ($\lambda_{\rm c} = 1.082$\,{\micron})\footnote{We measured this value from the {\sl HST} archived data (Programme ID: 7837, PI: S.R. Pottasch).}, 
and \mbox{c({\hb}) = 0.68} \citep{Hyung:1993aa}. $T_{\rm eff}$ and $L_{\ast}$ of LMC99 
(124\,000\,K and 5700\,L$_{\sun}$) are taken from \citet{Dopita:1991aa}.

The transition from the AGB to the PN involves stellar mass loss, involving a stellar wind change from a slow to a fast speed. During this transition period, the central star might experience the He-burning phase, responsible for the presently derived abundances.
As seen in Fig.\,\ref{F-track}, all the CSPN temperatures are presently increasing (not yet cooling down) along the post-AGB tracks toward higher temperature values. The locations seem to indicate that these PNe presumably originated from similar initial mass stars and exited the AGB phase a fairly long ago. Thus, we can safely compare their abundances with the theoretical predictions of \citet{Karakas:2018aa} for the initially 
$1.5-2.0$\,M$_{\sun}$ stars.

In Fig.\,\ref{F-agb2}, we plot the observed and the model-predicted abundances. For IC1747, NGC6567, and LMC99, we 
adopt the uncertainty of 0.05\,dex for $\epsilon$(C/N/O/Ne/S/Cl/Ar) and 0.15\,dex for $\epsilon$(Se/Kr/Rb), respectively. 
We find the following results: (1) [C,Ne/H] indicates that J900 and IC1747 evolved from the initially 
2.0\,M$_{\sun}$ stars as described in \S\,\ref{S-AGB}, 
and (2) NGC6567 and LMC99 from the stars whose initial mass could be between 
1.5 and 2.0\,M$_{\sun}$. Meanwhile, (3) [Se,Kr,Rb] suggests that all of these PNe are descendants of 2.0\,M$_{\sun}$ stars. As seen in the case of J900 (and see Fig.\,\ref{F-agb}(c)), [Kr/Rb] $< 0 $ in LMC99 
suggests that the initial mass of this PN does not exceed 2.5\,M$_{\sun}$. 

Although there is a slight indeterminacy in the initial mass, all of these PNe show very similar abundance patterns, which 
can be adequately explained by either 1.5\,M$_{\sun}$ or 2.0\,M$_{\sun}$ models. Based on the self-consistent model for the observables, 
we confirm that 
(1) the derived elemental abundances of J900 are not peculiar and but quite similar to IC1747, NGC6567, and LMC99;
(2) PMZ mass would be $(2.0-6.0)\times10^{-3}$\,M$_{\sun}$ required to match with the observed abundances of all of these PNe; and 
(3) the initial masses predicted from elemental abundances are consistent with those deduced from the locations on the post-AGB evolution tracks.

\section{Summary}
\label{S:summary}

We performed detailed spectroscopic analyses of J900 in order to characterise 
the properties of the CSPN and nebula. We obtained 17 elemental abundances. 
J900 is a C, F, and $n$-capture element rich PN. We investigated 
the physical conditions of H$_{2}$. The H$_{2}$ lines are likely to be 
emitted from the warm ($\sim$670\,K) and hot ($\sim$3200\,K) temperature regions. 
We constructed the SED model to be consistent with all the observed quantities 
of the CSPN and the dusty nebula. We found that about 67\,\% of the total dust 
and gas components exist beyond the ionisation front, indicating the presence 
of the neutral atomic and molecular gas-rich PDRs, critically important 
in the understanding of the stellar mass loss and also the recycling of galactic material. 
The best-fitting SED shows an excellent agreement with the observations except 
for the observed near-IR SED. The near-IR excess suggests the presence of 
a high-density structure near the central star. The best-fitting SED model 
indicates that the progenitor evolved from a star of initial mass 
$\sim$2.0\,M$_{\sun}$ had been in the course of the He-burning phase after AGB. 
The present age is likely to be $\sim$1\,Gyr after the progenitor star was formed. 
The derived elemental abundance pattern is consistent with that predicted 
by the AGB nucleosynthesis model for 2.0\,M$_{\sun}$ stars with $Z=0.003$
 and a PMZ $= 6.0(-3)$\,M$_{\sun}$. Other models without PMZ cannot accommodate 
the observed abundances of $n$-capture elements, strongly suggesting 
that the $^{13}$C pocket is likely to be 
formed in the He-intershell of the progenitor. We showed how critically important the 
physical properties of the CSPN and the nebula derived through multiwavelength 
data analysis are for understanding the origin (i.e., initial mass and age) and 
internal evolution of the PN progenitors. Accurately determined abundances 
(in particular, C/F/Ne/$n$-capture elements) and gas/dust masses are very 
helpful for these purposes.

\section*{Acknowledgements}

We are grateful to the anonymous referee for improving the paper. 
We thank Dr. Beth Sargent for her careful reading and valuable suggestions. 
We wish to acknowledge Dr. Toshiya Ueta for his constructive suggestions. 
We sincerely express our thanks to Dr. Seong-Jae Lee, who conducted
 BOES observations at the BOAO. 
MO was supported by JSPS Grants-in-Aid for Scientific Research(C) 
(JP19K03914), and the research fund 104-2811-M-001-138 and 104-2112-M-001-041-MY3 
from the Ministry of Science and Technology (MOST), Republic of China. 
SH would like to acknowledge support from the Basic Science Research Program 
through the National Research Foundation of Korea (NRF 2017R1D1A3B03029309). 
This work was partly based on archival 
data obtained with the \emph{Spitzer} Space Telescope,
 which is operated by the Jet Propulsion Laboratory, California
 Institute of Technology under a contract with NASA. This research
 is in part based on observations with
 \emph{AKARI}, a JAXA project with the participation of ESA.
 

\begin{thebibliography}{}
\makeatletter
\relax
\def\mn@urlcharsother{\let\do\@makeother \do\$\do\&\do\#\do\^\do\_\do\%\do\~}
\def\mn@doi{\begingroup\mn@urlcharsother \@ifnextchar [ {\mn@doi@}
  {\mn@doi@[]}}
\def\mn@doi@[#1]#2{\def\@tempa{#1}\ifx\@tempa\@empty \href
  {http://dx.doi.org/#2} {doi:#2}\else \href {http://dx.doi.org/#2} {#1}\fi
  \endgroup}
\def\mn@eprint#1#2{\mn@eprint@#1:#2::\@nil}
\def\mn@eprint@arXiv#1{\href {http://arxiv.org/abs/#1} {{\tt arXiv:#1}}}
\def\mn@eprint@dblp#1{\href {http://dblp.uni-trier.de/rec/bibtex/#1.xml}
  {dblp:#1}}
\def\mn@eprint@#1:#2:#3:#4\@nil{\def\@tempa {#1}\def\@tempb {#2}\def\@tempc
  {#3}\ifx \@tempc \@empty \let \@tempc \@tempb \let \@tempb \@tempa \fi \ifx
  \@tempb \@empty \def\@tempb {arXiv}\fi \@ifundefined
  {mn@eprint@\@tempb}{\@tempb:\@tempc}{\expandafter \expandafter \csname
  mn@eprint@\@tempb\endcsname \expandafter{\@tempc}}}

\bibitem[\protect\citeauthoryear{{Abia}, {Cunha}, {Cristallo}  \& {de
  Laverny}}{{Abia} et~al.}{2015}]{Abia:2015aa}
{Abia} C.,  {Cunha} K.,  {Cristallo} S.,   {de Laverny} P.,  2015, \mn@doi
  [\aap] {10.1051/0004-6361/201526586}, \href
  {http://adsabs.harvard.edu/abs/2015A%26A...581A..88A} {581, A88}

\bibitem[\protect\citeauthoryear{{Aitken} \& {Roche}}{{Aitken} \&
  {Roche}}{1982}]{Aitken:1982aa}
{Aitken} D.~K.,  {Roche} P.~F.,  1982, \mn@doi [\mnras]
  {10.1093/mnras/200.2.217}, \href
  {http://adsabs.harvard.edu/abs/1982MNRAS.200..217A} {200, 217}

\bibitem[\protect\citeauthoryear{{Allamandola}, {Tielens}  \&
  {Barker}}{{Allamandola} et~al.}{1987}]{Allamandola:1987aa}
{Allamandola} L.~J.,  {Tielens} A.~G.~G.~M.,   {Barker} J.~R.,  1987, in
  {Morfill} G.~E.,  {Scholer} M.,  eds, NATO ASIC Proc. 210: Physical Processes
  in Interstellar Clouds. pp 305--331

\bibitem[\protect\citeauthoryear{{Aller}}{{Aller}}{1984}]{Aller:1984aa}
{Aller} L.~H.,  ed. 1984, {Physics of thermal gaseous nebulae}  Astrophysics
  and Space Science Library Vol. 112, \mn@doi{10.1007/978-94-010-9639-3.
}

\bibitem[\protect\citeauthoryear{{Aller} \& {Czyzak}}{{Aller} \&
  {Czyzak}}{1983}]{Aller:1983aa}
{Aller} L.~H.,  {Czyzak} S.~J.,  1983, \mn@doi [\apjs] {10.1086/190846}, \href
  {http://adsabs.harvard.edu/abs/1983ApJS...51..211A} {51, 211}

\bibitem[\protect\citeauthoryear{{Asplund}, {Grevesse}, {Sauval}  \&
  {Scott}}{{Asplund} et~al.}{2009}]{Asplund:2009aa}
{Asplund} M.,  {Grevesse} N.,  {Sauval} A.~J.,   {Scott} P.,  2009, \mn@doi
  [\araa] {10.1146/annurev.astro.46.060407.145222}, \href
  {http://adsabs.harvard.edu/abs/2009ARA%26A..47..481A} {47, 481}

\bibitem[\protect\citeauthoryear{{Badnell} \& {Griffin}}{{Badnell} \&
  {Griffin}}{2000}]{Badnell:2000aa}
{Badnell} N.~R.,  {Griffin} D.~C.,  2000, \mn@doi [Journal of Physics B Atomic
  Molecular Physics] {10.1088/0953-4075/33/16/302}, \href
  {http://ads.nao.ac.jp/abs/2000JPhB...33.2955B} {33, 2955}

\bibitem[\protect\citeauthoryear{{Becker}, {Butler}  \& {Zeippen}}{{Becker}
  et~al.}{1989}]{Becker:1989aa}
{Becker} S.~R.,  {Butler} K.,   {Zeippen} C.~J.,  1989, \aap, \href
  {http://ads.nao.ac.jp/abs/1989A%26A...221..375B} {221, 375}

\bibitem[\protect\citeauthoryear{{Benjamin}, {Skillman}  \& {Smits}}{{Benjamin}
  et~al.}{1999}]{Benjamin:1999aa}
{Benjamin} R.~A.,  {Skillman} E.~D.,   {Smits} D.~P.,  1999, \mn@doi [\apj]
  {10.1086/306923}, \href {http://adsabs.harvard.edu/abs/1999ApJ...514..307B}
  {514, 307}

\bibitem[\protect\citeauthoryear{{Bergeson} \& {Lawler}}{{Bergeson} \&
  {Lawler}}{1993}]{Bergeson:1993aa}
{Bergeson} S.~D.,  {Lawler} J.~E.,  1993, \mn@doi [\apjl] {10.1086/187015},
  \href {http://ads.nao.ac.jp/abs/1993ApJ...414L.137B} {414, L137}

\bibitem[\protect\citeauthoryear{{Bernard-Salas}, {Peeters}, {Sloan},
  {Gutenkunst}, {Matsuura}, {Tielens}, {Zijlstra}  \& {Houck}}{{Bernard-Salas}
  et~al.}{2009}]{Bernard-Salas:2009aa}
{Bernard-Salas} J.,  {Peeters} E.,  {Sloan} G.~C.,  {Gutenkunst} S.,
  {Matsuura} M.,  {Tielens} A.~G.~G.~M.,  {Zijlstra} A.~A.,   {Houck} J.~R.,
  2009, \mn@doi [\apj] {10.1088/0004-637X/699/2/1541}, \href
  {http://adsabs.harvard.edu/abs/2009ApJ...699.1541B} {699, 1541}

\bibitem[\protect\citeauthoryear{{Berrington}, {Burke}, {Dufton}  \&
  {Kingston}}{{Berrington} et~al.}{1985}]{Berrington:1985aa}
{Berrington} K.~A.,  {Burke} P.~G.,  {Dufton} P.~L.,   {Kingston} A.~E.,  1985,
  \mn@doi [Atomic Data and Nuclear Data Tables] {10.1016/0092-640X(85)90001-4},
  \href {http://ads.nao.ac.jp/abs/1985ADNDT..33..195B} {33, 195}

\bibitem[\protect\citeauthoryear{{Bhatia} \& {Doschek}}{{Bhatia} \&
  {Doschek}}{1993}]{Bhatia:1993aa}
{Bhatia} A.~K.,  {Doschek} G.~A.,  1993, \mn@doi [Atomic Data and Nuclear Data
  Tables] {10.1006/adnd.1993.1024}, \href
  {http://ads.nao.ac.jp/abs/1993ADNDT..55..315B} {55, 315}

\bibitem[\protect\citeauthoryear{{Bhatia} \& {Kastner}}{{Bhatia} \&
  {Kastner}}{1988}]{Bhatia:1988aa}
{Bhatia} A.~K.,  {Kastner} S.~O.,  1988, \mn@doi [\apj] {10.1086/166713}, \href
  {http://ads.nao.ac.jp/abs/1988ApJ...332.1063B} {332, 1063}

\bibitem[\protect\citeauthoryear{{Bhatia} \& {Kastner}}{{Bhatia} \&
  {Kastner}}{1995}]{Bhatia:1995aa}
{Bhatia} A.~K.,  {Kastner} S.~O.,  1995, \mn@doi [\apjs] {10.1086/192121},
  \href {http://ads.nao.ac.jp/abs/1995ApJS...96..325B} {96, 325}

\bibitem[\protect\citeauthoryear{{Bi{\'e}mont} \& {Hansen}}{{Bi{\'e}mont} \&
  {Hansen}}{1986}]{Biemont:1986aa}
{Bi{\'e}mont} E.,  {Hansen} J.~E.,  1986, \mn@doi [\physscr]
  {10.1088/0031-8949/33/2/006}, \href
  {http://ads.nao.ac.jp/abs/1986PhyS...33..117B} {33, 117}

\bibitem[\protect\citeauthoryear{{Biemont}, {Hansen}, {Quinet}  \&
  {Zeippen}}{{Biemont} et~al.}{1995}]{Biemont:1995aa}
{Biemont} E.,  {Hansen} J.~E.,  {Quinet} P.,   {Zeippen} C.~J.,  1995, \aaps,
  \href {http://ads.nao.ac.jp/abs/1995A%26AS..111..333B} {111, 333}

\bibitem[\protect\citeauthoryear{{Blum} \& {Pradhan}}{{Blum} \&
  {Pradhan}}{1992}]{Blum:1992aa}
{Blum} R.~D.,  {Pradhan} A.~K.,  1992, \mn@doi [\apjs] {10.1086/191670}, \href
  {http://ads.nao.ac.jp/abs/1992ApJS...80..425B} {80, 425}

\bibitem[\protect\citeauthoryear{{Boersma}, {Bauschlicher}, {Allamandola},
  {Ricca}, {Peeters}  \& {Tielens}}{{Boersma} et~al.}{2010}]{Boersma:2010aa}
{Boersma} C.,  {Bauschlicher} C.~W.,  {Allamandola} L.~J.,  {Ricca} A.,
  {Peeters} E.,   {Tielens} A.~G.~G.~M.,  2010, \mn@doi [\aap]
  {10.1051/0004-6361/200912714}, \href
  {http://adsabs.harvard.edu/abs/2010A%26A...511A..32B} {511, A32}

\bibitem[\protect\citeauthoryear{{Busso}, {Gallino}  \& {Wasserburg}}{{Busso}
  et~al.}{1999}]{Busso:1999aa}
{Busso} M.,  {Gallino} R.,   {Wasserburg} G.~J.,  1999, \mn@doi [\araa]
  {10.1146/annurev.astro.37.1.239}, \href
  {http://adsabs.harvard.edu/abs/1999ARA%26A..37..239B} {37, 239}

\bibitem[\protect\citeauthoryear{{Butler} \& {Zeippen}}{{Butler} \&
  {Zeippen}}{1994}]{Butler:1994aa}
{Butler} K.,  {Zeippen} C.~J.,  1994, \aaps, \href
  {http://ads.nao.ac.jp/abs/1994A%26AS..108....1B} {108, 1}

\bibitem[\protect\citeauthoryear{{Calamai}, {Smith}  \& {Bergeson}}{{Calamai}
  et~al.}{1993}]{Calamai:1993aa}
{Calamai} A.~G.,  {Smith} P.~L.,   {Bergeson} S.~D.,  1993, \mn@doi [\apjl]
  {10.1086/187032}, \href {http://ads.nao.ac.jp/abs/1993ApJ...415L..59C} {415,
  L59}

\bibitem[\protect\citeauthoryear{{Cami}, {Bernard-Salas}, {Peeters}  \&
  {Malek}}{{Cami} et~al.}{2010}]{Cami:2010aa}
{Cami} J.,  {Bernard-Salas} J.,  {Peeters} E.,   {Malek} S.~E.,  2010, \mn@doi
  [Science] {10.1126/science.1192035}, \href
  {http://adsabs.harvard.edu/abs/2010Sci...329.1180C} {329, 1180}

\bibitem[\protect\citeauthoryear{{Cardelli}, {Clayton}  \& {Mathis}}{{Cardelli}
  et~al.}{1989}]{Cardelli:1989aa}
{Cardelli} J.~A.,  {Clayton} G.~C.,   {Mathis} J.~S.,  1989, \mn@doi [\apj]
  {10.1086/167900}, \href {http://adsabs.harvard.edu/abs/1989ApJ...345..245C}
  {345, 245}

\bibitem[\protect\citeauthoryear{{Cristallo} et~al.,}{{Cristallo}
  et~al.}{2011}]{Cristallo:2011aa}
{Cristallo} S.,  et~al., 2011, \mn@doi [\apjs] {10.1088/0067-0049/197/2/17},
  \href {https://ui.adsabs.harvard.edu/abs/2011ApJS..197...17C} {197, 17}

\bibitem[\protect\citeauthoryear{{Cristallo}, {Straniero}, {Piersanti}  \&
  {Gobrecht}}{{Cristallo} et~al.}{2015}]{Cristallo:2015aa}
{Cristallo} S.,  {Straniero} O.,  {Piersanti} L.,   {Gobrecht} D.,  2015,
  \mn@doi [\apjs] {10.1088/0067-0049/219/2/40}, \href
  {http://adsabs.harvard.edu/abs/2015ApJS..219...40C} {219, 40}

\bibitem[\protect\citeauthoryear{{Cutri}}{{Cutri}}{2014}]{Cutri:2014aa}
{Cutri} R.~M.~e.,  2014, VizieR Online Data Catalog, \href
  {http://adsabs.harvard.edu/abs/2014yCat.2328....0C} {2328}

\bibitem[\protect\citeauthoryear{{Dabrowski}}{{Dabrowski}}{1984}]{Dabrowski:1984aa}
{Dabrowski} I.,  1984, \mn@doi [Canadian Journal of Physics] {10.1139/p84-210},
  \href {http://adsabs.harvard.edu/abs/1984CaJPh..62.1639D} {62, 1639}

\bibitem[\protect\citeauthoryear{{Davey}, {Storey}  \& {Kisielius}}{{Davey}
  et~al.}{2000}]{Davey:2000aa}
{Davey} A.~R.,  {Storey} P.~J.,   {Kisielius} R.,  2000, \mn@doi [\aaps]
  {10.1051/aas:2000139}, \href {http://ads.nao.ac.jp/abs/2000A%26AS..142...85D}
  {142, 85}

\bibitem[\protect\citeauthoryear{{Delgado-Inglada} \&
  {Rodr{\'{\i}}guez}}{{Delgado-Inglada} \&
  {Rodr{\'{\i}}guez}}{2014}]{Delgado-Inglada:2014aa}
{Delgado-Inglada} G.,  {Rodr{\'{\i}}guez} M.,  2014, \mn@doi [\apj]
  {10.1088/0004-637X/784/2/173}, \href
  {http://adsabs.harvard.edu/abs/2014ApJ...784..173D} {784, 173}

\bibitem[\protect\citeauthoryear{{Delgado-Inglada}, {Rodr{\'{\i}}guez},
  {Peimbert}, {Stasi{\'n}ska}  \& {Morisset}}{{Delgado-Inglada}
  et~al.}{2015}]{Delgado-Inglada:2015aa}
{Delgado-Inglada} G.,  {Rodr{\'{\i}}guez} M.,  {Peimbert} M.,  {Stasi{\'n}ska}
  G.,   {Morisset} C.,  2015, \mn@doi [\mnras] {10.1093/mnras/stv388}, \href
  {http://adsabs.harvard.edu/abs/2015MNRAS.449.1797D} {449, 1797}

\bibitem[\protect\citeauthoryear{{Dinerstein}}{{Dinerstein}}{2001}]{Dinerstein:2001aa}
{Dinerstein} H.~L.,  2001, \mn@doi [\apjl] {10.1086/319645}, \href
  {http://adsabs.harvard.edu/abs/2001ApJ...550L.223D} {550, L223}

\bibitem[\protect\citeauthoryear{{Dopita} \& {Meatheringham}}{{Dopita} \&
  {Meatheringham}}{1991}]{Dopita:1991aa}
{Dopita} M.~A.,  {Meatheringham} S.~J.,  1991, \mn@doi [\apj] {10.1086/170377},
  \href {http://adsabs.harvard.edu/abs/1991ApJ...377..480D} {377, 480}

\bibitem[\protect\citeauthoryear{{Dopita}, {Mason}  \& {Robb}}{{Dopita}
  et~al.}{1976}]{Dopita:1976aa}
{Dopita} M.~A.,  {Mason} D.~J.,   {Robb} W.~D.,  1976, \mn@doi [\apj]
  {10.1086/154472}, \href {http://ads.nao.ac.jp/abs/1976ApJ...207..102D} {207,
  102}

\bibitem[\protect\citeauthoryear{{Draine} \& {Li}}{{Draine} \&
  {Li}}{2007}]{2007ApJ...657..810D}
{Draine} B.~T.,  {Li} A.,  2007, \mn@doi [\apj] {10.1086/511055}, \href
  {http://ads.nao.ac.jp/abs/2007ApJ...657..810D} {657, 810}

\bibitem[\protect\citeauthoryear{{Dufton} \& {Kingston}}{{Dufton} \&
  {Kingston}}{1991}]{Dufton:1991aa}
{Dufton} P.~L.,  {Kingston} A.~E.,  1991, \mn@doi [\mnras]
  {10.1093/mnras/248.4.827}, \href
  {http://ads.nao.ac.jp/abs/1991MNRAS.248..827D} {248, 827}

\bibitem[\protect\citeauthoryear{{Dufton}, {Hibbert}, {Kingston}  \&
  {Doschek}}{{Dufton} et~al.}{1982}]{Dufton:1982aa}
{Dufton} P.~L.,  {Hibbert} A.,  {Kingston} A.~E.,   {Doschek} G.~A.,  1982,
  \mn@doi [\apj] {10.1086/159992}, \href
  {http://ads.nao.ac.jp/abs/1982ApJ...257..338D} {257, 338}

\bibitem[\protect\citeauthoryear{{Egan} et~al.,}{{Egan}
  et~al.}{2003}]{Egan:2003aa}
{Egan} M.~P.,  et~al., 2003, VizieR Online Data Catalog, \href
  {http://adsabs.harvard.edu/abs/2003yCat.5114....0E} {5114}

\bibitem[\protect\citeauthoryear{{Ellis} \& {Martinson}}{{Ellis} \&
  {Martinson}}{1984}]{Ellis:1984aa}
{Ellis} D.~G.,  {Martinson} I.,  1984, \mn@doi [\physscr]
  {10.1088/0031-8949/30/4/007}, \href
  {http://ads.nao.ac.jp/abs/1984PhyS...30..255E} {30, 255}

\bibitem[\protect\citeauthoryear{{Fang} \& {Liu}}{{Fang} \&
  {Liu}}{2011}]{Fang:2011aa}
{Fang} X.,  {Liu} X.-W.,  2011, \mn@doi [\mnras]
  {10.1111/j.1365-2966.2011.18681.x}, \href
  {http://adsabs.harvard.edu/abs/2011MNRAS.415..181F} {415, 181}

\bibitem[\protect\citeauthoryear{{Ferland} et~al.,}{{Ferland}
  et~al.}{2013}]{Ferland:2013aa}
{Ferland} G.~J.,  et~al., 2013, \rmxaa, \href
  {http://adsabs.harvard.edu/abs/2013RMxAA..49..137F} {49, 137}

\bibitem[\protect\citeauthoryear{{Frew}, {Parker}  \& {Boji{\v
  c}i{\'c}}}{{Frew} et~al.}{2016}]{Frew:2016aa}
{Frew} D.~J.,  {Parker} Q.~A.,   {Boji{\v c}i{\'c}} I.~S.,  2016, \mn@doi
  [\mnras] {10.1093/mnras/stv1516}, \href
  {http://adsabs.harvard.edu/abs/2016MNRAS.455.1459F} {455, 1459}

\bibitem[\protect\citeauthoryear{{Froese Fischer}}{{Froese
  Fischer}}{1994}]{Froese:1994aa}
{Froese Fischer} C.,  1994, \mn@doi [\physscr] {10.1088/0031-8949/49/3/011},
  \href {http://ads.nao.ac.jp/abs/1994PhyS...49..323F} {49, 323}

\bibitem[\protect\citeauthoryear{{Froese Fischer} \& {Saha}}{{Froese Fischer}
  \& {Saha}}{1985}]{Froese:1985aa}
{Froese Fischer} C.,  {Saha} H.~P.,  1985, \mn@doi [\physscr]
  {10.1088/0031-8949/32/3/004}, \href
  {http://ads.nao.ac.jp/abs/1985PhyS...32..181F} {32, 181}

\bibitem[\protect\citeauthoryear{{Gaia Collaboration} et~al.,}{{Gaia
  Collaboration} et~al.}{2018}]{Gaia:2018aa}
{Gaia Collaboration} et~al., 2018, \mn@doi [\aap]
  {10.1051/0004-6361/201833051}, \href
  {https://ui.adsabs.harvard.edu/abs/2018A&A...616A...1G} {616, A1}

\bibitem[\protect\citeauthoryear{{Galavis}, {Mendoza}  \& {Zeippen}}{{Galavis}
  et~al.}{1995}]{Galavis:1995aa}
{Galavis} M.~E.,  {Mendoza} C.,   {Zeippen} C.~J.,  1995, \aaps, \href
  {http://ads.nao.ac.jp/abs/1995A%26AS..111..347G} {111, 347}

\bibitem[\protect\citeauthoryear{{Gallino}, {Arlandini}, {Busso}, {Lugaro},
  {Travaglio}, {Straniero}, {Chieffi}  \& {Limongi}}{{Gallino}
  et~al.}{1998}]{Gallino:1998aa}
{Gallino} R.,  {Arlandini} C.,  {Busso} M.,  {Lugaro} M.,  {Travaglio} C.,
  {Straniero} O.,  {Chieffi} A.,   {Limongi} M.,  1998, \mn@doi [\apj]
  {10.1086/305437}, \href {http://adsabs.harvard.edu/abs/1998ApJ...497..388G}
  {497, 388}

\bibitem[\protect\citeauthoryear{{Garc{\'{\i}}a-Rojas}, {Madonna}, {Luridiana},
  {Sterling}, {Morisset}, {Delgado-Inglada}  \& {Toribio San
  Cipriano}}{{Garc{\'{\i}}a-Rojas} et~al.}{2015}]{Garcia-Rojas:2015aa}
{Garc{\'{\i}}a-Rojas} J.,  {Madonna} S.,  {Luridiana} V.,  {Sterling} N.~C.,
  {Morisset} C.,  {Delgado-Inglada} G.,   {Toribio San Cipriano} L.,  2015,
  \mn@doi [\mnras] {10.1093/mnras/stv1415}, \href
  {http://adsabs.harvard.edu/abs/2015MNRAS.452.2606G} {452, 2606}

\bibitem[\protect\citeauthoryear{{Garstang}}{{Garstang}}{1951}]{Garstang:1951aa}
{Garstang} R.~H.,  1951, \mn@doi [\mnras] {10.1093/mnras/111.1.115}, \href
  {http://ads.nao.ac.jp/abs/1951MNRAS.111..115G} {111, 115}

\bibitem[\protect\citeauthoryear{{Garstang}}{{Garstang}}{1957}]{Garstang:1957aa}
{Garstang} R.~H.,  1957, \mn@doi [\mnras] {10.1093/mnras/117.4.393}, \href
  {http://ads.nao.ac.jp/abs/1957MNRAS.117..393G} {117, 393}

\bibitem[\protect\citeauthoryear{{Giammanco} et~al.,}{{Giammanco}
  et~al.}{2011}]{Giammanco:2011aa}
{Giammanco} C.,  et~al., 2011, \mn@doi [\aap] {10.1051/0004-6361/201014464},
  \href {https://ui.adsabs.harvard.edu/\#abs/2011A&A...525A..58G} {525, A58}

\bibitem[\protect\citeauthoryear{{Heise}, {Smith}  \& {Calamai}}{{Heise}
  et~al.}{1995}]{Heise:1995aa}
{Heise} C.,  {Smith} P.~L.,   {Calamai} A.~G.,  1995, \mn@doi [\apjl]
  {10.1086/309676}, \href {http://ads.nao.ac.jp/abs/1995ApJ...451L..41H} {451,
  L41}

\bibitem[\protect\citeauthoryear{{Henry}, {Kwitter}, {Jaskot}, {Balick},
  {Morrison}  \& {Milingo}}{{Henry} et~al.}{2010}]{Henry:2010aa}
{Henry} R.~B.~C.,  {Kwitter} K.~B.,  {Jaskot} A.~E.,  {Balick} B.,  {Morrison}
  M.~A.,   {Milingo} J.~B.,  2010, \mn@doi [\apj]
  {10.1088/0004-637X/724/1/748}, \href
  {http://adsabs.harvard.edu/abs/2010ApJ...724..748H} {724, 748}

\bibitem[\protect\citeauthoryear{{Higdon} et~al.,}{{Higdon}
  et~al.}{2004}]{Higdon:2004aa}
{Higdon} S.~J.~U.,  et~al., 2004, \mn@doi [\pasp] {10.1086/425083}, \href
  {http://adsabs.harvard.edu/abs/2004PASP..116..975H} {116, 975}

\bibitem[\protect\citeauthoryear{{Hora}, {Latter}  \& {Deutsch}}{{Hora}
  et~al.}{1999}]{Hora:1999aa}
{Hora} J.~L.,  {Latter} W.~B.,   {Deutsch} L.~K.,  1999, \mn@doi [\apjs]
  {10.1086/313256}, \href {http://adsabs.harvard.edu/abs/1999ApJS..124..195H}
  {124, 195}

\bibitem[\protect\citeauthoryear{{Houck} et~al.,}{{Houck}
  et~al.}{2004}]{Houck:2004aa}
{Houck} J.~R.,  et~al., 2004, \mn@doi [\apjs] {10.1086/423134}, \href
  {http://adsabs.harvard.edu/abs/2004ApJS..154...18H} {154, 18}

\bibitem[\protect\citeauthoryear{{Huggins}, {Bachiller}, {Cox}  \&
  {Forveille}}{{Huggins} et~al.}{1996}]{Huggins:1996aa}
{Huggins} P.~J.,  {Bachiller} R.,  {Cox} P.,   {Forveille} T.,  1996, \aap,
  \href {http://adsabs.harvard.edu/abs/1996A%26A...315..284H} {315, 284}

\bibitem[\protect\citeauthoryear{{Hyung}, {Aller}  \& {Feibelman}}{{Hyung}
  et~al.}{1993}]{Hyung:1993aa}
{Hyung} S.,  {Aller} L.~H.,   {Feibelman} W.~A.,  1993, \mn@doi [\pasp]
  {10.1086/133308}, \href {http://adsabs.harvard.edu/abs/1993PASP..105.1279H}
  {105, 1279}

\bibitem[\protect\citeauthoryear{{Isaacman}}{{Isaacman}}{1984a}]{Isaacman:1984aa}
{Isaacman} R.,  1984a, \aap, \href
  {http://adsabs.harvard.edu/abs/1984A%26A...130..151I} {130, 151}

\bibitem[\protect\citeauthoryear{{Isaacman}}{{Isaacman}}{1984b}]{Isaacman:1984ab}
{Isaacman} R.,  1984b, \mn@doi [\mnras] {10.1093/mnras/208.2.399}, \href
  {http://adsabs.harvard.edu/abs/1984MNRAS.208..399I} {208, 399}

\bibitem[\protect\citeauthoryear{{Ishihara} et~al.,}{{Ishihara}
  et~al.}{2010}]{Ishihara:2010aa}
{Ishihara} D.,  et~al., 2010, \mn@doi [\aap] {10.1051/0004-6361/200913811},
  \href {http://adsabs.harvard.edu/abs/2010A%26A...514A...1I} {514, A1}

\bibitem[\protect\citeauthoryear{{Johnson}, {Kingston}  \& {Dufton}}{{Johnson}
  et~al.}{1986}]{Johnson:1986aa}
{Johnson} C.~T.,  {Kingston} A.~E.,   {Dufton} P.~L.,  1986, \mn@doi [\mnras]
  {10.1093/mnras/220.1.155}, \href
  {http://ads.nao.ac.jp/abs/1986MNRAS.220..155J} {220, 155}

\bibitem[\protect\citeauthoryear{{Johnson}, {Burke}  \& {Kingston}}{{Johnson}
  et~al.}{1987}]{Johnson:1987aa}
{Johnson} C.~T.,  {Burke} P.~G.,   {Kingston} A.~E.,  1987, \mn@doi [Journal of
  Physics B Atomic Molecular Physics] {10.1088/0022-3700/20/11/022}, \href
  {http://ads.nao.ac.jp/abs/1987JPhB...20.2553J} {20, 2553}

\bibitem[\protect\citeauthoryear{{Karakas}}{{Karakas}}{2016}]{Karakas:2016ab}
{Karakas} A.~I.,  2016, \memsai, \href
  {http://adsabs.harvard.edu/abs/2016MmSAI..87..229K} {87, 229}

\bibitem[\protect\citeauthoryear{{Karakas} \& {Lattanzio}}{{Karakas} \&
  {Lattanzio}}{2014}]{Karakas:2014aa}
{Karakas} A.~I.,  {Lattanzio} J.~C.,  2014, \mn@doi [\pasa]
  {10.1017/pasa.2014.21}, \href
  {http://adsabs.harvard.edu/abs/2014PASA...31...30K} {31, e030}

\bibitem[\protect\citeauthoryear{{Karakas}, {van Raai}, {Lugaro}, {Sterling}
  \& {Dinerstein}}{{Karakas} et~al.}{2009}]{Karakas:2009aa}
{Karakas} A.~I.,  {van Raai} M.~A.,  {Lugaro} M.,  {Sterling} N.~C.,
  {Dinerstein} H.~L.,  2009, \mn@doi [\apj] {10.1088/0004-637X/690/2/1130},
  \href {http://adsabs.harvard.edu/abs/2009ApJ...690.1130K} {690, 1130}

\bibitem[\protect\citeauthoryear{{Karakas}, {Lugaro}, {Carlos}, {Cseh},
  {Kamath}  \& {Garc{\'{\i}}a-Hern{\'a}ndez}}{{Karakas}
  et~al.}{2018}]{Karakas:2018aa}
{Karakas} A.~I.,  {Lugaro} M.,  {Carlos} M.,  {Cseh} B.,  {Kamath} D.,
  {Garc{\'{\i}}a-Hern{\'a}ndez} D.~A.,  2018, \mn@doi [\mnras]
  {10.1093/mnras/sty625}, \href
  {http://adsabs.harvard.edu/abs/2018MNRAS.477..421K} {477, 421}

\bibitem[\protect\citeauthoryear{{Kaufman} \& {Sugar}}{{Kaufman} \&
  {Sugar}}{1986}]{Kaufman:1986aa}
{Kaufman} V.,  {Sugar} J.,  1986, \mn@doi [Journal of Physical and Chemical
  Reference Data] {10.1063/1.555775}, \href
  {http://ads.nao.ac.jp/abs/1986JPCRD..15..321K} {15, 321}

\bibitem[\protect\citeauthoryear{{Keenan}, {Hibbert}, {Ojha}  \&
  {Conlon}}{{Keenan} et~al.}{1993}]{Keenan:1993aa}
{Keenan} F.~P.,  {Hibbert} A.,  {Ojha} P.~C.,   {Conlon} E.~S.,  1993, \mn@doi
  [\physscr] {10.1088/0031-8949/48/2/001}, \href
  {http://ads.nao.ac.jp/abs/1993PhyS...48..129K} {48, 129}

\bibitem[\protect\citeauthoryear{{Kim} et~al.,}{{Kim}
  et~al.}{2002}]{Kim:2002aa}
{Kim} K.-M.,  et~al., 2002, \mn@doi [Journal of Korean Astronomical Society]
  {10.5303/JKAS.2002.35.4.221}, \href
  {http://adsabs.harvard.edu/abs/2002JKAS...35..221K} {35, 221}

\bibitem[\protect\citeauthoryear{{Kingsburgh} \& {Barlow}}{{Kingsburgh} \&
  {Barlow}}{1992}]{Kingsburgh:1992aa}
{Kingsburgh} R.~L.,  {Barlow} M.~J.,  1992, \mn@doi [\mnras]
  {10.1093/mnras/257.2.317}, \href
  {http://adsabs.harvard.edu/abs/1992MNRAS.257..317K} {257, 317}

\bibitem[\protect\citeauthoryear{{Kingsburgh} \& {Barlow}}{{Kingsburgh} \&
  {Barlow}}{1994}]{Kingsburgh:1994aa}
{Kingsburgh} R.~L.,  {Barlow} M.~J.,  1994, \mn@doi [\mnras]
  {10.1093/mnras/271.2.257}, \href
  {http://adsabs.harvard.edu/abs/1994MNRAS.271..257K} {271, 257}

\bibitem[\protect\citeauthoryear{{Kwitter}, {Henry}  \& {Milingo}}{{Kwitter}
  et~al.}{2003}]{Kwitter:2003aa}
{Kwitter} K.~B.,  {Henry} R.~B.~C.,   {Milingo} J.~B.,  2003, \mn@doi [\pasp]
  {10.1086/345108}, \href {http://adsabs.harvard.edu/abs/2003PASP..115...80K}
  {115, 80}

\bibitem[\protect\citeauthoryear{{LaJohn} \& {Luke}}{{LaJohn} \&
  {Luke}}{1993}]{LaJohn:1993aa}
{LaJohn} L.,  {Luke} T.~M.,  1993, Physica Scripta, 47, 542

\bibitem[\protect\citeauthoryear{{Lanzafame}}{{Lanzafame}}{1994}]{Lanzafame:1994aa}
{Lanzafame} A.~C.,  1994, \aap, \href
  {http://ads.nao.ac.jp/abs/1994A%26A...287..972L} {287, 972}

\bibitem[\protect\citeauthoryear{{Leisy} \& {Dennefeld}}{{Leisy} \&
  {Dennefeld}}{2006}]{Leisy:2006aa}
{Leisy} P.,  {Dennefeld} M.,  2006, \mn@doi [\aap]
  {10.1051/0004-6361:20053063}, \href
  {https://ui.adsabs.harvard.edu/\#abs/2006A&A...456..451L} {456, 451}

\bibitem[\protect\citeauthoryear{{Lennon} \& {Burke}}{{Lennon} \&
  {Burke}}{1994}]{Lennon:1994aa}
{Lennon} D.~J.,  {Burke} V.~M.,  1994, \aaps, \href
  {http://ads.nao.ac.jp/abs/1994A%26AS..103..273L} {103, 273}

\bibitem[\protect\citeauthoryear{{Liang}, {Badnell}  \& {Zhao}}{{Liang}
  et~al.}{2012}]{Liang:2012aa}
{Liang} G.~Y.,  {Badnell} N.~R.,   {Zhao} G.,  2012, \mn@doi [\aap]
  {10.1051/0004-6361/201220277}, \href
  {http://adsabs.harvard.edu/abs/2012A%26A...547A..87L} {547, A87}

\bibitem[\protect\citeauthoryear{{Liu}, {Storey}, {Barlow}, {Danziger}, {Cohen}
   \& {Bryce}}{{Liu} et~al.}{2000}]{Liu:2000aa}
{Liu} X.-W.,  {Storey} P.~J.,  {Barlow} M.~J.,  {Danziger} I.~J.,  {Cohen} M.,
   {Bryce} M.,  2000, \mn@doi [\mnras] {10.1046/j.1365-8711.2000.03167.x},
  \href {http://adsabs.harvard.edu/abs/2000MNRAS.312..585L} {312, 585}

\bibitem[\protect\citeauthoryear{{Lodders}}{{Lodders}}{2010}]{Lodders:2010aa}
{Lodders} K.,  2010, in {Goswami} A.,  {Reddy} B.~E.,  eds, Principles and
  Perspectives in Cosmochemistry. p.~379 (\mn@eprint {arXiv} {1010.2746}),
  \mn@doi{10.1007/978-3-642-10352-0_8}

\bibitem[\protect\citeauthoryear{{Lugaro}, {Karakas}, {Stancliffe}  \&
  {Rijs}}{{Lugaro} et~al.}{2012}]{Lugaro:2012aa}
{Lugaro} M.,  {Karakas} A.~I.,  {Stancliffe} R.~J.,   {Rijs} C.,  2012, \mn@doi
  [\apj] {10.1088/0004-637X/747/1/2}, \href
  {http://adsabs.harvard.edu/abs/2012ApJ...747....2L} {747, 2}

\bibitem[\protect\citeauthoryear{{Maciel} \& {Costa}}{{Maciel} \&
  {Costa}}{2010}]{Maciel:2010aa}
{Maciel} W.~J.,  {Costa} R.~D.~D.,  2010, in {Cunha} K.,  {Spite} M.,
  {Barbuy} B.,  eds,  IAU Symposium Vol. 265, Chemical Abundances in the
  Universe: Connecting First Stars to Planets. pp 317--324 (\mn@eprint {arXiv}
  {0911.3763}), \mn@doi{10.1017/S1743921310000803}

\bibitem[\protect\citeauthoryear{{Maciel} \& {Costa}}{{Maciel} \&
  {Costa}}{2013}]{Maciel:2013aa}
{Maciel} W.~J.,  {Costa} R.~D.~D.,  2013, \rmxaa, \href
  {http://adsabs.harvard.edu/abs/2013RMxAA..49..333M} {49, 333}

\bibitem[\protect\citeauthoryear{{Madonna}, {Garc{\'{\i}}a-Rojas}, {Sterling},
  {Delgado-Inglada}, {Mesa-Delgado}, {Luridiana}, {Roederer}  \&
  {Mashburn}}{{Madonna} et~al.}{2017}]{Madonna:2017aa}
{Madonna} S.,  {Garc{\'{\i}}a-Rojas} J.,  {Sterling} N.~C.,  {Delgado-Inglada}
  G.,  {Mesa-Delgado} A.,  {Luridiana} V.,  {Roederer} I.~U.,   {Mashburn}
  A.~L.,  2017, \mn@doi [\mnras] {10.1093/mnras/stx1585}, \href
  {http://adsabs.harvard.edu/abs/2017MNRAS.471.1341M} {471, 1341}

\bibitem[\protect\citeauthoryear{{Madonna} et~al.,}{{Madonna}
  et~al.}{2018}]{Madonna:2018aa}
{Madonna} S.,  et~al., 2018, \mn@doi [\apjl] {10.3847/2041-8213/aaccef}, \href
  {http://adsabs.harvard.edu/abs/2018ApJ...861L...8M} {861, L8}

\bibitem[\protect\citeauthoryear{{Mallik} \& {Peimbert}}{{Mallik} \&
  {Peimbert}}{1988}]{Mallik:1988aa}
{Mallik} D.~C.~V.,  {Peimbert} M.,  1988, \rmxaa, \href
  {http://adsabs.harvard.edu/abs/1988RMxAA..16..111M} {16, 111}

\bibitem[\protect\citeauthoryear{{Martin}, {Karwowski}, {Diercksen}  \&
  {Barrientos}}{{Martin} et~al.}{1993}]{Martin:1993aa}
{Martin} I.,  {Karwowski} J.,  {Diercksen} G.~H.~F.,   {Barrientos} C.,  1993,
  \aaps, \href {http://ads.nao.ac.jp/abs/1993A%26AS..100..595M} {100, 595}

\bibitem[\protect\citeauthoryear{{Mashburn}, {Sterling}, {Madonna},
  {Dinerstein}, {Roederer}  \& {Geballe}}{{Mashburn}
  et~al.}{2016}]{Mashburn:2016aa}
{Mashburn} A.~L.,  {Sterling} N.~C.,  {Madonna} S.,  {Dinerstein} H.~L.,
  {Roederer} I.~U.,   {Geballe} T.~R.,  2016, \mn@doi [\apjl]
  {10.3847/2041-8205/831/1/L3}, \href
  {http://adsabs.harvard.edu/abs/2016ApJ...831L...3M} {831, L3}

\bibitem[\protect\citeauthoryear{{Matsuura} et~al.,}{{Matsuura}
  et~al.}{2014}]{Matsuura:2014aa}
{Matsuura} M.,  et~al., 2014, \mn@doi [\mnras] {10.1093/mnras/stt2495}, \href
  {http://adsabs.harvard.edu/abs/2014MNRAS.439.1472M} {439, 1472}

\bibitem[\protect\citeauthoryear{{McElroy}, {Walsh}, {Markwick}, {Cordiner},
  {Smith}  \& {Millar}}{{McElroy} et~al.}{2013}]{McElroy:2013aa}
{McElroy} D.,  {Walsh} C.,  {Markwick} A.~J.,  {Cordiner} M.~A.,  {Smith} K.,
  {Millar} T.~J.,  2013, \mn@doi [\aap] {10.1051/0004-6361/201220465}, \href
  {https://ui.adsabs.harvard.edu/\#abs/2013A&A...550A..36M} {550, A36}

\bibitem[\protect\citeauthoryear{{McLaughlin} \& {Bell}}{{McLaughlin} \&
  {Bell}}{1993}]{McLaughlin:1993aa}
{McLaughlin} B.~M.,  {Bell} K.~L.,  1993, \mn@doi [\apj] {10.1086/172635},
  \href {http://ads.nao.ac.jp/abs/1993ApJ...408..753M} {408, 753}

\bibitem[\protect\citeauthoryear{{McLaughlin} \& {Bell}}{{McLaughlin} \&
  {Bell}}{2000}]{McLaughlin:2000aa}
{McLaughlin} B.~M.,  {Bell} K.~L.,  2000, \mn@doi [Journal of Physics B Atomic
  Molecular Physics] {10.1088/0953-4075/33/4/301}, \href
  {http://ads.nao.ac.jp/abs/2000JPhB...33..597M} {33, 597}

\bibitem[\protect\citeauthoryear{{Mendoza}}{{Mendoza}}{1983}]{Mendoza:1983aa}
{Mendoza} C.,  1983, in {Flower} D.~R.,  ed.,  IAU Symposium Vol. 103,
  Planetary Nebulae. pp 143--172

\bibitem[\protect\citeauthoryear{{Mendoza} \& {Zeippen}}{{Mendoza} \&
  {Zeippen}}{1982a}]{Mendoza:1982aa}
{Mendoza} C.,  {Zeippen} C.~J.,  1982a, \mn@doi [\mnras]
  {10.1093/mnras/198.1.127}, \href
  {http://ads.nao.ac.jp/abs/1982MNRAS.198..127M} {198, 127}

\bibitem[\protect\citeauthoryear{{Mendoza} \& {Zeippen}}{{Mendoza} \&
  {Zeippen}}{1982b}]{Mendoza:1982ab}
{Mendoza} C.,  {Zeippen} C.~J.,  1982b, \mn@doi [\mnras]
  {10.1093/mnras/199.4.1025}, \href
  {http://ads.nao.ac.jp/abs/1982MNRAS.199.1025M} {199, 1025}

\bibitem[\protect\citeauthoryear{{Miller Bertolami}}{{Miller
  Bertolami}}{2016}]{Miller-Bertolami:2016aa}
{Miller Bertolami} M.~M.,  2016, \mn@doi [\aap] {10.1051/0004-6361/201526577},
  \href {http://adsabs.harvard.edu/abs/2016A%26A...588A..25M} {588, A25}

\bibitem[\protect\citeauthoryear{{Milne} \& {Webster}}{{Milne} \&
  {Webster}}{1979}]{Milne:1979aa}
{Milne} D.~K.,  {Webster} B.~L.,  1979, \aaps, \href
  {http://adsabs.harvard.edu/abs/1979A%26AS...36..169M} {36, 169}

\bibitem[\protect\citeauthoryear{{Miszalski} et~al.,}{{Miszalski}
  et~al.}{2013}]{Miszalski:2013aa}
{Miszalski} B.,  et~al., 2013, \mn@doi [\mnras] {10.1093/mnras/stt1795}, \href
  {http://adsabs.harvard.edu/abs/2013MNRAS.436.3068M} {436, 3068}

\bibitem[\protect\citeauthoryear{{Nahar} \& {Pradhan}}{{Nahar} \&
  {Pradhan}}{1996}]{Nahar:1996aa}
{Nahar} S.~N.,  {Pradhan} A.~K.,  1996, \aaps, \href
  {http://ads.nao.ac.jp/abs/1996A%26AS..119..509N} {119, 509}

\bibitem[\protect\citeauthoryear{{Naqvi}}{{Naqvi}}{1951}]{Naqvi:1951aa}
{Naqvi} A.~M.,  1951, PhD thesis, Harvard Univ

\bibitem[\protect\citeauthoryear{{Nussbaumer}}{{Nussbaumer}}{1977}]{Nussbaumer:1977aa}
{Nussbaumer} H.,  1977, \aap, \href
  {http://ads.nao.ac.jp/abs/1977A%26A....58..291N} {58, 291}

\bibitem[\protect\citeauthoryear{{Nussbaumer} \& {Storey}}{{Nussbaumer} \&
  {Storey}}{1981}]{Nussbaumer:1981aa}
{Nussbaumer} H.,  {Storey} P.~J.,  1981, \aap, \href
  {http://ads.nao.ac.jp/abs/1981A%26A....96...91N} {96, 91}

\bibitem[\protect\citeauthoryear{{Ohsawa}, {Onaka}, {Sakon}, {Matsuura}  \&
  {Kaneda}}{{Ohsawa} et~al.}{2016}]{Ohsawa:2016aa}
{Ohsawa} R.,  {Onaka} T.,  {Sakon} I.,  {Matsuura} M.,   {Kaneda} H.,  2016,
  \mn@doi [\aj] {10.3847/0004-6256/151/4/93}, \href
  {http://adsabs.harvard.edu/abs/2016AJ....151...93O} {151, 93}

\bibitem[\protect\citeauthoryear{{Onaka} et~al.,}{{Onaka}
  et~al.}{2007}]{Onaka:2007aa}
{Onaka} T.,  et~al., 2007, \mn@doi [\pasj] {10.1093/pasj/59.sp2.S401}, \href
  {http://adsabs.harvard.edu/abs/2007PASJ...59S.401O} {59, S401}

\bibitem[\protect\citeauthoryear{{Otsuka}}{{Otsuka}}{2015}]{Otsuka:2015ab}
{Otsuka} M.,  2015, \mn@doi [\mnras] {10.1093/mnras/stv1553}, \href
  {http://adsabs.harvard.edu/abs/2015MNRAS.452.4070O} {452, 4070}

\bibitem[\protect\citeauthoryear{{Otsuka} \& {Tajitsu}}{{Otsuka} \&
  {Tajitsu}}{2013}]{Otsuka:2013ab}
{Otsuka} M.,  {Tajitsu} A.,  2013, \mn@doi [\apj]
  {10.1088/0004-637X/778/2/146}, \href
  {http://adsabs.harvard.edu/abs/2013ApJ...778..146O} {778, 146}

\bibitem[\protect\citeauthoryear{{Otsuka}, {Izumiura}, {Tajitsu}  \&
  {Hyung}}{{Otsuka} et~al.}{2008}]{Otsuka:2008aa}
{Otsuka} M.,  {Izumiura} H.,  {Tajitsu} A.,   {Hyung} S.,  2008, \mn@doi
  [\apjl] {10.1086/591147}, \href
  {http://adsabs.harvard.edu/abs/2008ApJ...682L.105O} {682, L105}

\bibitem[\protect\citeauthoryear{{Otsuka}, {Hyung}, {Lee}, {Izumiura}  \&
  {Tajitsu}}{{Otsuka} et~al.}{2009}]{Otsuka:2009aa}
{Otsuka} M.,  {Hyung} S.,  {Lee} S.-J.,  {Izumiura} H.,   {Tajitsu} A.,  2009,
  \mn@doi [\apj] {10.1088/0004-637X/705/1/509}, \href
  {http://adsabs.harvard.edu/abs/2009ApJ...705..509O} {705, 509}

\bibitem[\protect\citeauthoryear{{Otsuka}, {Tajitsu}, {Hyung}  \&
  {Izumiura}}{{Otsuka} et~al.}{2010}]{Otsuka:2010aa}
{Otsuka} M.,  {Tajitsu} A.,  {Hyung} S.,   {Izumiura} H.,  2010, \mn@doi [\apj]
  {10.1088/0004-637X/723/1/658}, \href
  {http://adsabs.harvard.edu/abs/2010ApJ...723..658O} {723, 658}

\bibitem[\protect\citeauthoryear{{Otsuka}, {Meixner}, {Riebel}, {Hyung},
  {Tajitsu}  \& {Izumiura}}{{Otsuka} et~al.}{2011}]{Otsuka:2011aa}
{Otsuka} M.,  {Meixner} M.,  {Riebel} D.,  {Hyung} S.,  {Tajitsu} A.,
  {Izumiura} H.,  2011, \mn@doi [\apj] {10.1088/0004-637X/729/1/39}, \href
  {http://adsabs.harvard.edu/abs/2011ApJ...729...39O} {729, 39}

\bibitem[\protect\citeauthoryear{{Otsuka}, {Kemper}, {Cami}, {Peeters}  \&
  {Bernard-Salas}}{{Otsuka} et~al.}{2014}]{Otsuka:2014aa}
{Otsuka} M.,  {Kemper} F.,  {Cami} J.,  {Peeters} E.,   {Bernard-Salas} J.,
  2014, \mn@doi [\mnras] {10.1093/mnras/stt2070}, \href
  {http://adsabs.harvard.edu/abs/2014MNRAS.437.2577O} {437, 2577}

\bibitem[\protect\citeauthoryear{{Otsuka}, {Hyung}  \& {Tajitsu}}{{Otsuka}
  et~al.}{2015}]{Otsuka:2015aa}
{Otsuka} M.,  {Hyung} S.,   {Tajitsu} A.,  2015, \mn@doi [\apjs]
  {10.1088/0067-0049/217/2/22}, \href
  {http://adsabs.harvard.edu/abs/2015ApJS..217...22O} {217, 22}

\bibitem[\protect\citeauthoryear{{Otsuka} et~al.,}{{Otsuka}
  et~al.}{2017}]{Otsuka:2017aa}
{Otsuka} M.,  et~al., 2017, \mn@doi [\apjs] {10.3847/1538-4365/aa8175}, \href
  {http://adsabs.harvard.edu/abs/2017ApJS..231...22O} {231, 22}

\bibitem[\protect\citeauthoryear{{Pazderska} et~al.,}{{Pazderska}
  et~al.}{2009}]{Pazderska:2009aa}
{Pazderska} B.~M.,  et~al., 2009, \mn@doi [\aap] {10.1051/0004-6361/200811369},
  \href {http://adsabs.harvard.edu/abs/2009A%26A...498..463P} {498, 463}

\bibitem[\protect\citeauthoryear{{Peeters}, {Tielens}, {Allamandola}  \&
  {Wolfire}}{{Peeters} et~al.}{2012}]{Peeters:2012aa}
{Peeters} E.,  {Tielens} A.~G.~G.~M.,  {Allamandola} L.~J.,   {Wolfire} M.~G.,
  2012, \mn@doi [\apj] {10.1088/0004-637X/747/1/44}, \href
  {http://adsabs.harvard.edu/abs/2012ApJ...747...44P} {747, 44}

\bibitem[\protect\citeauthoryear{{Peeters}, {Bauschlicher}, {Allamandola},
  {Tielens}, {Ricca}  \& {Wolfire}}{{Peeters} et~al.}{2017}]{Peeters:2017aa}
{Peeters} E.,  {Bauschlicher} Jr. C.~W.,  {Allamandola} L.~J.,  {Tielens}
  A.~G.~G.~M.,  {Ricca} A.,   {Wolfire} M.~G.,  2017, \mn@doi [\apj]
  {10.3847/1538-4357/836/2/198}, \href
  {http://adsabs.harvard.edu/abs/2017ApJ...836..198P} {836, 198}

\bibitem[\protect\citeauthoryear{{Pequignot} \& {Aldrovandi}}{{Pequignot} \&
  {Aldrovandi}}{1976}]{Pequignot:1976aa}
{Pequignot} D.,  {Aldrovandi} S.~M.~V.,  1976, \aap, \href
  {http://ads.nao.ac.jp/abs/1976A%26A....50..141P} {50, 141}

\bibitem[\protect\citeauthoryear{{Pequignot} \& {Baluteau}}{{Pequignot} \&
  {Baluteau}}{1994}]{Pequignot:1994aa}
{Pequignot} D.,  {Baluteau} J.-P.,  1994, \aap, \href
  {http://adsabs.harvard.edu/abs/1994A%26A...283..593P} {283, 593}

\bibitem[\protect\citeauthoryear{{Pequignot}, {Petitjean}  \&
  {Boisson}}{{Pequignot} et~al.}{1991}]{Pequignot:1991aa}
{Pequignot} D.,  {Petitjean} P.,   {Boisson} C.,  1991, \aap, \href
  {http://adsabs.harvard.edu/abs/1991A%26A...251..680P} {251, 680}

\bibitem[\protect\citeauthoryear{{Phillips}}{{Phillips}}{2004}]{Phillips:2004aa}
{Phillips} J.~P.,  2004, \mn@doi [\mnras] {10.1111/j.1365-2966.2004.08088.x},
  \href {https://ui.adsabs.harvard.edu/abs/2004MNRAS.353..589P} {353, 589}

\bibitem[\protect\citeauthoryear{{Pradhan}}{{Pradhan}}{1976}]{Pradhan:1976aa}
{Pradhan} A.~K.,  1976, \mn@doi [\mnras] {10.1093/mnras/177.1.31}, \href
  {http://ads.nao.ac.jp/abs/1976MNRAS.177...31P} {177, 31}

\bibitem[\protect\citeauthoryear{{Ramsbottom}, {Bell}  \&
  {Stafford}}{{Ramsbottom} et~al.}{1996}]{Ramsbottom:1996aa}
{Ramsbottom} C.~A.,  {Bell} K.~L.,   {Stafford} R.~P.,  1996, \mn@doi [Atomic
  Data and Nuclear Data Tables] {10.1006/adnd.1996.0009}, \href
  {http://ads.nao.ac.jp/abs/1996ADNDT..63...57R} {63, 57}

\bibitem[\protect\citeauthoryear{{Ramsbottom}, {Bell}  \&
  {Keenan}}{{Ramsbottom} et~al.}{1998}]{Ramsbottom:1998aa}
{Ramsbottom} C.~A.,  {Bell} K.~L.,   {Keenan} F.~P.,  1998, \mn@doi [\mnras]
  {10.1046/j.1365-8711.1998.01054.x}, \href
  {http://ads.nao.ac.jp/abs/1998MNRAS.293..233R} {293, 233}

\bibitem[\protect\citeauthoryear{{Ramsbottom}, {Bell}  \&
  {Keenan}}{{Ramsbottom} et~al.}{2001}]{Ramsbottom:2001aa}
{Ramsbottom} C.~A.,  {Bell} K.~L.,   {Keenan} F.~P.,  2001, \mn@doi [Atomic
  Data and Nuclear Data Tables] {10.1006/adnd.2000.0846}, \href
  {http://ads.nao.ac.jp/abs/2001ADNDT..77...57R} {77, 57}

\bibitem[\protect\citeauthoryear{{Rauch}}{{Rauch}}{2003}]{Rauch:2003aa}
{Rauch} T.,  2003, \mn@doi [\aap] {10.1051/0004-6361:20030412}, \href
  {http://adsabs.harvard.edu/abs/2003A%26A...403..709R} {403, 709}

\bibitem[\protect\citeauthoryear{{Ricca}, {Bauschlicher}, {Boersma}, {Tielens}
  \& {Allamandola}}{{Ricca} et~al.}{2012}]{Ricca:2012aa}
{Ricca} A.,  {Bauschlicher} Jr. C.~W.,  {Boersma} C.,  {Tielens} A.~G.~G.~M.,
  {Allamandola} L.~J.,  2012, \mn@doi [\apj] {10.1088/0004-637X/754/1/75},
  \href {http://adsabs.harvard.edu/abs/2012ApJ...754...75R} {754, 75}

\bibitem[\protect\citeauthoryear{{Rouleau} \& {Martin}}{{Rouleau} \&
  {Martin}}{1991}]{Rouleau:1991aa}
{Rouleau} F.,  {Martin} P.~G.,  1991, \mn@doi [\apj] {10.1086/170382}, \href
  {http://adsabs.harvard.edu/abs/1991ApJ...377..526R} {377, 526}

\bibitem[\protect\citeauthoryear{{Rynkun}, {J\"onsson}, {Gaigalas}  \&
  {Froese-Fischer}}{{Rynkun} et~al.}{2012}]{Rynkun:2012aa}
{Rynkun} P.,  {J\"onsson} P.,  {Gaigalas} G.,   {Froese-Fischer} C.,  2012,
  Atomic Data and Nuclear Data Tables, 98, 481

\bibitem[\protect\citeauthoryear{{Saraph} \& {Storey}}{{Saraph} \&
  {Storey}}{1999}]{Saraph:1999aa}
{Saraph} H.~E.,  {Storey} P.~J.,  1999, \mn@doi [\aaps] {10.1051/aas:1999437},
  \href {http://adsabs.harvard.edu/abs/1999A%26AS..134..369S} {134, 369}

\bibitem[\protect\citeauthoryear{{Saraph} \& {Tully}}{{Saraph} \&
  {Tully}}{1994}]{Saraph:1994aa}
{Saraph} H.~E.,  {Tully} J.~A.,  1994, \aaps, \href
  {http://ads.nao.ac.jp/abs/1994A%26AS..107...29S} {107, 29}

\bibitem[\protect\citeauthoryear{{Schoening} \& {Butler}}{{Schoening} \&
  {Butler}}{1998}]{Schoening:1998aa}
{Schoening} T.,  {Butler} K.,  1998, \mn@doi [\aaps] {10.1051/aas:1998165},
  \href {http://ads.nao.ac.jp/abs/1998A%26AS..128..581S} {128, 581}

\bibitem[\protect\citeauthoryear{{Schoning}}{{Schoning}}{1997}]{Schoning:1997aa}
{Schoning} T.,  1997, \mn@doi [\aaps] {10.1051/aas:1997133}, \href
  {http://ads.nao.ac.jp/abs/1997A%26AS..122..277S} {122, 277}

\bibitem[\protect\citeauthoryear{{Sharpee}, {Zhang}, {Williams}, {Pellegrini},
  {Cavagnolo}, {Baldwin}, {Phillips}  \& {Liu}}{{Sharpee}
  et~al.}{2007}]{Sharpee:2007aa}
{Sharpee} B.,  {Zhang} Y.,  {Williams} R.,  {Pellegrini} E.,  {Cavagnolo} K.,
  {Baldwin} J.~A.,  {Phillips} M.,   {Liu} X.-W.,  2007, \mn@doi [\apj]
  {10.1086/511665}, \href {http://adsabs.harvard.edu/abs/2007ApJ...659.1265S}
  {659, 1265}

\bibitem[\protect\citeauthoryear{{Shingles} \& {Karakas}}{{Shingles} \&
  {Karakas}}{2013}]{Shingles:2013aa}
{Shingles} L.~J.,  {Karakas} A.~I.,  2013, \mn@doi [\mnras]
  {10.1093/mnras/stt386}, \href
  {http://adsabs.harvard.edu/abs/2013MNRAS.431.2861S} {431, 2861}

\bibitem[\protect\citeauthoryear{{Shupe}, {Armus}, {Matthews}  \&
  {Soifer}}{{Shupe} et~al.}{1995}]{Shupe:1995aa}
{Shupe} D.~L.,  {Armus} L.,  {Matthews} K.,   {Soifer} B.~T.,  1995, \mn@doi
  [\aj] {10.1086/117350}, \href
  {http://adsabs.harvard.edu/abs/1995AJ....109.1173S} {109, 1173}

\bibitem[\protect\citeauthoryear{{Skrutskie} et~al.,}{{Skrutskie}
  et~al.}{2006}]{Skrutskie:2006aa}
{Skrutskie} M.~F.,  et~al., 2006, \mn@doi [\aj] {10.1086/498708}, \href
  {http://adsabs.harvard.edu/abs/2006AJ....131.1163S} {131, 1163}

\bibitem[\protect\citeauthoryear{{Sloan} et~al.,}{{Sloan}
  et~al.}{2014}]{Sloan:2014aa}
{Sloan} G.~C.,  et~al., 2014, \mn@doi [\apj] {10.1088/0004-637X/791/1/28},
  \href {http://adsabs.harvard.edu/abs/2014ApJ...791...28S} {791, 28}

\bibitem[\protect\citeauthoryear{{Smith} et~al.,}{{Smith}
  et~al.}{2002}]{Smith:2002aa}
{Smith} J.~A.,  et~al., 2002, \mn@doi [\aj] {10.1086/339311}, \href
  {http://adsabs.harvard.edu/abs/2002AJ....123.2121S} {123, 2121}

\bibitem[\protect\citeauthoryear{{Stanghellini} \& {Haywood}}{{Stanghellini} \&
  {Haywood}}{2010}]{Stanghellini:2010aa}
{Stanghellini} L.,  {Haywood} M.,  2010, \mn@doi [\apj]
  {10.1088/0004-637X/714/2/1096}, \href
  {http://adsabs.harvard.edu/abs/2010ApJ...714.1096S} {714, 1096}

\bibitem[\protect\citeauthoryear{{Stanghellini} \& {Haywood}}{{Stanghellini} \&
  {Haywood}}{2018}]{Stanghellini:2018aa}
{Stanghellini} L.,  {Haywood} M.,  2018, \mn@doi [\apj]
  {10.3847/1538-4357/aacaf8}, \href
  {http://adsabs.harvard.edu/abs/2018ApJ...862...45S} {862, 45}

\bibitem[\protect\citeauthoryear{{Stanghellini}, {Shaw}  \&
  {Villaver}}{{Stanghellini} et~al.}{2008}]{Stanghellini:2008aa}
{Stanghellini} L.,  {Shaw} R.~A.,   {Villaver} E.,  2008, \mn@doi [\apj]
  {10.1086/592395}, \href {http://adsabs.harvard.edu/abs/2008ApJ...689..194S}
  {689, 194}

\bibitem[\protect\citeauthoryear{{Stasi{\'n}ska} \& {Szczerba}}{{Stasi{\'n}ska}
  \& {Szczerba}}{1999}]{1999A&A...352..297S}
{Stasi{\'n}ska} G.,  {Szczerba} R.,  1999, \aap, \href
  {https://ui.adsabs.harvard.edu/abs/1999A&A...352..297S} {352, 297}

\bibitem[\protect\citeauthoryear{{Sterling} \& {Dinerstein}}{{Sterling} \&
  {Dinerstein}}{2008}]{Sterling:2008aa}
{Sterling} N.~C.,  {Dinerstein} H.~L.,  2008, \mn@doi [\apjs] {10.1086/520845},
  \href {http://adsabs.harvard.edu/abs/2008ApJS..174..158S} {174, 158}

\bibitem[\protect\citeauthoryear{{Sterling}, {Dinerstein}  \&
  {Bowers}}{{Sterling} et~al.}{2002}]{Sterling:2002aa}
{Sterling} N.~C.,  {Dinerstein} H.~L.,   {Bowers} C.~W.,  2002, \mn@doi [\apjl]
  {10.1086/344473}, \href {http://adsabs.harvard.edu/abs/2002ApJ...578L..55S}
  {578, L55}

\bibitem[\protect\citeauthoryear{{Sterling} et~al.,}{{Sterling}
  et~al.}{2009}]{Sterling:2009aa}
{Sterling} N.~C.,  et~al., 2009, \mn@doi [\pasa] {10.1071/AS08067}, \href
  {http://adsabs.harvard.edu/abs/2009PASA...26..339S} {26, 339}

\bibitem[\protect\citeauthoryear{{Sterling}, {Porter}  \&
  {Dinerstein}}{{Sterling} et~al.}{2015}]{Sterling:2015aa}
{Sterling} N.~C.,  {Porter} R.~L.,   {Dinerstein} H.~L.,  2015, \mn@doi [\apjs]
  {10.1088/0067-0049/218/2/25}, \href
  {http://adsabs.harvard.edu/abs/2015ApJS..218...25S} {218, 25}

\bibitem[\protect\citeauthoryear{{Sterling}, {Dinerstein}, {Kaplan}  \&
  {Bautista}}{{Sterling} et~al.}{2016}]{Sterling:2016aa}
{Sterling} N.~C.,  {Dinerstein} H.~L.,  {Kaplan} K.~F.,   {Bautista} M.~A.,
  2016, \mn@doi [\apjl] {10.3847/2041-8205/819/1/L9}, \href
  {http://adsabs.harvard.edu/abs/2016ApJ...819L...9S} {819, L9}

\bibitem[\protect\citeauthoryear{{Sterling}, {Madonna}, {Butler},
  {Garc{\'{\i}}a-Rojas}, {Mashburn}, {Morisset}, {Luridiana}  \&
  {Roederer}}{{Sterling} et~al.}{2017}]{Sterling:2017aa}
{Sterling} N.~C.,  {Madonna} S.,  {Butler} K.,  {Garc{\'{\i}}a-Rojas} J.,
  {Mashburn} A.~L.,  {Morisset} C.,  {Luridiana} V.,   {Roederer} I.~U.,  2017,
  \mn@doi [\apj] {10.3847/1538-4357/aa6c28}, \href
  {http://adsabs.harvard.edu/abs/2017ApJ...840...80S} {840, 80}

\bibitem[\protect\citeauthoryear{{Storey} \& {Hummer}}{{Storey} \&
  {Hummer}}{1995}]{Storey:1995aa}
{Storey} P.~J.,  {Hummer} D.~G.,  1995, \mn@doi [\mnras]
  {10.1093/mnras/272.1.41}, \href
  {http://adsabs.harvard.edu/abs/1995MNRAS.272...41S} {272, 41}

\bibitem[\protect\citeauthoryear{{Storey} \& {Zeippen}}{{Storey} \&
  {Zeippen}}{2000}]{Storey:2000aa}
{Storey} P.~J.,  {Zeippen} C.~J.,  2000, \mn@doi [\mnras]
  {10.1046/j.1365-8711.2000.03184.x}, \href
  {http://ads.nao.ac.jp/abs/2000MNRAS.312..813S} {312, 813}

\bibitem[\protect\citeauthoryear{{Takeda}, {Kaneko}, {Matsumoto}, {Oshino},
  {Ito}  \& {Shibuya}}{{Takeda} et~al.}{2009}]{Takeda:2009aa}
{Takeda} Y.,  {Kaneko} H.,  {Matsumoto} N.,  {Oshino} S.,  {Ito} H.,
  {Shibuya} T.,  2009, \mn@doi [\pasj] {10.1093/pasj/61.3.563}, \href
  {http://adsabs.harvard.edu/abs/2009PASJ...61..563T} {61, 563}

\bibitem[\protect\citeauthoryear{{Tayal}}{{Tayal}}{2004a}]{Tayal:2004aa}
{Tayal} S.~S.,  2004a, \mn@doi [\apjs] {10.1086/380784}, \href
  {http://adsabs.harvard.edu/abs/2004ApJS..150..465T} {150, 465}

\bibitem[\protect\citeauthoryear{{Tayal}}{{Tayal}}{2004b}]{Tayal:2004ab}
{Tayal} S.~S.,  2004b, \mn@doi [\aap] {10.1051/0004-6361:20041557}, \href
  {http://adsabs.harvard.edu/abs/2004A%26A...426..717T} {426, 717}

\bibitem[\protect\citeauthoryear{{Turner}, {Kirby-Docken}  \&
  {Dalgarno}}{{Turner} et~al.}{1977}]{Turner:1977aa}
{Turner} J.,  {Kirby-Docken} K.,   {Dalgarno} A.,  1977, \mn@doi [\apjs]
  {10.1086/190481}, \href {http://adsabs.harvard.edu/abs/1977ApJS...35..281T}
  {35, 281}

\bibitem[\protect\citeauthoryear{{Tylenda}, {Acker}, {Stenholm}, {Gleizes}  \&
  {Raytchev}}{{Tylenda} et~al.}{1991}]{Tylenda:1991aa}
{Tylenda} R.,  {Acker} A.,  {Stenholm} B.,  {Gleizes} F.,   {Raytchev} B.,
  1991, \aaps, \href {https://ui.adsabs.harvard.edu/abs/1991A&AS...89...77T}
  {89, 77}

\bibitem[\protect\citeauthoryear{{Umana}, {Leto}, {Trigilio}, {Buemi},
  {Manzitto}, {Toscano}, {Dolei}  \& {Cerrigone}}{{Umana}
  et~al.}{2008}]{Umana:2008aa}
{Umana} G.,  {Leto} P.,  {Trigilio} C.,  {Buemi} C.~S.,  {Manzitto} P.,
  {Toscano} S.,  {Dolei} S.,   {Cerrigone} L.,  2008, \mn@doi [\aap]
  {10.1051/0004-6361:20078796}, \href
  {http://adsabs.harvard.edu/abs/2008A%26A...482..529U} {482, 529}

\bibitem[\protect\citeauthoryear{{Vassiliadis} \& {Wood}}{{Vassiliadis} \&
  {Wood}}{1994}]{Vassiliadis:1994ab}
{Vassiliadis} E.,  {Wood} P.~R.,  1994, \mn@doi [\apjs] {10.1086/191962}, \href
  {http://adsabs.harvard.edu/abs/1994ApJS...92..125V} {92, 125}

\bibitem[\protect\citeauthoryear{{Verner}, {Verner}  \& {Ferland}}{{Verner}
  et~al.}{1996}]{Verner:1996aa}
{Verner} D.~A.,  {Verner} E.~M.,   {Ferland} G.~J.,  1996, \mn@doi [Atomic Data
  and Nuclear Data Tables] {10.1006/adnd.1996.0018}, \href
  {http://ads.nao.ac.jp/abs/1996ADNDT..64....1V} {64, 1}

\bibitem[\protect\citeauthoryear{{Vollmer} et~al.,}{{Vollmer}
  et~al.}{2010}]{Vollmer:2010aa}
{Vollmer} B.,  et~al., 2010, \mn@doi [\aap] {10.1051/0004-6361/200913460},
  \href {https://ui.adsabs.harvard.edu/\#abs/2010A&A...511A..53V} {511, A53}

\bibitem[\protect\citeauthoryear{{Wesson}, {Liu}  \& {Barlow}}{{Wesson}
  et~al.}{2005}]{Wesson:2005aa}
{Wesson} R.,  {Liu} X.-W.,   {Barlow} M.~J.,  2005, \mn@doi [\mnras]
  {10.1111/j.1365-2966.2005.09325.x}, \href
  {http://adsabs.harvard.edu/abs/2005MNRAS.362..424W} {362, 424}

\bibitem[\protect\citeauthoryear{{Wiese}, {Fuhr}  \& {Deters}}{{Wiese}
  et~al.}{1996}]{Wiese:1996aa}
{Wiese} W.~L.,  {Fuhr} J.~R.,   {Deters} T.~M.,  1996, {Atomic transition
  probabilities of carbon, nitrogen, and oxygen : a critical data compilation}.
Springer

\bibitem[\protect\citeauthoryear{{Yamamura}, {Makiuti}, {Ikeda}, {Fukuda},
  {Oyabu}, {Koga}  \& {White}}{{Yamamura} et~al.}{2010}]{Yamamura:2010aa}
{Yamamura} I.,  {Makiuti} S.,  {Ikeda} N.,  {Fukuda} Y.,  {Oyabu} S.,  {Koga}
  T.,   {White} G.~J.,  2010, VizieR Online Data Catalog, \href
  {http://adsabs.harvard.edu/abs/2010yCat.2298....0Y} {2298}

\bibitem[\protect\citeauthoryear{{Zeippen}, {Butler}  \& {Le
  Bourlot}}{{Zeippen} et~al.}{1987}]{Zeippen:1987aa}
{Zeippen} C.~J.,  {Butler} K.,   {Le Bourlot} J.,  1987, \aap, \href
  {http://ads.nao.ac.jp/abs/1987A%26A...188..251Z} {188, 251}

\bibitem[\protect\citeauthoryear{{Zhang}}{{Zhang}}{1996}]{Zhang:1996aa}
{Zhang} H.,  1996, \aaps, \href
  {http://ads.nao.ac.jp/abs/1996A%26AS..119..523Z} {119, 523}

\bibitem[\protect\citeauthoryear{{Zhang} \& {Kwok}}{{Zhang} \&
  {Kwok}}{1991}]{1991A&A...250..179Z}
{Zhang} C.~Y.,  {Kwok} S.,  1991, \aap, \href
  {https://ui.adsabs.harvard.edu/abs/1991A&A...250..179Z} {250, 179}

\bibitem[\protect\citeauthoryear{{Zhang} \& {Liu}}{{Zhang} \&
  {Liu}}{2005}]{Zhang:2005ab}
{Zhang} Y.,  {Liu} X.-W.,  2005, \mn@doi [\apjl] {10.1086/497113}, \href
  {http://adsabs.harvard.edu/abs/2005ApJ...631L..61Z} {631, L61}

\bibitem[\protect\citeauthoryear{{van Raai}, {Lugaro}, {Karakas},
  {Garc{\'{\i}}a-Hern{\'a}ndez}  \& {Yong}}{{van Raai}
  et~al.}{2012}]{Raai:2012aa}
{van Raai} M.~A.,  {Lugaro} M.,  {Karakas} A.~I.,
  {Garc{\'{\i}}a-Hern{\'a}ndez} D.~A.,   {Yong} D.,  2012, \mn@doi [\aap]
  {10.1051/0004-6361/201117896}, \href
  {http://adsabs.harvard.edu/abs/2012A%26A...540A..44V} {540, A44}

\makeatother
\end{thebibliography}

\appendix

\section{Supporting Results}

\begin{table*}
\centering
\renewcommand{\thetable}{A\arabic{table}}
\caption{\label{T-emission} %
Identified emission lines in J900.}
\renewcommand{\arraystretch}{0.85}
\begin{tabularx}{\textwidth}{@{}@{\extracolsep{\fill}}
D{.}{.}{-1}@{\hspace{-2pt}}c@{\hspace{1pt}}D{.}{.}{-1}@{\hspace{2pt}}r@{\hspace{2pt}}r
D{.}{.}{-1}@{\hspace{-2pt}}c@{\hspace{1pt}}D{.}{.}{-1}@{\hspace{2pt}}r@{\hspace{2pt}}r
D{.}{.}{-1}@{\hspace{-2pt}}c@{\hspace{1pt}}D{.}{.}{-1}@{\hspace{2pt}}r@{\hspace{2pt}}r@{}}
\midrule
\multicolumn{1}{c}{$\lambda_{\rm lab.}$ ({\AA})}&Line&\multicolumn{1}{c}{$f(\lambda)$}&$I$($\lambda$)&$\delta$\,$I$($\lambda$)&
\multicolumn{1}{c}{$\lambda_{\rm lab.}$ ({\AA})}&Line&\multicolumn{1}{c}{$f(\lambda)$}&$I$($\lambda$)&$\delta$\,$I$($\lambda$)&
\multicolumn{1}{c}{$\lambda_{\rm lab.}$ ({\AA})}&Line&\multicolumn{1}{c}{$f(\lambda)$}&$I$($\lambda$)&$\delta$\,$I$($\lambda$)\\
\midrule
\multicolumn{1}{l}{1548/51}    &    C\,{\sc iv}    & 1.238 & 624.365 & 13.594 & 5047.74 &   {\hei}          & -0.048 & 0.134 & 0.010 & 7135.80 &   {\ariii}      & -0.374 & 8.470 & 0.173 \\ 
1640.42 &    {\heii}    & 1.177 & 235.081 & 4.276 & 5191.82 &   {\ariii}      & -0.081 & 0.084 & 0.006 & 7170.50 &   {\ariv}       & -0.378 & 0.086 & 0.003 \\ 
\multicolumn{1}{l}{1749/54}    &    N\,{\sc iii}]    & 1.154 & 19.214 & 5.870 & 5197.90 &   {\NI}        & -0.082 & 0.345 & 0.009 & 7177.50 &   {\heii}       & -0.379 & 0.395 & 0.009 \\ 
\multicolumn{1}{l}{1906/09}    &    C\,{\sc iii}]    & 1.258 & 1026.996 & 22.841 & 5200.26 &   {\NI}         & -0.083 & 0.257 & 0.008 & 7236.42 &   C\,{\sc ii}        & -0.387 & 0.414 & 0.012 \\ 
\multicolumn{1}{l}{2325/26/28}    &    [C\,{\sc ii}]    & 1.372 & 87.207 & 3.980 & 5342.43 &   C\,{\sc ii}         & -0.112 & 0.056 & 0.009 & 7262.70 &   {\ariv}       & -0.391 & 0.070 & 0.004 \\ 
\multicolumn{1}{l}{2422/25}    &    [Ne\,{\sc iv}]        & 1.127 & 48.638 & 1.998 & 5355.87 &   {\feiii}    & -0.115 & 0.033 & 0.004 & 7281.35 &   {\hei}          & -0.393 & 0.639 & 0.015 \\ 
2733.30 &   {\heii}   & 0.720 & 7.631 & 0.725 & 5411.52 &   {\heii}         & -0.126 & 2.785 & 0.029 & 7318.92 &   {\oii}       & -0.398 & 5.782 & 0.099 \\ 
3721.94 &   {\hi}         & 0.323 & 2.750 & 0.387 & 5470.68 &   C\,{\sc iv}         & -0.137 & 0.057 & 0.005 & 7329.66 &   {\oii}       & -0.400 & 5.022 & 0.111 \\ 
3726.03 &   {\oii}   & 0.322 & 75.920 & 1.385 & 5480.96 &    Fe\,{\sc ii}          & -0.139 & 0.032 & 0.002 & 7486.80 &   C\,{\sc iii}        & -0.421 & 0.020 & 0.002 \\ 
3728.81 &   {\oii}   & 0.322 & 42.811 & 1.403 & 5483.35 &    C\,{\sc ii}         & -0.139 & 0.021 & 0.002 & 7499.85 &   {\hei}        & -0.422 & 0.040 & 0.003 \\ 
3734.37 &   {\hi}         & 0.321 & 2.969 & 0.176 & 5517.72 &    {\cliii}       & -0.145 & 0.298 & 0.008 & 7530.80 &   {\cliv}       & -0.426 & 0.180 & 0.005 \\ 
3750.15 &   {\hi}         & 0.317 & 2.532 & 0.277 & 5519.53 &    Fe\,{\sc iii}    & -0.145 & 0.035 & 0.005 & 7534.88 &   Ca\,{\sc i}   & -0.427 & 0.041 & 0.004 \\ 
3759.87 &   O\,{\sc iii}   & 0.315 & 3.154 & 0.189 & 5537.89 &    {\cliii}       & -0.149 & 0.436 & 0.007 & 7722.58 &   Mg\,{\sc i}]         & -0.452 & 0.011 & 0.001 \\ 
3770.63 &   {\hi}         & 0.313 & 3.439 & 0.164 & 5577.34 &    {\oi}         & -0.156 & 0.111 & 0.007 & 7725.00 &   N\,{\sc ii}        & -0.452 & 0.019 & 0.003 \\ 
3797.90 &   {\hi}         & 0.307 & 5.243 & 0.192 & 5592.25 &   O\,{\sc iii}         & -0.158 & 0.054 & 0.005 & 7725.90 &   C\,{\sc iv}          & -0.452 & 0.241 & 0.007 \\ 
3835.38 &   {\hi}         & 0.299 & 7.654 & 0.183 & 5709.11 &   {\xeiv}   & -0.177 & 0.030 & 0.006 & 7736.00 &   C\,{\sc iv}          & -0.453 & 0.026 & 0.002 \\ 
3868.76 &   {\neiii}   & 0.291 & 117.595 & 2.176 & 5710.77 &   N\,{\sc ii}         & -0.178 & 0.061 & 0.004 & 7751.10 &    {\ariii}       & -0.455 & 2.069 & 0.052 \\ 
3889.05 &   {\hi}         & 0.286 & 19.620 & 0.356 & 5754.64 &   {\nii}        & -0.185 & 1.243 & 0.015 & 7816.13 &    {\hei}         & -0.464 & 0.044 & 0.004 \\ 
3964.73 &   {\hei}   & 0.267 & 0.610 & 0.064 & 5759.43 &   {\heii}$^{\rm a}$   & -0.185 & 0.006 & 0.002 & 7875.99 &    {\pii}         & -0.471 & 0.067 & 0.004 \\ 
3967.47 &   {\neiii}   & 0.267 & 35.354 & 0.622 & 5759.64 &   {\rbiv}   & -0.185 & 0.014 & 0.004 & 8113.58 &    P\,{\sc iii}    & -0.501 & 0.052 & 0.002 \\ 
3970.07 &   {\hi}         & 0.266 & 17.820 & 0.279 & 5801.33 &   C\,{\sc iv}        & -0.192 & 0.310 & 0.007 & 8236.79 &    {\heii}          & -0.515 & 0.752 & 0.022 \\ 
4025.60 &   {\heii}       & 0.251 & 0.520 & 0.070 & 5811.97 &   C\,{\sc iv}         & -0.193 & 0.160 & 0.007 & 8247.73 &    {\hi}            & -0.516 & 0.049 & 0.006 \\ 
4026.20 &   {\hei}   & 0.251 & 1.936 & 0.081 & 5846.66 &   {\heii}$^{\rm b}$  & -0.199 & 0.014 & 0.004 & 8249.97 &   {\hi}           & -0.517 & 0.061 & 0.010 \\ 
4045.70 &   Ca\,{\sc i}   & 0.246 & 0.456 & 0.042 & 5846.67 &   {\xeiii}   & -0.199 & 0.024 & 0.007 & 8252.40 &   {\hi}           & -0.517 & 0.072 & 0.007 \\ 
4059.90 &   {\fiv}   & 0.241 & 0.072 & 0.014 & 5857.26 &   {\heii}         & -0.200 & 0.015 & 0.004 & 8260.93 &   {\hi}           & -0.518 & 0.031 & 0.004 \\ 
4060.20 &   N\,{\sc ii}   & 0.241 & 0.129 & 0.021 & 5867.74 &   {\kriv}       & -0.202 & 0.146 & 0.005 & 8262.33 &   Ca\,{\sc i}   & -0.518 & 0.027 & 0.004 \\ 
4067.94 &   C\,{\sc iii}       & 0.239 & 0.492 & 0.116 & 5875.62 &   {\hei}          & -0.203 & 11.723 & 0.135 & 8264.28 &   {\hi}         & -0.518 & 0.063 & 0.005 \\ 
4068.60 &   {\sii}   & 0.239 & 1.961 & 0.164 & 5913.26 &   {\heii}       & -0.209 & 0.022 & 0.003 & 8267.94 &   {\hi}         & -0.519 & 0.064 & 0.004 \\ 
4070.31 &   C\,{\sc iii}   & 0.239 & 0.537 & 0.049 & 5931.83 &   {\heii}         & -0.211 & 0.057 & 0.004 & 8271.93 &   {\hi}         & -0.519 & 0.049 & 0.004 \\ 
4076.35 &   {\sii}   & 0.237 & 0.823 & 0.065 & 5977.03 &   {\heii}         & -0.218 & 0.040 & 0.004 & 8276.31 &   {\hi}           & -0.520 & 0.088 & 0.006 \\ 
4089.29 &   O\,{\sc ii}         & 0.233 & 0.095 & 0.032 & 6004.72 &   {\heii}         & -0.222 & 0.056 & 0.006 & 8281.12 &   {\hi}         & -0.520 & 0.093 & 0.005 \\ 
4097.38 &   N\,{\sc ii}        & 0.231 & 0.712 & 0.028 & 6036.78 &   {\heii}       & -0.226 & 0.063 & 0.002 & 8292.31 &   {\hi}           & -0.521 & 0.067 & 0.003 \\ 
4101.73 &   {\hi}          & 0.230 & 29.172 & 0.415 & 6074.19 &   {\heii}         & -0.232 & 0.079 & 0.003 & 8298.83 &   {\hi}           & -0.522 & 0.096 & 0.004 \\ 
4120.81 &   {\hei}          & 0.224 & 0.292 & 0.039 & 6086.99 &   Fe\,{\sc i}?   & -0.233 & 0.022 & 0.003 & 8306.11 &   {\hi}           & -0.523 & 0.120 & 0.004 \\ 
4128.66 &   N\,{\sc ii}         & 0.222 & 0.143 & 0.017 & 6101.79 &   {\kiv}       & -0.235 & 0.115 & 0.004 & 8312.10 &   C\,{\sc iii}         & -0.524 & 0.056 & 0.003 \\ 
4143.76 &   {\hei}          & 0.217 & 0.285 & 0.024 & 6118.26 &   {\heii}         & -0.238 & 0.083 & 0.003 & 8314.26 &   {\hi}           & -0.524 & 0.116 & 0.004 \\ 
4189.79 &   O\,{\sc ii}         & 0.204 & 0.491 & 0.035 & 6151.27 &   C\,{\sc ii}         & -0.242 & 0.049 & 0.005 & 8323.42 &   {\hi}           & -0.525 & 0.145 & 0.005 \\ 
4199.83 &   {\heii}         & 0.200 & 0.847 & 0.032 & 6159.89 &   Fe\,{\sc iii}?   & -0.243 & 0.129 & 0.006 & 8333.78 &   {\hi}           & -0.526 & 0.189 & 0.007 \\ 
4238.29 &   C\,{\sc iii}        & 0.189 & 0.129 & 0.018 & 6170.60 &   {\heii}       & -0.245 & 0.084 & 0.004 & 8342.20 &   C\,{\sc iii}         & -0.527 & 0.040 & 0.002 \\ 
4254.12 &   O\,{\sc ii}        & 0.184 & 0.085 & 0.016 & 6174.94 &   [Br\,{\sc iv}]?   & -0.246 & 0.016 & 0.002 & 8345.47 &   {\hi}         & -0.527 & 0.158 & 0.005 \\ 
4267.18 &   C\,{\sc ii}         & 0.180 & 1.183 & 0.035 & 6209.27 &   Ca\,{\sc i}   & -0.250 & 0.018 & 0.001 & 8359.00 &   {\hi}           & -0.529 & 0.196 & 0.006 \\ 
4338.67 &   {\heii}         & 0.157 & 0.964 & 0.018 & 6233.82 &   {\heii}         & -0.254 & 0.113 & 0.005 & 8361.73 &   {\hei}          & -0.529 & 0.071 & 0.003 \\ 
4340.46 &   {\hi}          & 0.157 & 49.523 & 0.434 & 6238.39 &   Fe\,{\sc ii}       & -0.254 & 0.013 & 0.002 & 8374.48 &   {\hi}           & -0.531 & 0.201 & 0.006 \\ 
4363.21 &   {\oiii}   & 0.149 & 14.442 & 0.218 & 6256.52 &   C\,{\sc ii}         & -0.257 & 0.021 & 0.002 & 8386.40 &   N\,{\sc iii}         & -0.532 & 0.017 & 0.003 \\ 
4387.93 &   {\hei}          & 0.142 & 0.566 & 0.023 & 6257.62 &   O\,{\sc i}         & -0.257 & 0.015 & 0.002 & 8392.40 &   {\hi}           & -0.533 & 0.239 & 0.008 \\ 
4471.47 &   {\hei}          & 0.115 & 3.799 & 0.033 & 6300.30 &   {\oi}        & -0.263 & 7.974 & 0.358 & 8413.32 &   {\hi}           & -0.535 & 0.271 & 0.009 \\ 
4541.59 &   {\heii}         & 0.093 & 1.250 & 0.015 & 6310.80 &   {\heii}       & -0.264 & 0.125 & 0.010 & 8437.95 &   {\hi}           & -0.537 & 0.331 & 0.010 \\ 
4571.10 &   Mg\,{\sc i}]         & 0.084 & 0.323 & 0.018 & 6312.10 &   {\siii}      & -0.264 & 1.249 & 0.019 & 8446.48 &   O\,{\sc i}           & -0.538 & 0.080 & 0.004 \\ 
4634.12 &   N\,{\sc iii}         & 0.065 & 0.426 & 0.016 & 6363.78 &   {\oi}        & -0.271 & 2.742 & 0.042 & 8467.25 &   {\hi}           & -0.541 & 0.404 & 0.012 \\ 
4640.64 &   N\,{\sc iii}        & 0.063 & 0.950 & 0.015 & 6406.38 &   {\heii}         & -0.277 & 0.156 & 0.004 & 8480.79 &   {\hei}          & -0.542 & 0.017 & 0.002 \\ 
4641.81 &   O\,{\sc ii}        & 0.063 & 0.137 & 0.010 & 6434.73 &   {\arv}   & -0.281 & 0.207 & 0.008 & 8502.48 &   {\hi}           & -0.544 & 0.424 & 0.013 \\ 
4647.42 &   C\,{\sc iii}        & 0.061 & 0.458 & 0.011 & 6461.95 &   C\,{\sc ii}        & -0.284 & 0.103 & 0.003 & 8519.35 &   {\heii}         & -0.546 & 0.028 & 0.003 \\ 
4649.13 &   O\,{\sc ii}         & 0.061 & 0.080 & 0.010 & 6471.66 &   Ca\,{\sc i}   & -0.285 & 0.169 & 0.005 & 8541.88 &   {\heii}         & -0.548 & 0.044 & 0.007 \\ 
4650.25 &   C\,{\sc iii}       & 0.060 & 0.296 & 0.015 & 6515.90 &   N\,{\sc iii}        & -0.291 & 0.037 & 0.003 & 8545.38 &   {\hi}         & -0.549 & 0.556 & 0.017 \\ 
4651.47 &   C\,{\sc iii}        & 0.060 & 0.120 & 0.014 & 6527.24 &    {\nii}      & -0.293 & 0.207 & 0.005 & 8578.69 &   {\clii}       & -0.552 & 0.139 & 0.005 \\ 
4658.05 &   {\feiii}      & 0.058 & 0.184 & 0.012 & 6548.04 &   {\nii}       & -0.296 & 17.678 & 0.709 & 8582.61 &   {\hei}   & -0.552 & 0.050 & 0.005 \\ 
4658.30 &   C\,{\sc iv}        & 0.058 & 0.487 & 0.014 & 6560.00 &   {\heii}   & -0.297 & 5.875 & 0.691 & 8598.39 &   {\hi}           & -0.554 & 0.674 & 0.020 \\ 
4685.71 &   {\heii}         & 0.050 & 35.847 & 0.173 & 6562.80 &   {\hi}   & -0.298 & 286.007 & 4.772 & 8616.59 &   Ca\,{\sc i}   & -0.556 & 0.070 & 0.005 \\ 
4711.37 &   {\ariv}       & 0.042 & 1.866 & 0.025 & 6578.05 &   C\,{\sc ii}         & -0.300 & 0.453 & 0.014 & 8626.19 &   {\heii}         & -0.557 & 0.030 & 0.003 \\ 
4713.14 &   {\hei}          & 0.042 & 0.582 & 0.020 & 6583.46 &   {\nii}       & -0.300 & 53.335 & 0.902 & 8665.02 &   {\hi}           & -0.560 & 0.858 & 0.026 \\ 
4714.17 &   {\neiv}       & 0.041 & 0.195 & 0.018 & 6656.00 &   Fe\,{\sc ii}]      & -0.310 & 0.145 & 0.004 & 8727.12 &   [C\,{\sc i}]         & -0.566 & 0.241 & 0.010 \\ 
4715.61 &   {\neiv}       & 0.041 & 0.074 & 0.010 & 6678.15 &   {\hei}          & -0.313 & 3.127 & 0.053 & 8746.89 &   {\heii}         & -0.568 & 0.024 & 0.004 \\ 
4724.17 &   {\neiv}       & 0.038 & 0.202 & 0.009 & 6683.20 &   {\heii}         & -0.313 & 0.262 & 0.009 & 8750.47 &   {\hi}           & -0.568 & 1.072 & 0.033 \\ 
4725.64 &   {\neiv}       & 0.038 & 0.158 & 0.009 & 6716.44 &   {\sii}       & -0.318 & 3.263 & 0.058 & 8845.37 &   {\hei}        & -0.576 & 0.042 & 0.004 \\ 
4740.16 &   {\ariv}       & 0.034 & 2.124 & 0.028 & 6730.81 &   {\sii}        & -0.320 & 5.588 & 0.100 & 8859.15 &   {\heii}         & -0.578 & 0.043 & 0.007 \\ 
4751.59 &   {\feiii}      & 0.030 & 0.069 & 0.009 & 6744.10 &   {\hei}          & -0.322 & 0.028 & 0.002 & 8862.78 &   {\hi}           & -0.578 & 1.350 & 0.042 \\ 
4760.87 &   O\,{\sc i}         & 0.001 & 2.160 & 0.116 & 6779.94 &   C\,{\sc ii}          & -0.326 & 0.051 & 0.003 & 8929.11 &   {\heii}         & -0.583 & 0.041 & 0.003 \\ 
4861.33 &   {\hi}   & 0.000 & 100.000 & 0.381 & 6791.47 &   C\,{\sc ii}         & -0.328 & 0.027 & 0.003 & 9011.21 &   {\heii}         & -0.590 & 0.014 & 0.010 \\ 
4906.83 &   O\,{\sc ii}        & -0.012 & 0.157 & 0.010 & 6795.10 &   {\kiv}      & -0.328 & 0.031 & 0.002 & 9014.91 &   {\hi}         & -0.590 & 0.323 & 0.017 \\ 
4921.93 &   {\hei}          & -0.016 & 1.018 & 0.013 & 6821.16 &   [Mn\,{\sc iii}]      & -0.332 & 0.009 & 0.002 & 9068.60 &   {\siii}       & -0.594 & 11.272 & 0.307 \\ 
4931.23 &   {\oiii}      & -0.019 & 0.126 & 0.008 & 6826.70 &   {\kriii}      & -0.333 & 0.026 & 0.006 & 9108.54 &   {\heii}         & -0.597 & 0.056 & 0.003 \\ 
4959.00 &   {\oiii}   & -0.026 & 357.906 & 3.090 & 7062.28 &   {\hei}          & -0.364 & 0.079 & 0.003 & 9123.60 &   {\clii}       & -0.598 & 0.040 & 0.003 \\ 
5007.00 &   {\oiii}   & -0.038 & 1048.258 & 6.313 & 7065.71 &   {\hei}        & -0.364 & 5.210 & 0.105 & 9210.34 &   {\hei}          & -0.604 & 0.067 & 0.006 \\ 
5015.68 &   {\hei}          & -0.040 & 1.429 & 0.012 & 7099.80 &   [Pb\,{\sc ii}]?       & -0.369 & 0.037 & 0.004 &    &    &    &    &     \\ 
\midrule
\end{tabularx}
\end{table*}

\setcounter{table}{0}
 \begin{table}
     \renewcommand{\thetable}{A\arabic{table}}
   \renewcommand{\arraystretch}{0.85}
  \centering
  \caption{Continued.}
  \begin{threeparttable}
\begin{tabularx}{\columnwidth}{@{}@{\extracolsep{\fill}}
D{.}{.}{-1}@{\hspace{-2pt}}c@{\hspace{1pt}}D{.}{.}{-1}@{\hspace{2pt}}r@{\hspace{2pt}}r@{}}
 \midrule
  \multicolumn{1}{c}{$\lambda_{\rm lab.}$ ({\micron})}&Line&&$I$($\lambda$)&$\delta$\,$I$($\lambda$)\\
\midrule
3.04 & {\hi} 5-10 &  & 0.328 & 0.070 \\ 
3.10 & {\heii} &  & 1.028 & 0.056 \\ 
3.74 & {\hi} 5-8 &  & 0.970 & 0.082 \\ 
3.84 & H$_{2}$ $v=0-0$ S(13) &  & 0.215 & 0.057 \\
4.05 & {\hi} 4-5 &  & 7.781 & 0.107 \\ 
4.49 & {\mgiv} &  & 6.128 & 0.147 \\ 
4.65 & {\hi} 5-7 &  & 1.147 & 0.141 \\ 
5.61 & {\mgv} &  & 4.509 & 0.524 \\ 
6.91 & H$_{2}$ $v=0-0$ S(5) &  & 0.762 & 0.312 \\
6.98 & {\arii} &  & 0.867 & 0.221 \\ 
8.99 & {\ariii} &  & 6.293 & 0.903 \\ 
9.68 & H$_{2}$ $v=0-0$ S(3) &  & 1.017 & 0.220 \\ 
10.51 & {\siv} &  & 58.937 & 3.466 \\ 
11.76 & {\cliv} &  & 0.418 & 0.042 \\ 
12.28 & H$_{2}$ $v=0-0$ S(2) &  & 0.189 & 0.057 \\ 
12.37 & {\hi} &  & 1.043 & 0.043 \\ 
12.82 & {\neii} &  & 4.006 & 0.249 \\ 
13.10 & {\arv} &  & 1.684 & 0.118 \\ 
13.43 & {\fv} &  & 0.309 & 0.023 \\ 
13.52 & {\mgv} &  & 0.370 & 0.026 \\ 
14.07 & {\hei}? &  & 0.362 & 0.048 \\ 
14.32 & {\nev} &  & 44.563 & 2.690 \\ 
15.56 & {\neiii} &  & 113.924 & 6.646 \\ 
17.62 & {\hi}? &  & 1.356 & 0.096 \\ 
17.89 & {\piii} &  & 0.724 & 0.074 \\ 
18.72 & {\siii} &  & 15.049 & 0.890 \\ 
19.07 & {\hi} 7-8 &  & 0.522 & 0.102 \\ 
20.31 & {\cliv} &  & 0.307 & 0.026 \\ 
21.83 & {\ariii} &  & 0.566 & 0.052 \\ 
22.34 & {\feii} &  & 0.602 & 0.047 \\ 
24.32 & {\nev} &  & 25.480 & 1.491 \\ 
25.90 & {\oiv} &  & 282.948 & 16.450 \\ 
33.49 & {\siii} &  & 5.215 & 0.319 \\ 
34.84 & {\Siii} &  & 1.427 & 0.190 \\ 
36.02 & {\neiii} &  & 6.710 & 0.429 \\
\midrule
 \end{tabularx}
\begin{tablenotes}[flushleft] 
   \item[a] A predicted $I$({\heii}\,5759.64\,{\AA}) is estimated to be
	      $0.006 ~\pm~ 0.002$ using the theoretical 
 {\heii}\,$I$(5759\,{\AA})/$I$(5857\,{\AA}) = 0.390 by
	      \citet{Storey:1995aa} under Case B assumption.   
   \item[b] A predicted $I$({\heii}\,5846\,{\AA}) is estimated to be
	      $0.014 ~\pm~ 0.004$ using the theoretical 
 {\heii}\,$I$(5846\,{\AA})/$I$(5857\,{\AA}) = 0.914 by
	      \citet{Storey:1995aa} under Case B assumption.
\end{tablenotes}
  \end{threeparttable}
 \end{table}

\begin{table}
 \renewcommand{\arraystretch}{0.85}
    \renewcommand{\thetable}{A\arabic{table}} 
\caption{Broadband flux density of J900. We assume 3\,$\%$ uncertainty because
	      $MSX$ data do not give uncertainty.}
 \label{T-photo}
\begin{tabularx}{\columnwidth}{@{\extracolsep{\fill}}@{}D{.}{.}{-1}cD{p}{\pm}{-1}@{}}
 \midrule
\multicolumn{1}{c}{$\lambda_{\rm c}$ ({\micron})}&Band&\multicolumn{1}{c}{$I_{\lambda}$ (erg\,s$^{-1}$\,cm$^{-2}$\,{\micron}$^{-1}$)}\\
 \midrule
0.4640 & WFC/$g$           & 1.17(-8)~~ ~p~ 7.69(-11) \\ 
0.6122 & WFC/$R$           & 2.97(-9)~~ ~p~ 3.98(-11) \\ 
0.5443 & $HST$/F555W (CSPN)&6.98(-12) ~p~ 4.57(-13) \\ 
0.5443 & $HST$/F555W       & 1.96(-9)~~ ~p~ 1.48(-11) \\ 
1.2350 & WFCAM/$J$         & 9.92(-11) ~p~ 4.03(-12) \\ 
1.6620 & WFCAM/$H$         & 4.10(-11) ~p~ 1.90(-12) \\ 
2.1590 & WFCAM/$K$         & 4.20(-11) ~p~ 2.06(-12) \\ 
3.3526 & $WISE$/W1         & 3.39(-11) ~p~ 1.75(-12) \\ 
4.6028 & $WISE$/W2         & 2.55(-11) ~p~ 1.38(-12) \\ 
9.0000 & $AKARI$/S9W       & 5.43(-11) ~p~ 3.05(-12) \\ 
11.5608 & $WISE$/W3        & 4.60(-11) ~p~ 2.48(-12) \\ 
12.1300 & $MSX$/C          & 4.70(-11) ~p~ 1.41(-12) \\ 
14.6500 & $MSX$/D          & 4.82(-11) ~p~ 1.45(-12) \\ 
18.0000 & $AKARI$/L18W     & 4.61(-11) ~p~ 2.52(-12) \\ 
21.3400 & $MSX$/E          & 4.03(-11) ~p~ 1.21(-12) \\ 
22.0883 & $WISE$/W4        & 4.85(-11) ~p~ 2.61(-12) \\ 
65.0000 & $AKARI$/N60      & 3.95(-12) ~p~ 5.44(-13) \\ 
90.0000 & $AKARI$/WIDE-S   & 1.55(-12) ~p~ 1.28(-13) \\ 
140.0000 & $AKARI$/WIDE-L & 4.50(-13) ~p~ 1.73(-13) \\
\midrule
\multicolumn{1}{c}{$\nu_{\rm c}$ (GHz)}&&\multicolumn{1}{c}{$F_{\nu}$ (Jy)}\\
\midrule
43.0 & &0.090 ~p~ 0.020\\
30.0 & &0.086 ~p~ 0.005\\
20.0 & &0.119 ~p~ 0.012\\
6.0  & &0.109 ~p~ 0.011\\
5.0  & &0.121 ~p~ 0.036\\
2.7  & &0.180 ~p~ 0.060\\
1.4  & &0.108 ~p~ 0.022\\
\midrule
\end{tabularx}
\end{table}

\begin{table*}
\centering
\renewcommand{\thetable}{A\arabic{table}}
\caption{Atomic data adopted in CEL analysis of J900.}
\label{atomf}
\begin{threeparttable}
\begin{tabularx}{\textwidth}{@{}@{\extracolsep{\fill}}lll@{}}
\midrule
Line &transition probability $A_{ji}$ &effective collision strength $\Psi(T)$\\
\midrule
$[${\ci}$]$  &\citet{Froese:1985aa}&\citet{Johnson:1987aa}; \citet{Pequignot:1976aa}\\
$[${\cii}$]$ &\citet{Nussbaumer:1981aa}; \citet{Froese:1994aa}  &\citet{Blum:1992aa}\\
{\ciii}$]$   &\citet{Wiese:1996aa} &\citet{Berrington:1985aa}\\
C\,{\sc iv}  &\citet{Wiese:1996aa} &\citet{Badnell:2000aa}; \citet{Martin:1993aa}\\
{\NI}        &\citet{Wiese:1996aa} &\citet{Pequignot:1976aa}; \citet{Dopita:1976aa}\\
{\nii}       &\citet{Wiese:1996aa} &\citet{Lennon:1994aa}\\
{\niii}      &\citet{Wiese:1996aa} &\citet{Blum:1992aa}\\
{\oi}        &\citet{Wiese:1996aa} &\citet{Bhatia:1995aa}\\
{\oii}       &\citet{Wiese:1996aa} &\citet{McLaughlin:1993aa}; \citet{Pradhan:1976aa}\\
{\oiii}      &\citet{Wiese:1996aa} &\citet{Lennon:1994aa}\\
{\oiv}       &\citet{Wiese:1996aa} &\citet{Blum:1992aa}\\
{\fiv}       &\citet{Garstang:1951aa}; \citet{Storey:2000aa}& \citet{Lennon:1994aa}\\
{\fv}        &\citet{Rynkun:2012aa} &\citet{Liang:2012aa}\\
{\neii}      &\citet{Saraph:1994aa} & \citet{Saraph:1994aa}\\
{\neiii}     &\citet{Mendoza:1983aa}; \citet{Kaufman:1986aa} &\citet{McLaughlin:2000aa}\\
{\neiv}      &\citet{Becker:1989aa}; \citet{Bhatia:1988aa}&\citet{Ramsbottom:1998aa}\\
{\nev}       &\citet{Kaufman:1986aa}; \citet{Bhatia:1993aa}&\citet{Lennon:1994aa}\\
{\mgiv}      &CHIANTI Data Base$^{\rm a}$ &CHIANTI Data Base$^{\rm a}$\\
{\mgv}       &\citet{Mendoza:1983aa}; \citet{Kaufman:1986aa}&\citet{Butler:1994aa}\\
{\Siii}      &\citet{Calamai:1993aa}; \citet{Bergeson:1993aa}; \citet{Lanzafame:1994aa};              &\citet{Dufton:1991aa}\\
             &\citet{Nussbaumer:1977aa}\\
{\pii}       &\citet{Mendoza:1982ab} &\citet{Tayal:2004aa}\\
{\piii}      &\citet{Naqvi:1951aa}&\citet{Saraph:1999aa}\\
{\sii}       &\citet{Verner:1996aa}; \citet{Keenan:1993aa}&\citet{Ramsbottom:1996aa}\\
{\siii}      &\citet{Mendoza:1982ab}; \citet{LaJohn:1993aa}; \citet{Kaufman:1986aa}; &\citet{Galavis:1995aa}\\
             &\citet{Heise:1995aa}\\
{\siv}       &\citet{Johnson:1986aa}; \citet{Dufton:1982aa}; \citet{Verner:1996aa}&\citet{Dufton:1982aa}            \\
{\clii}      &CHIANTI Data Base$^{\rm a}$ &\citet{Tayal:2004ab}\\
{\cliii} &\citet{Mendoza:1982aa}; \citet{Kaufman:1986aa} &\citet{Ramsbottom:2001aa}\\
{\cliv}  &\citet{Mendoza:1982ab}; \citet{Ellis:1984aa}&\citet{Galavis:1995aa}\\
         &\citet{Kaufman:1986aa}\\
{\ariii} &\citet{Mendoza:1983aa}; \citet{Kaufman:1986aa}&\citet{Galavis:1995aa} \\
{\ariv}  &\citet{Mendoza:1982aa}; \citet{Kaufman:1986aa} &\citet{Zeippen:1987aa}\\
{\arv}   &\citet{Mendoza:1982ab}; \citet{Kaufman:1986aa}; \citet{LaJohn:1993aa}&\citet{Galavis:1995aa}\\
{\kiv}   &\citet{Mendoza:1983aa}; \citet{Kaufman:1986aa}&\citet{Galavis:1995aa}\\
{\feiii} &\citet{Garstang:1957aa}; \citet{Nahar:1996aa}& \citet{Zhang:1996aa}\\
{\kriii}  &\citet{Biemont:1986aa} & \citet{Schoning:1997aa}             \\
{\kriv}  &\citet{Biemont:1986aa} & \citet{Schoning:1997aa}              \\
{\rbiv}  &\citet{Sterling:2016aa}&\citet{Sterling:2016aa}\\
{\xeiii} & \citet{Biemont:1995aa}  & \citet{Schoening:1998aa}  \\
{\xeiv}  & \citet{Biemont:1995aa}  & \citet{Schoening:1998aa}\\
\midrule
\end{tabularx}
 \begin{tablenotes}[flushleft] 
\item[a] \url{http://www.chiantidatabase.org}
\end{tablenotes}
\end{threeparttable}
\end{table*}

\begin{table}
\centering
\renewcommand{\thetable}{A\arabic{table}}
\caption{Adopted effective recombination coefficient of RLs in analysis of J900.}
 \begin{tabularx}{\columnwidth}{@{\extracolsep{\fill}}ll@{}}
\midrule
Line &References\\
\midrule
H\,{\sc i}& \citet{Aller:1984aa}; \citet{Storey:1995aa}\\
He\,{\sc i}&\citet{Benjamin:1999aa}; \citet{Pequignot:1991aa}\\
He\,{\sc ii}&\citet{Benjamin:1999aa}; \citet{Storey:1995aa}\\
C\,{\sc ii}&\citet{Davey:2000aa}\\
C\,{\sc iii}, C\,{\sc iv}&\citet{Pequignot:1991aa}\\
\midrule
\end{tabularx}
\label{T-R-atomic}
\end{table}

 \begin{table*}
     \renewcommand{\thetable}{A\arabic{table}}
  \centering
\caption{Adopted {\te}--{\Ne} pair for the ionic abundance
  calculations in J900. \label{T-Atene}}
  \begin{threeparttable}
 \begin{tabularx}{\textwidth}{@{\extracolsep{\fill}}@{}
  lcclcc@{}}
  \midrule
 Ion &\multicolumn{1}{c}{{\te} (K)}&\multicolumn{1}{c}{{\Ne} (cm$^{-3}$)}&Ion &\multicolumn{1}{c}{{\te} (K)}&\multicolumn{1}{c}{{\Ne} (cm$^{-3}$)}\\
 \midrule
He$^{+}$ & {\te}({\hei})& \multicolumn{1}{c}{10\,000} &Mg$^{3+}$ & {\te}({\neiv})& {\Ne}({\nev})  \\ 
He$^{2+}$ & {\te}({\hei}) & \multicolumn{1}{c}{10\,000} &Mg$^{4+}$ & {\te}({\neiv})& {\Ne}({\nev}) \\
C$^{2+}$(RL) & {\te}({\hei}) & \multicolumn{1}{c}{10\,000} &Si$^{+}$ & 12\,530 $\pm$ 180$^{\rm a}$&{\Ne}({\sii}) \\
C$^{3+}$(RL) & {\te}({\hei}) & \multicolumn{1}{c}{10\,000} & P$^{+}$ & {\te}({\sii})&{\Ne}({\sii}) \\
C$^{4+}$(RL) & {\te}({\hei}) & \multicolumn{1}{c}{10\,000} & P$^{2+}$ & 11\,800 $\pm$ 490$^{\rm b}$& 6430 $\pm$ 650$^{\rm c}$ \\ 
C$^{0}$(CEL) & \multicolumn{1}{c}{{\te}({\oi})} & \multicolumn{1}{c}{{\Ne}({\NI})} & S$^{+}$ & {\te}({\sii})&{\Ne}({\sii}) \\
C$^{+}$(CEL) & 12\,530 $\pm$ 180 & {\Ne}({\sii})&S$^{2+}$ &12\,060 $\pm$ 550$^{\rm d}$ &{\Ne}({\siii}) \\ 
C$^{2+}$(CEL) & 11\,800 $\pm$ 490 & 6430 $\pm$ 650& S$^{3+}$ & \multicolumn{1}{c}{{\te}({\oiii})}& 5280 $\pm$ 970$^{\rm e}$  \\ 
C$^{3+}$(CEL) & {\te}({\arv})& {\Ne}({\nev})&Cl$^{+}$ & 12\,530 $\pm$ 180& {\Ne}({\oii})  \\ 
N$^{0}$ & \multicolumn{1}{c}{{\te}({\oi})}& \multicolumn{1}{c}{{\Ne}({\NI})}&Cl$^{2+}$ & 11\,800 $\pm$ 490&{\Ne}({\cliii}) \\ 
N$^{+}$ & {\te}({\nii})& {\Ne}({\oii})&Cl$^{3+}$ & {\te}({\cliv})&{\Ne}({\cliv}) \\ 
N$^{2+}$ & \multicolumn{1}{c}{{\te}({\oiii})}& 5280 $\pm$ 970&Ar$^{+}$ &12\,530 $\pm$ 180& {\Ne}({\oii}) \\ 
O$^{0}$ & \multicolumn{1}{c}{{\te}({\oi})}&\multicolumn{1}{c}{{\Ne}({\NI})}&Ar$^{2+}$ &\multicolumn{1}{c}{{\te}({\ariii})}& 6430 $\pm$ 650 \\
O$^{+}$ & {\te}({\oii})& {\Ne}({\oii})&Ar$^{3+}$ & {\te}({\ariv})& {\Ne}({\ariv})  \\ 
O$^{2+}$ & \multicolumn{1}{c}{{\te}({\oiii})}& 5280 $\pm$ 970&Ar$^{4+}$ & {\te}({\arv})& {\Ne}({\nev})  \\ 
O$^{3+}$ & {\te}({\arv})& {\Ne}({\nev})&K$^{3+}$ &13\,890 $\pm$ 1950$^{\rm f}$& {\Ne}({\ariv}) \\
F$^{3+}$ & {\te}({\neiv})& {\Ne}({\nev})& Fe$^{2+}$ & 12\,530 $\pm$ 180& {\Ne}({\oii}) \\ 
F$^{4+}$ & {\te}({\neiv})& {\Ne}({\nev})&Kr$^{2+}$ & \multicolumn{1}{c}{{\te}({\ariii})}& 6430 $\pm$ 650 \\ 
Ne$^{+}$ & 11\,800 $\pm$ 490& 6430 $\pm$ 650&Kr$^{3+}$ & \multicolumn{1}{c}{{\te}({\oiii})}& {\Ne}({\ariv}) \\ 
Ne$^{2+}$ & {\te}({\neiii})& 5280 $\pm$ 970&Rb$^{3+}$ & 15\,020 $\pm$ 580$^{\rm g}$& 5280 $\pm$ 970 \\ 
Ne$^{3+}$ & {\te}({\neiv})& {\Ne}({\nev})&Xe$^{2+}$ & \multicolumn{1}{c}{{\te}({\ariii})}& 6430 $\pm$ 650 \\
Ne$^{4+}$ & {\te}({\neiv})& {\Ne}({\nev})&Xe$^{3+}$ & \multicolumn{1}{c}{{\te}({\oiii})}& {\Ne}({\ariv})\\
 \midrule
 \end{tabularx}
 \begin{tablenotes}[flushleft] 
\item[a] The average between {\te}({\nii}) and {\te}({\oii}).
\item[b] The average between amongst {\te}({\siii}) and {\te}({\ariii}).
\item[c] The average between {\Ne}({\siii}) and {\Ne}({\cliii}).
\item[d] The average between two {\te}({\siii}).
\item[e] The average between {\Ne}({\cliv}) and {\Ne}({\arv}).
\item[f] The average amongst {\te}({\cliv}), {\te}({\ariv}), and {\te}({\neiii}).
\item[g] The average between {\te}({\cliv}) and {\te}({\ariv}).  
 \end{tablenotes}
\end{threeparttable}
 \end{table*}

\begin{table*}
\centering
   \renewcommand{\arraystretch}{0.70}
\caption{
 \label{T-CEL-RL-abund}
CEL and RL ionic abundances in J900.}
\begin{tabularx}{\textwidth}{@{}@{\extracolsep{\fill}}lcD{.}{.}{-1}D{.}{.}{-1}lllcD{.}{.}{-1}D{.}{.}{-1}ll@{}}
 \midrule
X$^{\rm m+}$ &$\lambda_{\rm lab.}$ &\multicolumn{1}{c}{$I$($\lambda$)}
 &\multicolumn{1}{c}{$\delta\,I$($\lambda$)}&X$^{\rm m+}$/H$^{+}$
 &$\delta$(X$^{\rm m+}$/H$^{+}$)&
X$^{\rm m+}$ &$\lambda_{\rm lab.}$ &\multicolumn{1}{c}{$I$($\lambda$)}
 &\multicolumn{1}{c}{$\delta\,I$($\lambda$)}&X$^{\rm m+}$/H$^{+}$
 &$\delta$(X$^{\rm m+}$/H$^{+}$)\\
 \midrule
C$^{0}$ & 8727.12\,{\AA} & 0.241 & 0.010 & 2.45(--6) & 3.04(--7) & S$^{+}$ & 4068.60\,{\AA} & 1.961 & 0.164 & 2.26(--7) & 6.34(--8) \\ 
C$^{+}$ & 2328.00\,{\AA} & 87.207 & 3.980 & 4.49(--5) & 3.95(--6) &  & 4076.35\,{\AA} & 0.823 & 0.065 & 2.93(--7) & 8.18(--8) \\ 
C$^{2+}$ & 1906/09\,{\AA} & 1026.996 & 22.841 & 8.16(--4) & 2.24(--4) &  & 6716.44\,{\AA} & 3.263 & 0.058 & 2.47(--7) & 1.75(--8) \\ 
C$^{3+}$ & 1548/51\,{\AA} & 624.365 & 13.594 & 1.10(--4) & 3.43(--5) &  & 6730.81\,{\AA} & 5.588 & 0.100 & 2.49(--7) & 2.01(--8) \\ 
N$^{0}$ & 5197.90\,{\AA} & 0.345 & 0.009 & 1.00(--6) & 1.05(--7) &  &  &  &  & {\bf 2.48(--7)} & {\bf 1.28(--8)} \\ 
 & 5200.26\,{\AA} & 0.257 & 0.008 & 1.00(--6) & 6.40(--8) & S$^{2+}$ & 6312.10\,{\AA} & 1.249 & 0.019 & 1.43(--6) & 2.40(--7) \\ 
 &  &  &  & {\bf 1.00(--6)} & {\bf 5.47(--8)} &  & 9068.60\,{\AA} & 11.272 & 0.307 & 1.99(--6) & 1.71(--7) \\ 
N$^{+}$ & 5754.64\,{\AA} & 1.243 & 0.015 & 6.82(--6) & 3.68(--7) &  & 18.72\,{\micron} & 15.049 & 0.890 & 1.76(--6) & 1.04(--7) \\ 
 & 6548.04\,{\AA} & 17.678 & 0.709 & 6.71(--6) & 2.96(--7) &  & 33.49\,{\micron} & 5.215 & 0.319 & 1.76(--6) & 1.82(--7) \\ 
 & 6583.46\,{\AA} & 53.335 & 0.902 & 6.83(--6) & 1.70(--7) &  &  &  &  & {\bf 1.77(--6)} & {\bf 7.57(--8)} \\ 
 &  &  &  & {\bf 6.80(--6)} & {\bf 1.37(--7)} & S$^{3+}$ & 10.51\,{\micron} & 58.937 & 3.466 & 1.41(--6) & 1.00(--7) \\ 
N$^{2+}$ & 1749/54\,{\AA} &19.214  &5.870  &3.83(--5)  &1.17(--5)  & Cl$^{+}$ & 8578.69\,{\AA} & 0.139 & 0.005 & 4.16(--9) & 1.93(--10) \\ 
O$^{0}$ & 5577.34\,{\AA} & 0.111 & 0.007 & 3.05(--5) & 5.68(--6) &  & 9123.60\,{\AA} & 0.040 & 0.003 & 4.61(--9) & 4.08(--10) \\ 
 & 6300.30\,{\AA} & 7.974 & 0.358 & 2.97(--5) & 3.24(--6) &  &  &  &  & {\bf 4.39(--9)} & {\bf 3.18(--10)} \\ 
 & 6363.78\,{\AA} & 2.742 & 0.042 & 3.20(--5) & 3.22(--6) & Cl$^{2+}$ & 5517.72\,{\AA} & 0.298 & 0.008 & 3.38(--8) & 2.76(--9) \\ 
 &  &  &  & {\bf 3.08(--5)} & {\bf 2.12(--6)} &  & 5537.89\,{\AA} & 0.436 & 0.007 & 3.39(--8) & 3.59(--9) \\ 
O$^{+}$ & 3726.03\,{\AA} & 75.920 & 1.385 & 2.47(--5) & 1.10(--6) &  &  &  &  & {\bf 3.38(--8)} & {\bf 2.19(--9)} \\ 
 & 3728.81\,{\AA} & 42.811 & 1.403 & 2.46(--5) & 8.18(--7) & Cl$^{3+}$ & 7530.80\,{\AA} & 0.180 & 0.005 & 2.06(--8) & 2.06(--9) \\ 
 & 7320/30\,{\AA} & 10.606 & 0.149 & 2.49(--5) & 3.12(--6) &  & 11.76\,{\micron} & 0.418 & 0.042 & 2.02(--8) & 2.84(--9) \\ 
 &  &  &  & {\bf 2.47(--5)} & {\bf 6.41(--7)} &  & 20.31\,{\micron} & 0.307 & 0.026 & 2.16(--8) & 1.38(--9) \\ 
O$^{2+}$ & 4363.21\,{\AA} & 14.281 & 0.219 & 1.70(--4) & 4.08(--6) &  &  &  &  & {\bf 2.11(--8)} & {\bf 1.06(--9)} \\ 
 & 4931.23\,{\AA} & 0.126 & 0.008 & 1.49(--4) & 1.03(--5) & Ar$^{+}$ & 6.98\,{\micron} & 0.867 & 0.221 & 6.43(--8) & 1.64(--8) \\ 
 & 4959.00\,{\AA} & 357.906 & 3.090 & 1.66(--4) & 3.12(--6) & Ar$^{2+}$ & 5191.82\,{\AA} & 0.084 & 0.006 & 6.00(--7) & 9.89(--8) \\ 
 & 5007.00\,{\AA} & 1048.258 & 6.313 & 1.68(--4) & 3.01(--6) &  & 7135.80\,{\AA} & 8.470 & 0.173 & 5.95(--7) & 4.26(--8) \\ 
 &  &  &  & {\bf 1.67(--4)} & {\bf 1.88(--6)} &  & 7751.10\,{\AA} & 2.069 & 0.052 & 6.06(--7) & 4.43(--8) \\ 
O$^{3+}$ & 25.90\,{\micron} & 282.948 & 16.450 & 1.03(--4) & 1.09(--5) &  & 8.99\,{\micron} & 6.293 & 0.903 & 6.24(--7) & 9.00(--8) \\ 
F$^{3+}$ & 4059.90\,{\AA} & 0.072 & 0.014 & 7.26(--9) & 1.48(--9) &  & 21.83\,{\micron} & 0.566 & 0.052 & 9.16(--7) & 8.42(--8) \\ 
F$^{4+}$ & 13.43\,{\micron} & 0.309 & 0.023 & 1.26(--8) & 1.19(--9) &  &  &  &  & {\bf 6.00(--7)} & {\bf 2.93(--8)} \\ 
Ne$^{+}$ & 12.82\,{\micron} & 4.006 & 0.249 & 4.97(--6) & 3.28(--7) & Ar$^{3+}$ & 4711.37\,{\AA} & 1.866 & 0.025 & 1.81(--7) & 6.08(--9) \\ 
Ne$^{2+}$ & 3868.76\,{\AA} & 117.595 & 2.176 & 7.21(--5) & 5.34(--6) &  & 4740.16\,{\AA} & 2.124 & 0.028 & 1.78(--7) & 1.37(--8) \\ 
 & 3967.47\,{\AA} & 35.354 & 0.622 & 7.19(--5) & 5.31(--6) &  & 7170.50\,{\AA} & 0.086 & 0.003 & 1.84(--7) & 2.17(--8) \\ 
 & 15.56\,{\micron} & 113.924 & 6.646 & 7.13(--5) & 4.16(--6) &  & 7262.70\,{\AA} & 0.070 & 0.004 & 1.70(--7) & 2.17(--8) \\ 
 &  &  &  & {\bf 7.17(--5)} & {\bf 2.79(--6)} &  &  &  &  & {\bf 1.80(--7)} & {\bf 5.23(--9)} \\ 
Ne$^{3+}$ & 2422/25\,{\AA} & 48.638 & 1.998 & 8.37(--6) & 6.12(--7) & Ar$^{4+}$ & 6434.73\,{\AA} & 0.207 & 0.008 & 4.73(--8) & 5.13(--9) \\ 
 & 4714.17\,{\AA} & 0.195 & 0.018 & 9.39(--6) & 1.32(--6) &  & 13.10\,{\micron} & 1.684 & 0.118 & 4.71(--8) & 3.52(--9) \\ 
 & 4715.61\,{\AA} & 0.074 & 0.010 & 1.11(--5) & 2.70(--6) &  &  &  &  & {\bf 4.72(--8)} & {\bf 2.90(--9)} \\ 
 & 4724.17\,{\AA} & 0.202 & 0.009 & 8.03(--6) & 1.67(--6) & K$^{3+}$ & 6795.10\,{\AA} & 0.031 & 0.002 & 6.13(--9) & 1.86(--9) \\ 
 & 4725.64\,{\AA} & 0.158 & 0.009 & 6.68(--6) & 1.43(--6) &  & 6101.79\,{\AA} & 0.115 & 0.004 & 4.89(--9) & 1.46(--9) \\ 
 &  &  &  & {\bf 8.71(--6)} & {\bf 1.64(--6)} &  &  &  &  & {\bf 5.51(--9)} & {\bf 8.73(--10)} \\ 
Ne$^{4+}$ & 14.32\,{\micron} & 44.563 & 2.690 & 3.52(--6) & 2.34(--7) & Fe$^{2+}$ & 4658.05\,{\AA} & 0.184 & 0.012 & 3.72(--8) & 2.65(--9) \\ 
 & 24.32\,{\micron} & 25.480 & 1.491 & 3.43(--6) & 3.89(--7) &  & 4751.59\,{\AA} & 0.069 & 0.009 & 7.48(--8) & 1.03(--8) \\ 
 &  &  &  & {\bf 3.50(--6)} & {\bf 2.00(--7)} &  &  &  &  & {\bf 5.60(--8)} & {\bf 2.66(--8)} \\ 
Mg$^{3+}$ & 4.49\,{\micron} & 6.128 & 0.147 & 2.25(--6) & 6.92(--8) & Kr$^{2+}$ & 6826.70\,{\AA} & 0.026 & 0.006 & 1.82(--9) & 4.07(--10) \\ 
Mg$^{4+}$ & 13.52\,{\micron} & 0.370 & 0.026 & 7.93(--7) & 5.75(--8) & Kr$^{3+}$ & 5867.74\,{\AA} & 0.146 & 0.005 & 2.37(--9) & 9.36(--11) \\ 
Si$^{+}$ & 34.84\,{\micron} & 1.427 & 0.190 & 9.18(--7) & 1.55(--7) & Rb$^{3+}$ & 5759.43\,{\AA} & 0.014 & 0.004 & 4.13(--10) & 1.22(--10) \\ 
P$^{+}$ & 7875.99\,{\AA} & 0.067 & 0.004 & 3.01(--8) & 7.83(--9) & Xe$^{2+}$ & 5846.67\,{\AA} & 0.024 & 0.007 & 4.02(--10) & 1.25(--10) \\ 
P$^{2+}$ & 17.89\,{\micron} & 0.724 & 0.074 & 1.50(--7) & 1.61(--8) & Xe$^{3+}$ & 5709.11\,{\AA} & 0.030 & 0.006 & 1.14(--9) & 2.35(--10) \\ 
\midrule
He$^{+}$  &  4026.20\,{\AA}  & 1.936 & 0.081 &  8.46(--2)  &  5.17(--3)   & C$^{2+}$  &  4267.18\,{\AA}  & 1.183 & 0.035 &  1.14(--3)  &  4.61(--5)   \\ 
  &  4387.93\,{\AA}  & 0.566 & 0.023 &  9.34(--2)  &  5.68(--3)   &   &  5342.43\,{\AA}  & 0.056 & 0.009 &  1.01(--3)  &  1.65(--4)   \\ 
  &  4471.47\,{\AA}  & 3.799 & 0.033 &  7.47(--2)  &  2.82(--3)   &   &  6151.27\,{\AA}  & 0.049 & 0.005 &  1.12(--3)  &  1.15(--4)   \\ 
  &  4713.14\,{\AA}  & 0.582 & 0.02 &  8.47(--2)  &  5.47(--3)   &   &  6256.52\,{\AA}  & 0.015 & 0.002 &  1.20(--3)  &  1.85(--4)   \\ 
  &  4921.93\,{\AA}  & 1.018 & 0.013 &  7.58(--2)  &  2.88(--3)   &   &  6461.95\,{\AA}  & 0.103 & 0.003 &  9.74(--4)  &  4.30(--5)   \\ 
  &  5047.74\,{\AA}  & 0.134 & 0.01 &  6.00(--2)  &  4.96(--3)   &   &  6578.05\,{\AA}  & 0.453 & 0.014 &  8.72(--4)  &  4.00(--5)   \\ 
  &  5875.62\,{\AA}  & 11.723 & 0.135 &  7.90(--2)  &  3.36(--3)   &   &    &    &    &  {\bf 1.05(--3)}  &  {\bf 1.23(--4)}   \\ 
  &  6678.15\,{\AA}  & 3.127 & 0.053 &  7.77(--2)  &  3.17(--3)   & C$^{3+}$  &  4067.94\,{\AA}  & 0.492 & 0.116 &  5.10(--4)  &  1.20(--4)   \\ 
  &  7281.35\,{\AA}  & 0.639 & 0.015 &  6.66(--2)  &  2.53(--3)   &   &  4070.31\,{\AA}  & 0.537 & 0.049 &  7.40(--4)  &  7.10(--5)   \\ 
  &    &    &    &  {\bf 7.74(--2)}  &  {\bf 4.00(--3)}   &   &  4647.42\,{\AA}  & 0.458 & 0.011 &  7.74(--4)  &  2.47(--5)   \\ 
He$^{2+}$  &  1640.42\,{\AA}  & 235.081 & 4.276 &  3.02(--2)  &  5.64(--4)   &   &  4650.25\,{\AA}  & 0.296 & 0.015 &  8.35(--4)  &  4.34(--5)   \\ 
  &  2733.30\,{\AA}  & 7.631 & 0.725 &  3.03(--2)  &  2.88(--3)   &   &  4651.47\,{\AA}  & 0.120 & 0.014 &  1.01(--3)  &  1.21(--4)   \\ 
  &  4338.67\,{\AA}  & 0.964 & 0.018 &  3.45(--2)  &  6.64(--4)   &   &    &    &    &  {\bf 7.75(--4)}  &  {\bf 1.82(--4)}   \\ 
  &  4541.59\,{\AA}  & 1.25 & 0.015 &  3.19(--2)  &  4.10(--4)   & C$^{4+}$  &  4658.30\,{\AA}  & 0.487 & 0.014 &  1.11(--4)  &  4.29(--6)   \\ 
  &  4685.71\,{\AA}  & 35.847 & 0.173 &  2.98(--2)  &  1.92(--4)   &    &  7725.90\,{\AA}  & 0.241 & 0.007 &  9.13(--5)  &  3.48(--6)   \\ 
  &  5411.52\,{\AA}  & 2.785 & 0.029 &  3.03(--2)  &  3.40(--4)   &   &    &    &    &  {\bf 1.01(--4)}  &  {\bf 1.40(--5)}   \\ 
  &  6223.82\,{\AA}  & 0.113 & 0.005 &  3.22(--2)  &  1.36(--3)   &  &  &  &  &  &  \\ 
  &  6310.80\,{\AA}  & 0.125 & 0.01 &  2.95(--2)  &  2.41(--3)   &  &  &  &  &  &  \\ 
  &  6406.38\,{\AA}  & 0.156 & 0.004 &  3.04(--2)  &  8.52(--4)   &  &  &  &  &  &  \\ 
  &  6560.10\,{\AA}  & 5.875 & 0.691 &  3.63(--2)  &  4.28(--3)   &  &  &  &  &  &  \\ 
  &  7177.50\,{\AA}  & 0.395 & 0.009 &  2.90(--2)  &  6.96(--4)   &  &  &  &  &  &  \\ 
  &  8236.79\,{\AA}  & 0.752 & 0.022 &  2.83(--2)  &  8.24(--4)   &  &  &  &  &  &  \\ 
  &    &    &    &  {\bf 3.11(--2)}  &  {\bf 1.29(--3)}   &  &  &  &  &  &  \\ 
\midrule
\end{tabularx}
\end{table*}

\begin{table*}
\centering
\renewcommand{\arraystretch}{0.70}
\caption{
\label{T-cloudy}
Comparison of the observed line intensities, band fluxes, and flux densities between the {\sc Cloudy} 
model and the observation. The band width for the integrated band flux is as follows; 
0.139, 0.162, 0.251, 0.260, 0.626, 1.042, 20.17, and 39.9\,{\micron} in WFC-$g$, WFCAM-$J$/$H$/$K$, 
{\sl WISE}-W1/W2, and FIS65/90, respectively. 
0.10\,{\micron} in IRS-11, 0.30\,{\micron} in IRS-09/10/12, 0.8\,{\micron} in IRS-01/08, 
1.0\,{\micron} in IRS-23/24/25, and 0.50\,{\micron} in the other IRS bands, respectively.}
\begin{tabularx}{\textwidth}{@{}@{\extracolsep{\fill}}
c@{\hspace{8pt}}l@{\hspace{-3pt}}D{.}{.}{-1}@{\hspace{-3pt}}D{.}{.}{-1}@{\hspace{5pt}}
c@{\hspace{8pt}}l@{\hspace{-3pt}}D{.}{.}{-1}@{\hspace{-3pt}}D{.}{.}{-1}@{\hspace{5pt}}
c@{\hspace{8pt}}l@{\hspace{-3pt}}D{.}{.}{-1}@{\hspace{-3pt}}D{.}{.}{-1}@{}}
\midrule
$\lambda_{\rm lab.}$&Line&\multicolumn1c{$I$(Obs)}&\multicolumn1c{$I$(Model)}&
$\lambda_{\rm lab.}$&Line&\multicolumn1c{$I$(Obs)}&\multicolumn1c{$I$(Model)}&
$\lambda_{\rm lab.}$&Line&\multicolumn1c{$I$(Obs)}&\multicolumn1c{$I$(Model)}\\
\midrule
1548/51\,{\AA}   &   {\civ}   & 624.365 & 850.810 &   5197.90\,{\AA}   &    {\NI}         & 0.345 & 0.512 &   8437.95\,{\AA}   &    {\hi}            & 0.331 & 0.308 \\ 
1640.42\,{\AA}   &   {\heii}   & 235.081 & 234.814 &   5200.26\,{\AA}   &    {\NI}          & 0.257 & 0.369 &   8467.25\,{\AA}   &    {\hi}            & 0.404 & 0.364 \\ 
1749/54\,{\AA}  &  N\,{\sc iii}$]$  & 19.214 & 10.289 &   5411.52\,{\AA}   &    {\heii}          & 2.785 & 2.589 &   8502.48\,{\AA}   &    {\hi}            & 0.424 & 0.434 \\ 
1906/09\,{\AA}   &   {\ciii}$]$   & 1026.996 & 1014.349 &   5517.72\,{\AA}   &    {\cliii}       & 0.298 & 0.377 &   8545.38\,{\AA}   &    {\hi}          & 0.556 & 0.524 \\ 
2325-28\,{\AA}   &   $[${\cii}$]$   & 87.207 & 95.408 &   5537.89\,{\AA}   &    {\cliii}       & 0.436 & 0.440 &   8578.69\,{\AA}   &    {\clii}        & 0.139 & 0.151 \\ 
2422/25\,{\AA}   &   {\neiv}       & 48.638 & 54.425 &   5577.34\,{\AA}   &    {\oi}         & 0.111 & 0.205 &   8598.39\,{\AA}   &    {\hi}            & 0.674 & 0.643 \\ 
2733.30\,{\AA}   &    {\heii}    & 7.631 & 7.400 &   5754.64\,{\AA}   &    {\nii}         & 1.243 & 1.151 &   8665.02\,{\AA}   &    {\hi}            & 0.858 & 0.803 \\ 
3721.94\,{\AA}   &    {\hi}          & 2.750 & 1.931 &   5875.62\,{\AA}   &    {\hei}           & 11.723 & 10.754 &   8727.12\,{\AA}   &    [C\,{\sc i}]          & 0.241 & 0.152 \\ 
3726.03\,{\AA}   &    {\oii}    & 75.920 & 74.124 &   5977.03\,{\AA}   &    {\heii}          & 0.040 & 0.040 &   8746.89\,{\AA}   &    {\heii}          & 0.024 & 0.024 \\ 
3728.81\,{\AA}   &    {\oii}    & 42.811 & 39.029 &   6004.72\,{\AA}   &    {\heii}          & 0.056 & 0.046 &   8750.47\,{\AA}   &    {\hi}            & 1.072 & 1.021 \\ 
3734.37\,{\AA}   &    {\hi}          & 2.969 & 2.402 &   6036.78\,{\AA}   &    {\heii}        & 0.063 & 0.053 &   8859.15\,{\AA}   &    {\heii}          & 0.043 & 0.031 \\ 
3750.15\,{\AA}   &    {\hi}          & 2.532 & 3.049 &   6074.19\,{\AA}   &    {\heii}          & 0.079 & 0.061 &   8862.78\,{\AA}   &    {\hi}            & 1.350 & 1.327 \\ 
3770.63\,{\AA}   &    {\hi}          & 3.439 & 3.959 &   6101.79\,{\AA}   &    {\kiv}        & 0.115 & 0.117 &   8929.11\,{\AA}   &    {\heii}          & 0.041 & 0.035 \\ 
3797.90\,{\AA}   &    {\hi}          & 5.243 & 5.315 &   6118.26\,{\AA}   &    {\heii}          & 0.083 & 0.071 &   9011.21\,{\AA}   &    {\heii}          & 0.014 & 0.041 \\ 
3835.38\,{\AA}   &    {\hi}          & 7.654 & 7.322 &   6170.60\,{\AA}   &    {\heii}        & 0.084 & 0.084 &   9068.60\,{\AA}   &    {\siii}        & 11.272 & 13.391 \\ 
3868.76\,{\AA}   &    {\neiii}    & 117.595 & 147.475 &   6233.82\,{\AA}   &    {\heii}          & 0.113 & 0.100 &   9108.54\,{\AA}   &    {\heii}          & 0.056 & 0.048 \\ 
3889.05\,{\AA}   &    {\hi}+{\hei}          & 19.620 & 17.056 &   6300.30\,{\AA}   &    {\oi}         & 7.974 & 9.413 &   9123.60\,{\AA}   &    {\clii}        & 0.040 & 0.039 \\ 
3964.73\,{\AA}   &    {\hei}    & 0.610 & 0.749 &   6310.80\,{\AA}   &    {\heii}        & 0.125 & 0.120 &   3.04\,{\micron}   &    {\hi} 5-10    & 0.328 & 0.486 \\ 
3967.47\,{\AA}   &    {\neiii}    & 35.354 & 44.447 &   6312.10\,{\AA}   &    {\siii}       & 1.249 & 1.574 &   3.10\,{\micron}   &    {\heii}    & 1.028 & 1.187 \\ 
3970.07\,{\AA}   &    {\hi}          & 17.820 & 15.940 &   6363.78\,{\AA}   &    {\oi}         & 2.742 & 3.002 &   3.74\,{\micron}   &    {\hi} 5-8    & 0.970 & 0.972 \\ 
4025.60\,{\AA}   &    {\heii}        & 0.520 & 0.360 &   6406.38\,{\AA}   &    {\heii}          & 0.156 & 0.146 &   4.05\,{\micron}   &    {\hi} 4-5    & 7.781 & 7.488 \\ 
4026.20\,{\AA}   &    {\hei}    & 1.936 & 1.591 &   6434.73\,{\AA}   &    {\arv}    & 0.207 & 0.507 &   4.49\,{\micron}   &    {\mgiv}    & 6.128 & 6.629 \\ 
4059.90\,{\AA}   &    {\fiv}    & 0.072 & 0.125 &   6548.04\,{\AA}   &    {\nii}        & 17.678 & 16.077 &   4.65\,{\micron}   &    {\hi} 5-7    & 1.147 & 1.480 \\ 
4068.60\,{\AA}   &    {\sii}    & 1.961 & 2.311 &   6560.00\,{\AA}   &    {\heii}    & 5.875 & 4.527 &   5.61\,{\micron}   &    {\mgv}    & 4.509 & 3.870 \\ 
4076.35\,{\AA}   &    {\sii}    & 0.823 & 0.764 &   6562.80\,{\AA}   &    {\hi}    & 286.007 & 286.491 &   6.91\,{\micron}   &    H$_{2}$ $v=0-0$ S(5)    & 0.762 & 0.554 \\ 
4101.73\,{\AA}   &    {\hi}           & 29.172 & 25.989 &   6583.46\,{\AA}   &    {\nii}        & 53.335 & 47.442 &   6.98\,{\micron}   &    {\arii}    & 0.867 & 0.221 \\ 
4120.81\,{\AA}   &    {\hei}           & 0.292 & 0.180 &   6678.15\,{\AA}   &    {\hei}           & 3.127 & 2.758 &   8.99\,{\micron}   &    {\ariii}    & 6.293 & 5.005 \\ 
4143.76\,{\AA}   &    {\hei}           & 0.285 & 0.265 &   6683.20\,{\AA}   &    {\heii}          & 0.262 & 0.227 &   9.68\,{\micron}   &    H$_{2}$ $v=0-0$ S(3)    & 1.017 & 0.768 \\ 
4199.83\,{\AA}   &    {\heii}          & 0.847 & 0.603 &   6716.44\,{\AA}   &    {\sii}        & 3.263 & 3.519 &   10.51\,{\micron}   &    {\siv}    & 58.937 & 52.114 \\ 
4338.67\,{\AA}   &    {\heii}          & 0.964 & 0.812 &   6730.81\,{\AA}   &    {\sii}         & 5.588 & 6.113 &   11.76\,{\micron}   &    {\cliv}    & 0.418 & 0.590 \\ 
4340.46\,{\AA}   &    {\hi}           & 49.523 & 47.064 &   6795.10\,{\AA}   &    {\kiv}       & 0.031 & 0.025 &   12.28\,{\micron}   &    H$_{2}$ $v=0-0$ S(2)    & 0.189 & 0.128 \\ 
4363.21\,{\AA}   &    {\oiii}    & 14.442 & 14.558 &   7065.71\,{\AA}   &    {\hei}         & 5.210 & 5.131 &   12.37\,{\micron}   &    {\hi}    & 1.043 & 1.027 \\ 
4387.93\,{\AA}   &    {\hei}           & 0.566 & 0.423 &   7135.80\,{\AA}   &    {\ariii}       & 8.470 & 7.495 &   12.82\,{\micron}   &    {\neii}    & 4.006 & 2.295 \\ 
4471.47\,{\AA}   &    {\hei}           & 3.799 & 3.457 &   7170.50\,{\AA}   &    {\ariv}        & 0.086 & 0.094 &   13.10\,{\micron}   &    {\arv}    & 1.684 & 3.404 \\ 
4541.59\,{\AA}   &    {\heii}          & 1.250 & 1.120 &   7177.50\,{\AA}   &    {\heii}        & 0.395 & 0.384 &   13.43\,{\micron}   &    {\fv}    & 0.309 & 0.264 \\ 
4658.05\,{\AA}   &    {\feiii}       & 0.184 & 0.208 &   7262.70\,{\AA}   &    {\ariv}        & 0.070 & 0.058 &   13.52\,{\micron}   &    {\mgv}    & 0.370 & 0.317 \\ 
4685.71\,{\AA}   &    {\heii}          & 35.847 & 35.073 &   7281.35\,{\AA}   &    {\hei}           & 0.639 & 0.692 &   14.32\,{\micron}   &    {\nev}    & 44.563 & 62.770 \\ 
4711.37\,{\AA}   &    {\ariv}        & 1.866 & 2.273 &   7318.92\,{\AA}   &    {\oii}        & 5.782 & 5.957 &   15.56\,{\micron}   &    {\neiii}    & 113.924 & 97.124 \\ 
4713.14\,{\AA}   &    {\hei}           & 0.582 & 0.541 &   7329.66\,{\AA}   &    {\oii}        & 5.022 & 4.753 &   17.89\,{\micron}   &    {\piii}    & 0.724 & 0.767 \\ 
4714-26\,{\AA}   &    {\neiv}        & 0.629 & 0.512 &   7530.80\,{\AA}   &    {\cliv}        & 0.180 & 0.125 &   18.72\,{\micron}   &    {\siii}    & 15.049 & 12.672 \\ 
4740.16\,{\AA}   &    {\ariv}        & 2.124 & 2.397 &   7751.10\,{\AA}   &    {\ariii}       & 2.069 & 1.809 &   19.07\,{\micron}   &    {\hi} 7-8    & 0.522 & 0.408 \\ 
4751.59\,{\AA}   &    {\feiii}       & 0.069 & 0.038 &   7875.99\,{\AA}   &    {\pii}         & 0.067 & 0.057 &   20.31\,{\micron}   &    {\cliv}    & 0.307 & 0.366 \\ 
4861.33\,{\AA}   &    {\hi}    & 100.000 & 100.000 &   8236.79\,{\AA}   &    {\heii}          & 0.752 & 0.731 &   21.83\,{\micron}   &    {\ariii}    & 0.566 & 0.344 \\ 
4921.93\,{\AA}   &    {\hei}           & 1.018 & 0.911 &   8333.78\,{\AA}   &    {\hi}            & 0.189 & 0.140 &   24.32\,{\micron}   &    {\nev}    & 25.480 & 43.568 \\ 
4931.23\,{\AA}   &    {\oiii}       & 0.126 & 0.151 &   8345.47\,{\AA}   &    {\hi}          & 0.158 & 0.157 &   25.90\,{\micron}   &    {\oiv}    & 282.948 & 204.883 \\ 
4959.00\,{\AA}   &    {\oiii}    & 357.906 & 368.237 &   8359.00\,{\AA}   &    {\hi}            & 0.196 & 0.177 &   33.49\,{\micron}   &    {\siii}    & 5.215 & 4.294 \\ 
5007.00\,{\AA}   &    {\oiii}    & 1048.258 & 1108.393 &   8361.73\,{\AA}   &    {\hei}           & 0.071 & 0.068 &   34.84\,{\micron}   &    {\Siii}    & 1.427 & 1.951 \\ 
5015.68\,{\AA}   &    {\hei}           & 1.429 & 1.928 &   8374.48\,{\AA}   &    {\hi}            & 0.201 & 0.201 &   36.02\,{\micron}   &    {\neiii}    & 6.710 & 7.699 \\ 
5047.74\,{\AA}   &    {\hei}           & 0.134 & 0.168 &   8392.40\,{\AA}   &    {\hi}            & 0.239 & 0.229 &    &    &    &  \\ 
5191.82\,{\AA}   &    {\ariii}       & 0.084 & 0.098 &   8413.32\,{\AA}   &    {\hi}            & 0.271 & 0.265 &    &    &    &   \\ 
\midrule
$\lambda_{\rm c}$&Band&\multicolumn1c{$I$(Obs)}&\multicolumn1c{$I$(Model)}&
$\lambda_{\rm c}$&Band&\multicolumn1c{$I$(Obs)}&\multicolumn1c{$I$(Model)}&
$\lambda_{\rm c}$&Band&\multicolumn1c{$I$(Obs)}&\multicolumn1c{$I$(Model)}\\
\midrule
0.464\,{\micron}  &  WFC-$g$  & 3720.229 & 3824.605 &  9.500\,{\micron}   &   IRS06   & 74.723 & 72.805 &   27.00\,{\micron}   &   IRS17   & 61.671 & 60.972 \\ 
1.235\,{\micron}   &   WFCAM-$J$  & 72.463 & 85.611 &   10.00\,{\micron}   &   IRS07   & 62.611 & 67.794 &   27.50\,{\micron}   &   IRS18   & 60.111 & 58.978 \\ 
1.662\,{\micron}   &   WFCAM-$H$   & 49.660 & 49.122 &   11.30\,{\micron}   &   IRS08   & 213.509 & 332.018 &   28.00\,{\micron}   &   IRS19   & 58.459 & 57.181 \\ 
2.159\,{\micron}   &   WFCAM-$K$   & 51.576 & 39.521 &   12.00\,{\micron}   &   IRS09   & 59.355 & 74.988 &   28.50\,{\micron}   &   IRS20   & 56.770 & 55.198 \\ 
3.353\,{\micron}   &   $WISE$-W1   & 100.892 & 78.021 &   12.60\,{\micron}   &   IRS10   & 67.810 & 99.365 &   29.00\,{\micron}   &   IRS21   & 54.978 & 53.530 \\ 
4.603\,{\micron}   &   $WISE$-W2   & 119.646 & 58.326 &   13.30\,{\micron}   &   IRS11   & 15.893 & 17.488 &   29.50\,{\micron}   &   IRS22   & 53.779 & 51.828 \\ 
6.200\,{\micron}   &   IRS01   & 187.986 & 245.140 &   13.80\,{\micron}   &   IRS12   & 46.911 & 49.176 &   30.50\,{\micron}   &   IRS23   & 102.652 & 97.124 \\ 
7.000\,{\micron}   &   IRS02   & 104.256 & 101.323 &   15.00\,{\micron}   &   IRS13   & 79.516 & 85.157 &   31.50\,{\micron}   &   IRS24   & 95.304 & 90.993 \\ 
7.500\,{\micron}   &   IRS03   & 180.350 & 307.521 &   23.50\,{\micron}   &   IRS14   & 82.930 & 82.879 &   32.50\,{\micron}   &   IRS25   & 91.513 & 85.800 \\ 
8.000\,{\micron}   &   IRS04   & 274.346 & 272.462 &   25.00\,{\micron}   &   IRS15   & 73.460 & 69.834 &   66.70\,{\micron}   &   FIS65   & 358.193 & 358.538 \\ 
8.500\,{\micron}   &   IRS05   & 174.588 & 183.329 &   26.50\,{\micron}   &   IRS16   & 63.824 & 63.227 &   89.20\,{\micron}   &   FIS90   & 278.541 & 213.408 \\ 
\midrule
$\lambda_{\rm c}$&Band&\multicolumn1c{$F_{\nu}$(Obs) (Jy)}&\multicolumn1c{$F_{\nu}$(Model) (Jy)}&
$\nu_{\rm c}$&Band&\multicolumn1c{$F_{\nu}$(Obs)}&\multicolumn1c{$F_{\nu}$(Model)}\\
\midrule
5.460\,{\micron}   &   IR01   & 0.191 & 0.110 &   43\,GHz   &   Radio01   & 0.090 & 0.114 &      &      &      &   \\ 
10.40\,{\micron}   &   IR02   & 0.981 & 1.150 &   30\,GHz   &   Radio02   & 0.086 & 0.119 &      &      &      &   \\ 
14.80\,{\micron}   &   IR03   & 2.580 & 2.762 &   20\,GHz   &   Radio03   & 0.119 & 0.144 &      &      &      &   \\ 
20.00\,{\micron}   &   IR04   & 5.631 & 5.386 &   6.0\,GHz   &   Radio04   & 0.109 & 0.153 &      &      &      &   \\ 
25.00\,{\micron}   &   IR05   & 6.817 & 6.578 &   5.0\,GHz   &   Radio05   & 0.121 & 0.124 &      &      &      &   \\ 
30.00\,{\micron}   &   IR06   & 7.057 & 6.791 &   2.7\,GHz   &   Radio06   & 0.180 & 0.142 &      &      &      &   \\ 
140.0\,{\micron}   &   IR07   & 2.940 & 1.062 &   1.4\,GHz   &   Radio07   & 0.108 & 0.150 &      &      &      &   \\ 
\midrule
\end{tabularx}
\end{table*}

\label{lastpage}
\bsp  

\end{document}